\documentclass[journal,twocolumn]{IEEEtran}

\usepackage{headerEPI}
\usepackage{newcommandsEPI}

\newcommand{\modifica}[1]{#1}               

\begin{document}

\title{The multimode conditional quantum Entropy Power Inequality and the squashed entanglement of the multimode extreme bosonic Gaussian channels}

\author{
    Alessandro~Falco\orcidlink{0009-0004-3586-9491} and Giacomo~De~Palma\orcidlink{0000-0002-5064-8695}
    \thanks{Alessandro Falco was with the Scuola Normale Superiore, 56127 Pisa, Italy. He is now with the Institute for Quantum Inspired and Quantum Optimization, Technische Universität Hamburg, Germany and also with the Quantum Research Centre, Technology Innovation Institute, Abu Dhabi, United Arab Emirates (e-mail: \href{mailto:alessandro.falco@sns.it}{alessandro.falco@sns.it}).}
    \thanks{Giacomo De Palma is with the Department of Mathematics, University of Bologna, Bologna, Italy (e-mail: \href{mailto:giacomo.depalma@unibo.it}{giacomo.depalma@unibo.it}).}
    \thanks{This paper was presented in part at Quantum Information Processing (QIP) 2024 and Beyond IID in Information Theory 12 (2024).}
}

\maketitle

\begin{abstract}
    We prove the multimode conditional quantum Entropy Power Inequality for bosonic quantum systems. This inequality determines the minimum conditional von Neumann entropy of the output of the most general linear mixing of bosonic quantum modes among all the input states of the modes with given conditional entropies. Bosonic quantum systems constitute the mathematical model for the electromagnetic radiation in the quantum regime, which provides the most promising platform for quantum communication and quantum key distribution.
We apply our multimode conditional quantum Entropy Power Inequality to determine new lower bounds to the squashed entanglement of a large family of bosonic Gaussian states. The squashed entanglement is one of the main entanglement measures in quantum communication theory, providing the best known upper bound to the distillable key.
Exploiting this result, we determine a new lower bound to the squashed entanglement of the multimode bosonic Gaussian channels that are extreme, \emph{i.e.}, that cannot be decomposed as a non-trivial convex combination of quantum channels. The squashed entanglement of a quantum channel provides an upper bound to its secret-key capacity, \emph{i.e.}, the capacity to generate a secret key shared between the sender and the receiver.
Lower bounds to the squashed entanglement are notoriously hard to prove. Our results contribute to break this barrier and will stimulate further research in the field of quantum communication with bosonic quantum systems.
\end{abstract}


\section{Introduction}\label{sec:intro}

\IEEEPARstart{T}{he} Shannon differential entropy of a random variable $X$ with values in $\mathbb{R}^n$ and probability density $p(\mathbf{x})\,\mathrm{d}^nx$ is \cite{cover2006elements}
\begin{equation}
\modifica{
    S(X) := -\int_{\mathbb{R}^n}\ln(p(\mathbf{x}))\,\mathrm{d}p(\mathbf{x})\,,
}
\end{equation}
and quantifies the noise or the information contained in $X$.
The classical Entropy Power Inequality (EPI) \cite{dembo1991information,stam1959some,shannon1948mathematical} determines the minimum Shannon differential entropy of the sum of two independent random variables $X_1$ and $X_2$ with values in $\mathbb{R}^n$ and given Shannon differential entropies:
\begin{equation}\label{eq:cEPI}
\modifica{
    \exp(\frac{2\,S(X_1+X_2)}{n}) \ge \exp(\frac{2\,S(X_1)}{n}) + \exp(\frac{2\,S(X_2)}{n})\,,
}
\end{equation}
and is a fundamental element of classical information theory \cite{cover2006elements}.
For any random variable $X$ with values in $\mathbb{R}^n$ and any invertible $M\in\mathbb{R}^{n\times n}$ we have
\begin{equation}\label{eq:SM}
\modifica{    
    S(MX) = S(X) + \frac{1}{2}\ln(\det(M))\,.
}
\end{equation}
Let $X_1,\,\ldots,\,X_K$ be independent random variables taking values in $\mathbb{R}^n$, and let $M_1,\,\ldots,\,M_K\in\mathbb{R}^{n\times n}$.
Let us consider the linear combination
\begin{equation}\label{eq:linc}
    Y := \sum_{k=1}^K M_k\,X_k\,.
\end{equation}
The property \eqref{eq:SM} implies the following multimode generalization of the EPI \eqref{eq:cEPI}:
\modifica{
\begin{equation}\label{eq:mEPI}   
    \exp(\frac{2\,S(Y)}{n}) 
    \ge
    \sum_{k=1}^K \left|\det(M_k)\right|^\frac{2}{n} \exp(\frac{2\,S(X_k)}{n})\,.
\end{equation}
}
Let now $Z$ be an arbitrary random variable, and let $X_1,\,\ldots,\,X_K$ be random variables conditionally independent given $Z$ and with values in $\mathbb{R}^n$.
The entropy of the random variable $X$ conditioned on the random variable $Z$ is the expectation value with respect to $Z$ of the entropy of $X$ given the value of $Z$:
\begin{equation}\label{eq:X|Z}
    S(X|Z) = \int S(X|Z=z)\,\mathrm{d}p(z)\,.
\end{equation}
Jensen's inequality implies the following conditional version of the EPI \eqref{eq:mEPI}:
\modifica{
\begin{equation}\label{eq:mcEPI}
    \exp(\frac{2\,S(Y|Z)}{n})
    \ge
    \sum_{k=1}^K\left|\det(M_k)\right|^\frac{2}{n}\exp(\frac{2\,S(X_k|Z)}{n})\,.
\end{equation}
}

The first result of this paper is to prove a quantum generalization of the EPI \eqref{eq:mcEPI}.
The quantum counterpart of probability measures are quantum states, which are positive-semidefinite linear operators with unit trace acting on the Hilbert space of a quantum system.
The counterpart of the probability measures on $\mathbb{R}^n$ with even $n$ are the states of a bosonic quantum system with \modifica{$m=n/2$} modes, and the counterpart of the coordinates of $\mathbb{R}^n$ are self-adjoint linear operators called quadratures.
Bosonic quantum systems \modifica{\cite{holevo2019quantum,holevo2015gaussian, ptaszynski2023quantum, lee2023quantum, lami2019extendibility}}, model electromagnetic waves in the quantum regime.
Electromagnetic waves traveling through optical fibers or free space provide the most promising platform for quantum communication and quantum key distribution \cite{weedbrook2012gaussian}.
Bosonic quantum systems then play a key role in quantum communication and quantum cryptography and provide the model to determine the maximum communication and key distribution rates achievable in principle by quantum communication devices.
The quantum counterpart of the linear combination \eqref{eq:linc} is a symplectic unitary operation, which implements a symplectic linear redefinition of the quadratures.
Symplectic unitary operations are the fundamental elements of quantum optics and model the attenuation and the amplification of electromagnetic signals.
The quantum counterpart of the Shannon differential entropy is the von Neumann entropy of a quantum state \cite{holevo2019quantum,wilde2017quantum}
\begin{equation}
\modifica{
    S(\oprho) := -\mathrm{tr}\left[\oprho\ln(\oprho)\right]\,.
}
\end{equation}

Let $A_1,\,\ldots,\,A_K$ be $m$-mode bosonic quantum systems such that each $A_k$ has quadratures $\opR^{(k)}_1,\,\ldots,\,\opR^{(k)}_{2m}$.
Let $M_1$,\,$\ldots$,\,$M_K$$\in\mathbb{R}^{2m\times2m}$ such that the rectangular matrix $\left(M_1\,\ldots\,M_K\right)$ constitutes the first $2\,m$ rows of the symplectic matrix $S\in\mathrm{Sp}(2K m,\mathbb{R})$.
Let $\opU$ be the unitary operator that implements $S$, \emph{i.e.}, such that
\begin{equation}
    \opU^\dag\,\opR^{(1)}_i\,\opU = \sum_{k=1}^K \sum_{j=1}^{2m} (M_k)_{ij}\,\opR^{(k)}_j\,,
    \quad i=1,\,
    \ldots,\,2\,m\,.
\end{equation}
Let $M$ be an arbitrary quantum system.
Let us consider a joint quantum input state $\oprho_{A_1\ldots A_KM}$ such that $A_1\ldots A_K$ are conditionally independent given $M$, \emph{i.e.}, such that
\begin{equation}\label{eq:1.10}
    S(A_1\ldots A_K|M) = \sum_{k=1}^K S(A_k|M)\,,
\end{equation}
where
\begin{equation}
    S(X|M):=S(XM)-S(M)
\end{equation}
is the conditional quantum entropy,
and let us define the output state
\begin{equation}
    \oprho_{BM} = \mathrm{Tr}_{A_2\ldots A_K}\left[\left(\opU\otimes\I_M\right)\oprho_{A_1\ldots A_KM}\left(\opU^\dag\otimes\I_M\right)\right]\,,
\end{equation}
where we denote with $B$ the bosonic quantum system $A_1$ after the application of $\opU$.
The multimode conditional quantum EPI determines a lower bound to the conditional quantum entropy of the output among all the input states as above with given conditional quantum entropies:
\begin{equation}\label{eq:EPIm}
\modifica{    
    \exp(\frac{S(B|M)}{m}) \ge \sum_{k=1}^K\left|\det(M_k)\right|^\frac{1}{m}\exp(\frac{S(A_k|M)}{m})\,.
}
\end{equation}

The quantum EPI has been first proved in the unconditional case (\emph{i.e.}, when $M$ is not present) when $\opU$ implements the $50:50$ beamsplitter \cite{koenig2014entropy,koenig2016corrections} and has later been extended to any beamsplitter and amplifier \cite{de2014generalization} and to the most general symplectic unitary transformation \cite{de2015multimode,depalma2017gaussian}.
The conditional generalization of the quantum EPI for the beamsplitter and the amplifier has been conjectured in \cite{koenig2015conditional} and proved in \cite{de2018conditional,huber2018conditional}.
The conditional generalization of the EPI for arbitrary symplectic unitary transformations was still an open conjecture, which is settled by this paper.
While in the classical setting the multimode conditional EPI \eqref{eq:mcEPI} is an easy consequence of the EPI \eqref{eq:cEPI}, in the quantum setting the generalization is highly nontrivial and requires to rebuild the proof from scratch.

\vspace{\baselineskip}

We apply the multimode conditional quantum EPI to determine lower bounds to the squashed entanglement of any bipartite bosonic Gaussian state whose covariance matrix has half of the symplectic eigenvalues equal to \modifica{$1/2$}.
Furthermore, we apply the above bounds to determine lower bounds to the squashed entanglement of any multimode extreme bosonic Gaussian channel, \emph{i.e.}, any multimode bosonic Gaussian channel that cannot be expressed as a nontrivial convex combination of quantum channels.

The squashed entanglement of a bipartite quantum state $\oprho_{AB}$ is the infimum over all its possible extensions $\oprho_{ABR}$ of half of the quantum mutual information between the quantum systems $A$ and $B$ conditioned on the quantum system $R$ \cite{tucci1999quantum,tucci2000separability,tucci2000entanglement,tucci2001relaxation,tucci2001entanglement,tucci2002entanglement,christandl2004squashed,brandao2011faithful,seshadreesan2015renyi}:
\begin{equation}\label{eq:defEsq}
    E_{\mathrm{sq}}(\oprho_{AB}) = \frac{1}{2}\inf_{\oprho_{ABR}}\left\{I(A:B|R)_{\oprho_{ABR}}:\mathrm{tr}_R\oprho_{ABR}=\oprho_{AB}\right\}\,,
\end{equation}
where the conditional quantum mutual information is defined as \cite{wilde2017quantum}
\begin{equation}
    I(A:B|R) = S(A|R) + S(B|R) - S(AB|R)\,.
\end{equation}
One of the main properties of squashed entanglement is its additivity for tensor products of states \cite[Proposition 4]{christandl2004squashed}.
The squashed entanglement is one of the two main entanglement measures in quantum communication theory: together with the relative entropy of entanglement \cite{audenaert2001asymptotic,vedral2002role}, it provides the best known upper bound to the length of a shared secret key that can be generated by two parties holding many copies of the quantum state \cite{christandl2004squashed,christandl2007unifying,li2014relative,wilde2016squashed}.
Moreover, it has applications in recoverability theory \cite{seshadreesan2015fidelity,li2018squashed} and multiparty information theory \cite{adesso2007coexistence,avis2008distributed,yang2009squashed}.
Lower bounds to the squashed entanglement are notoriously difficult to prove, since the optimization in \eqref{eq:defEsq} over all the possible extensions of the quantum state is almost never analytically treatable.
The multimode conditional quantum EPI allows us to overcome the difficulty.

Any entanglement measure for quantum states can be extended to quantum channels defining it as the maximum entanglement achievable between sender and receiver.
The relative entropy of entanglement of several quantum channels has been determined in \cite{pirandola2017fundamental}.
The squashed entanglement of a quantum channel $\Phi$ \cite{takeoka2014squashed} is the maximum squashed entanglement achievable between sender and receiver:
\begin{equation}\label{eq:ChannelSqEnt}
    E_{\mathrm{sq}}(\Phi) = \sup_{\oprho_{AB}}E_{\mathrm{sq}}\left((\1_A\otimes\Phi)(\oprho_{AB})\right)\,,
\end{equation}
where the sender generates the bipartite quantum state $\oprho_{AB}$, keeps the quantum system $A$ and sends the quantum system $B$ to the receiver through $\Phi$.
Similarly to what happens for the states, the squashed entanglement is additive for the tensor product of channels \cite[Corollary 8]{takeoka2014squashed}.
In the same way as the squashed entanglement of a quantum state is an upper bound to the distillable key of the state, the squashed entanglement of a quantum channel is an upper bound to the capacity of the channel to generate a secret key shared between sender and receiver \cite{berta2018amortization,takeoka2014squashed}.
The results of this paper significantly extends the results of Ref. \cite{de2019squashed}, where the squashed entanglement of the noiseless bosonic Gaussian attenuator and amplifier have been determined.

\vspace{\baselineskip}

The paper is structured as follows. In \secref{sec:BQS} we present bosonic quantum systems. In \secref{sec:EPI} we prove the multimode conditional quantum Entropy Power Inequality. In \secref{sec:squashed} we apply such inequality to determine new lower bounds to the squashed entanglement of a large family of bosonic Gaussian states. In \secref{sec:channels}, we provide a theoretical method for finding lower and upper bounds to the multimode extreme bosonic Gaussian channels, \emph{i.e.}, the bosonic Gaussian channels that cannot be decomposed as non-trivial convex combinations of quantum channels. In \secref{sec:simulations} we show a numerical example for the calculation of upper and lower bounds to the squashed entanglement of a given extremal bosonic Gaussian channel. We discuss our results in \secref{sec:disc}. \aref{app:MathNotes} contains some results used in the main text, while \aref{app:lem} contains the proofs of the auxiliary lemmas.
Lastly, in \aref{app:NumSim}, we report some numerical simulations of the squashed entanglement bounds of a given multimode extreme bosonic Gaussian channel.
\section{Bosonic quantum systems}\label{sec:BQS}

In this section we briefly present bosonic quantum systems. For more details, the reader can refer to the books \cite{ferraro2005gaussian}, \cite{serafini2017quantum} and \cite[Chapter 12]{holevo2019quantum}.
A one-mode bosonic quantum system is the mathematical model for a harmonic oscillator or for one mode of the electromagnetic radiation. An $n$-mode bosonic quantum system is the union of $n$ one-mode bosonic quantum systems, and its Hilbert space is the $n$-th tensor power of the Hilbert space of a one-mode bosonic quantum system. The Hilbert space of a $n$-mode bosonic quantum system is the irreducible representation of the canonical commutation relations
\begin{equation}\label{eq:canonicalRel}
    \qty[\opQ_j,\opQ_k]=\qty[\opP_j,\opP_k]=0\,,
    \qquad
    \qty[\opQ_j,\opP_k]=i\delta_{jk}\I\,,
\end{equation}
for $j,k=1,\ldots,n$, where $\opQ_j$ and $\opP_j$ are the \emph{quadrature operators}, which for the harmonic oscillator represent the position and momentum operators. Defining the vector
\begin{equation}
    \R=\qty(\opQ_1,\opP_1,\ldots,\opQ_n,\opP_n)\,,
\end{equation}
where $\op{R}_{2k-1} = \opQ_k$, $\op{R}_{2k} = \opP_k$, for $k=1,\ldots,n$, Equation \eqref{eq:canonicalRel} becomes
\begin{equation}\label{eq:commut-canon}
    \qty[\opR_i,\opR_j] = i\Delta_{ij}\I\,,
\end{equation}
for $i,j=1,\ldots,2n$, \modifica{where $\Delta_{ij}$ are the entries of the matrix
\begin{equation}\label{eq:formOmega}
    \Delta = \bigoplus_{k=1}^n\mqty(0&1\\-1&0)\,,
\end{equation}
which is known as \emph{symplectic form}.} More generally, a symplectic form is any antisymmetric and invertible matrix $\Delta'$. In fact, there always exists a matrix $M$ such that $M\,\Delta'M^T=\Delta$, the canonical form. The Hamiltonian that counts the number of excitations or photons is
\begin{equation}
    \opH = \sum_{i=1}^n\aa_i^\dagger\,\aa_i = \sum_{k=1}^{2n}\qty(\opR_k)^2-\frac{n}{2}\I
    \,,
\end{equation}
where
\begin{equation}
    \aa_i = \frac{\opQ_i+i\opP_i}{\sqrt{2}}
\end{equation}
is the ladder operator. The vector annihilated by $\aa$ is the vacuum and is denoted by $\ket{0}$.

\begin{definition}[displacement operator]
    Given a vector $\vb{r}\in\mathbb{R}^{2n}$, the \emph{displacement operator} of parameter $\vb{r}$ is defined as
    \begin{equation}\label{eq:op-trasl}
        \dr = \exp(i\, \vb{r}^T \Delta^{-1} \R)\,.
    \end{equation}
\end{definition}
\begin{observation}[unitarity of $\dr$]
    Since $\vb{r}^T \Delta^{-1} \R$ is a self-adjoint operator, $\dr$ is unitary and in particular it holds
    \begin{equation}
        \drd = \dr^{-1} = \op{D}_{-\vb{r}}\,.
    \end{equation}
\end{observation}
The name ``displacement operator'' is due to the following
\begin{proposition}\label{prop:RTrasl}
    In Heisenberg's representation it holds
    \begin{equation}\label{eq:trasl}
        \drd \,\R\, \dr = \R + \vb{r}\I\,.
    \end{equation}	
\end{proposition}

For regularity reasons, in this paper we will consider only quantum states with finite average energy.
We stress that such restriction is not relevant from an experimental perspective, since any quantum state that can be realized in an actual experiment must have finite average energy.
Formally, we have the following definition:
\begin{definition}(finite average energy)\label{def:finite_average_energy}
    Let $\hat{\rho}$ be a state of an $n$-mode bosonic quantum system with quadratures $\hat{R}_1,\,\ldots,\,\hat{R}_{2n}$, and let
    \modifica{
    \begin{equation}
        \hat{\rho} = \sum_{k=0}^{+\infty} p_k |\psi_k\rangle\langle\psi_k|
    \end{equation}
    }
    be an eigendecomposition of $\hat{\rho}$ with eigenvalues $\{p_k:k\in\mathbb{N}\}$ and eigenvectors $\{|\psi_k\rangle:k\in\mathbb{N}\}$.
    We say that $\hat{\rho}$ has finite average energy if each $|\psi_k\rangle$ belongs to the domain of each $\hat{R}_i$ and
    \modifica{
    \begin{equation}
        \sum_{k=0}^{+\infty} p_k \norm{\hat{R}_i|\psi_k\rangle}^2 < +\infty\,,
        \quad
        \forall\,i=1,\,\ldots,\,2n\,.
    \end{equation}
    }
\end{definition}

The quantum counterparts of the first and second moments of a random variable with values in $\RR^n$ are defined as follows:
\begin{definition}(first moments)
    Given a quantum state $\oprho$ with finite average energy, the vector of the \emph{first moments} $\opsr(\oprho)$ is defined as the expectation value of the quadratures $\R$,
    \begin{equation}\label{eq:MomentoPrimo}
        \opsr(\oprho) = \tr\!\qty[\oprho\,\R]\,.
    \end{equation}
\end{definition}

\begin{definition}(second moments)
    Given a quantum state $\oprho$, the \emph{covariance matrix} $\sigma(\oprho)$ is defined as
    \begin{equation}\label{eq:sigma}
        \sigma(\oprho)_{ij} = \frac12 \tr\left[\left(\hat{R}_i-\bar{r}_i\right)\oprho\left(\hat{R}_j-\bar{r}_j\right) + \left(\hat{R}_j-\bar{r}_j\right)\oprho\left(\hat{R}_i-\bar{r}_i\right)\right]\,,
    \end{equation}
    where $\opsr=\opsr(\oprho)$\,.
\end{definition}
The eigenvalues of the matrix $\Delta^{-1}\sigma$ are pure imaginary and pairwise opposite.
\begin{definition}[symplectic eigenvalues]\label{def:symplectic_eigenvalues}
    The \emph{symplectic eigenvalues} of a real positive matrix $\sigma$ are the absolute values of the eigenvalues of $\Delta^{-1}\sigma$\,.
\end{definition}

\begin{remark}
    The average number of photons of a quantum state $\oprho$ is related to its covariance matrix $\sigma$ by
    \begin{equation}\label{eq:number-of-photons}
        N=\frac12\tr\left(\sigma-\frac{I}{2}\right)\,.
    \end{equation}
\end{remark}

\subsection{Symplectic group}\label{sec:prel-SimplGroup}

Given a symplectic matrix $S$, \emph{i.e.}, a matrix that satisfies
\begin{equation}
    S\,\Delta\,S^T=\Delta\,,
\end{equation}
$\opU_{\!S}$ is a unitary operator on the Hilbert space that applies the matrix $S$ on the quadratures.
\begin{proposition}[$\opU_{\!S}$ operator]\label{prop:opUsimpl}
    Given a symplectic matrix $S\in\mathrm{Sp}(2n,\mathbb{R})$, there exists a unitary operator $\opU_{\!S}$ acting on the Hilbert space of an $n$-mode bosonic quantum system such that
    \begin{equation}\label{eq:trasf-simpl-R}
        \opU_{\!S}^\dagger\, \R\, \opU_{\!S} = S\,\R\,.
    \end{equation}
\end{proposition}
\begin{proof}
    See \cite{serafini2017quantum}.
\end{proof}

\paragraph{Properties of $\opsr$ and $\sigma$\,.}
Equations \eqref{eq:trasl} and \eqref{eq:trasf-simpl-R} imply the following properties for the first and second moments $\opsr$ and $\sigma$\,. The displacement operators and the symplectic unitaries transform the first moments vector as
\begin{equation}\label{eq:TraslMomPrim}
    \opsr(\drrdrd) = \opsr(\oprho) + \r\,,
    \qquad
    \opsr(\opU_S\, \oprho\, \opU_S^\dagger) = S\,\opsr(\oprho)\,,
\end{equation}
while for the covariance matrix it holds
\begin{equation}\label{eq:TraslMomSec}
    \sigma(\drrdrd) = \sigma(\oprho)\,,
    \qquad
    \susrusd = S\,\sigma(\oprho)\,S^T,
\end{equation}
where $\r\in\RR^{2n}$ and $S\in\text{Sp}(2n,\RR)$.

\subsection{Bosonic Gaussian states}\label{sec:GaussianStates}

The quantum analogue of real Gaussian probability distributions are the bosonic Gaussian states.
\begin{definition}[Gaussian states]
    \emph{Bosonic Gaussian states} are density operators proportional to the exponential of a quadratic Hamiltonian in the quadratures together with the projectors on its ground states. In formulae, Gaussian states can be written as
    \modifica{
    \begin{equation}\label{eq:gaussState}
        \begin{split}
        &\oprho_\beta \propto \exp(-\frac{\beta}{2}(\R-\opsr_0)^Th(\R-\opsr_0))\,,\\
        &\oprho_0=\lim_{\beta\to{+\infty}}\oprho_\beta\,,
        \end{split}
    \end{equation}
    }
    where $h$ is a positive-definite real symmetric matrix.
\end{definition}

\begin{definition}[thermal states]\label{def:therm-state}
    A \emph{thermal Gaussian state} is a bosonic Gaussian state, as in Equation \eqref{eq:gaussState}, with $\bar{\vb{r}}_0=0$ and $h\propto I$\,.  
\end{definition}

\begin{proposition}[characterisation of Gaussian states]\label{prop:RSigmaGaussianStates}
    Bosonic Gaussian states are uniquely determined by their first and second moments. In other words, given a first moment $\vb{r}'$ and a covariance matrix $\sigma'$, there exists a unique bosonic Gaussian state $\oprho$ such that $\bar{\vb{r}}(\oprho)=\vb{r}'$ and $\sigma\qty(\oprho)=\sigma'$.
\end{proposition}
\begin{proof}
    See \cite{serafini2017quantum}.
\end{proof}

\begin{proposition}\label{prop:7.4}
    The von Neumann entropy of an $n$-mode bosonic Gaussian state is
    \begin{equation}\label{eq:entropiaGauss}
        S = \sum_{k=1}^n g\qty(\nu_k-\frac12)\,,
    \end{equation}
    where $\nu_1,\ldots,\nu_n$ are the symplectic eigenvalues of its covariance matrix and the function $g$ is defined as
    \begin{equation}\label{eq:g-funct}
        g(x) \coloneqq (x+1)\ln(x+1) - x\ln(x)\,.
    \end{equation}
\end{proposition}
\begin{proof}
    See \cite[Equation (3.92)]{serafini2017quantum}.
\end{proof}
One-mode thermal Gaussian states can be written as
\begin{equation}
    \op\omega(\En) = \frac{1}{\En+1}\qty(\frac{\En}{\En+1})^{\ada},
\end{equation}
where $\En\geq0$ is the average number of photons:
\begin{equation}
    \tr[\op\omega(\En)\,\ada] = \En\,.
\end{equation}
The covariance matrix of these states is given by
\begin{equation}
    \sigma(\op\omega(\En)) = \qty(\En+\frac12)I_2\,.
\end{equation}
We notice that $\op\omega(0)=\ketbra{0}$ is the vacuum state. The von Neumann entropy of $\op\omega(\En)$ is
\begin{equation}\label{eq:gdef}
    S(\op\omega(\En)) = (\En+1)\ln(\En+1) - \En\ln(\En) \eqqcolon g(\En)\,.
\end{equation}

\subsection{Bosonic Gaussian channels}\label{sec:qGchannels}

Bosonic Gaussian channels are those quantum channels that map Gaussian states into Gaussian states. They play a key role in quantum communication since they model the attenuation and the noise that affect light pulses travelling through optical fibres and electromagnetic signals travelling through free space. The most important families of bosonic Gaussian channels are the beam splitter, the squeezing and the bosonic Gaussian attenuators and amplifiers. The first two are the quantum analogue of classical linear mixing of random variables and are the main transformations of quantum optics. In the physical representation, a bosonic Gaussian channel can be seen as an ancillary system, the environment, that is in a Gaussian initial state and interacts with the state of the system of interest via a quadratic Hamiltonian coupling. Such a transformation preserves the Gaussian character of the state of the system of interest. The bosonic Gaussian channels are briefly and exhaustively characterised as follows.

\begin{proposition}[bosonic Gaussian channels]\label{prop:GaussianCharact}
    An $n$-mode bosonic Gaussian channels is completely characterized by two $2n \times 2n$ real matrices $K$ and $\alpha$ satisfying
    \begin{equation}\label{eq:3.144}
        \alpha \geq \pm\frac{i}{2}\qty(\Delta - K\,\Delta\, K^T)\,.
    \end{equation}
    Such channel acts on the moments of an $n$-mode input Gaussian state with vector of first moments $\bar{\vb{r}}$ and covariance matrix $\sigma$ as follows:
    \begin{equation}
        \begin{split}
        \bar{\vb{r}} &\longmapsto K\,\bar{\vb{r}}\,,\\
        \sigma &\longmapsto K\,\sigma\, K^T + \alpha \,.
        \end{split}
    \end{equation}
\end{proposition}
\begin{proof}
    See \cite[Section 5.3]{serafini2017quantum}.
\end{proof}

In what follows, we will refer to a bosonic Gaussian channel characterised by the matrices $K$ and $\alpha$, as in \pref{prop:GaussianCharact}, as a \textit{$(K,\alpha)$ Gaussian channel}. The space of quantum channels with fixed input and output systems is convex. A prominent role is played by the following class of quantum channels.

\begin{definition}[extreme quantum channels]
    A quantum channel is said to be \emph{extreme} if it cannot be decomposed as a non-trivial convex combination of quantum channels \cite{holevo2013extreme}.
\end{definition}

The following \pref{prop:extreme} provides an easy characterization of the extreme bosonic Gaussian channels.
\begin{proposition}[extreme bosonic Gaussian channels]\label{prop:extreme}
    A bosonic Gaussian channel characterized by the matrices $(K,\alpha)$, as in \pref{prop:GaussianCharact}, is extreme iff $\alpha$ is a minimal solution of inequality \eqref{eq:3.144}.
\end{proposition}
\begin{proof}
    See \cite[Corollary 1]{holevo2013extreme} and \cite[Equation (12.138)]{holevo2019quantum}.
\end{proof}
\begin{remark}
    In other words, a $(K,\alpha)$ Gaussian channel is extreme if there is no $(K,\alpha')$ Gaussian channel such that $\alpha'\leq\alpha$ and $\alpha'\neq\alpha$. That is, if the only $\alpha'\in\mathbb{R}^{2n\times2n}$ such that $	\alpha\ge\alpha'\ge\pm\frac{i}{2}(\Delta - K\,\Delta\,K^T)$ is $\alpha'=\alpha$\,.
\end{remark}

\subsubsection{Beam-splitter and squeezing.}

Let us now briefly introduce two relevant examples of extreme bosonic Gaussian channels \cite{serafini2017quantum}. Given the one-mode bosonic quantum systems $A,B$, and $C,D$, the \emph{beam splitter} with transmissivity coefficient $0\leq\eta\leq1$ is the mixing unitary operator $\opU_\eta:\H_A\otimes\H_B\to\H_C\otimes\H_D$ acting on the quadratures, as in Equation \eqref{eq:trasf-simpl-R}, through the symplectic matrix
\begin{equation}\label{eq:S_beam_splitter}
    S_\eta = \mqty(\sqrt{\eta}\,I_2&\sqrt{1-\eta}\,I_2\\-\sqrt{1-\eta}\,I_2&\sqrt{\eta}\,I_2)\,.
\end{equation}
Similarly, the \emph{squeezing} unitary operator $\opU_\kappa:\H_A\otimes\H_B\to\H_C\otimes\H_D$ with squeezing parameter $\kappa>1$ acts on the quadratures through the symplectic matrix
\begin{equation}\label{eq:S_squeezing}
    S_\kappa = \mqty(\sqrt{\kappa}\,I_2&\sqrt{\kappa-1}\,Z_2\\\sqrt{\kappa-1}\,Z_2&\sqrt{\kappa}\,I_2)\,,
\end{equation}
where $I_2$ is the $2\times2$ identity matrix and $Z_2$ is the $\sigma_z$ Pauli matrix.
\section{The multimode Entropy Power Inequality}\label{sec:EPI}

Entropic inequalities are the main tool to prove upper bounds to quantum communication rates \cite{holevo2019quantum,wilde2017quantum} and to prove the security of quantum key distribution protocols \cite{coles2017entropic}. In these scenarios, a prominent role is played by entropic inequalities in the presence of quantum memory, where the entropies are conditioned on the knowledge of an external observer holding a memory quantum system.

\subsection{The problem}\label{sec:MCQEPI-prob}

The aim of this section is to provide a proof of a new entropic inequality, the \emph{multimode conditional quantum Entropy Power Inequality} for bosonic quantum systems (\tref{theo:MCQEPI}). Let us first define the action of a quantum channel on input systems.
\begin{definition}[linear mixing of quantum modes]\label{def:PhiB}
    Let $B=\mqty(B_1&\cdots&B_K)\in\mathbb{R}^{2n\times2Kn}$ be a real matrix of rank $2n$ formed by $K$ blocks $B_i\in\mathbb{R}^{2n\times2n}$, satisfying the condition
    \begin{equation}\label{eq:5.8}
        \sum_{i=1}^K B_i\, \Delta\, B_i^T = \Delta\,.
    \end{equation}
    Let $S_B\in\mathrm{Sp}(2Kn,\mathbb{R})$ such that its first $2n$ rows are given by $B$. The quantum analogue of the linear transformation \eqref{eq:linc} is given by
    \begin{equation}\label{eq:5.6}
        \ry = \Phi_X^B(\op\rho_X) \coloneqq \tr_Z\!\qty\Big[\opU_{S_B}\, \rx\, \opU_{S_B}^\dagger]\,,
    \end{equation}
    where $\opU_{S_B}:\H_X\to\H_Y\otimes\H_Z$ (see \pref{prop:opUsimpl}) is an isometry between the input Hilbert space $\H_X$ and the tensor product of the output Hilbert space $\H_Y$ with an ancillary Hilbert space $\H_Z$, satisfying
    \begin{equation}\label{eq:USBTransorfmation}
        \opU_{S_B}^\dagger\, \R^Y \opU_{S_B} = B\, \R^X = \sum_{i=1}^K B_i\, \R^{X_i},
    \end{equation}
    with $\R^{X_i}$ and $\R^Y$ respectively the quadratures of the systems $X_i$ and $Y$. 
\end{definition}

\begin{remark}
    Condition \eqref{eq:5.8} is necessary to preserve the canonical commutation relations and ensures the existence of $S_B$. The isometry $\opU_{S_B}$ does not necessarily conserve energy, \emph{i.e.}, it can contain active elements, so that even if the input $\op\rho_X$ is the vacuum state on all its $K$ modes, the output $\op\rho_Y$ can be a thermal state with a non-zero temperature.
\end{remark}

\begin{example}[beam-splitter and quantum amplifier]\label{ex:5.1}
    For $K=2$, the beam splitter with attenuation parameter $0\leq\eta\leq1$ is easily recovered by defining
    \begin{equation}
	B_1 = \sqrt{\eta}\,I_{2n}\,,
	\qquad
	B_2 = \sqrt{1-\eta}\,I_{2n}\,.
    \end{equation}
    To get the quantum amplifier of parameter $\kappa\geq1$, one must instead take
    \begin{equation}
	B_1 = \sqrt{\kappa}\,I_{2n}\,,
	\qquad
	B_2 = \sqrt{\kappa-1}\,Z_{2n}\,,
    \end{equation}
    where $Z_{2n}$ is the $n$-mode time reversal matrix $Z_{2n} \coloneqq \bigoplus_{k=1}^n \text{diag}(1,-1)$.
\end{example}

We are going to study the output system given by $K$ conditionally independent input systems which interact through the linear mixing process given by $\Phi_B$ in presence of a memory quantum system. Let $\{X_i\}_{i=1,\ldots,K}$ be the $n$-mode input bosonic quantum systems, and let $Y$ be the $n$-mode output bosonic quantum system. Let us consider a joint quantum input state $\op\rho_{XM}$, where $X=X_1\cdots X_K$ is the overall input system, such that $\{X_i\}_{i=1,\ldots,K}$ are conditionally independent given the memory system $M$, \emph{i.e.},
\begin{equation}\label{eq:3.2}
    S(X_1\ldots X_K|M) = \sum_{i=1}^K S(X_i|M)\,.
\end{equation}

Let $\hat{\rho}_{YM} = (\Phi_X^B\otimes\mathbbm{1}_M)(\hat{\rho}_{XM})$ be the output state of the linear mixing process. The multimode conditional quantum Entropy Power Inequality determines the minimum conditional quantum entropy $S(Y|M)$ of the output state among all the input quantum states $\op\rho_{XM}$ as above and with given conditional quantum entropies $\{S(X_i|M)\}_{i=1,\ldots,K}$:
\begin{equation}
    \exp(\frac{S(Y|M)_{\op\rho_{YM}}}{n}) \geq \sum_{i=1}^{K} b_i \exp(\frac{S(X_i|M)_{\op\rho_{X_iM}}}{n})\,,
\end{equation}
where $b_i$ are constants dependent on $B$.

\paragraph{Outline of the proof.}
The proof proceeds as follows. We evolve the input system with the heat semigroup time evolution (\dref{def:heat-semigroup}). Employing the quantum de Brujin identity (\pref{pro:gdbi}) and a new conditional quantum Stam inequality (\secref{sec:QCSI}), we prove that such evolution degrades the Entropy Power Inequality. We conclude by proving that the inequality holds in the limit of infinite time via asymptotic estimates of the time scaling of the conditional entropies.

\subsection{Preliminary results}\label{sec:MCQEPI-PrelRes}

In this section we prove some important intermediate results that will play a key role in the proof of the multimode conditional quantum Entropy Power Inequality (MCQEPI).

\begin{definition}[heat semigroup]\label{def:heat-semigroup}
    The \emph{heat-semigroup} time evolution is generated by the convex combination of displacement operators according to a centred Gaussian distribution with covariance matrix $\alpha$. For any quantum state $\op\rho$
    \begin{equation}\label{eq:heat-semigroup}
        \N(\alpha)(\op\rho) \coloneqq \int_{\mathbb{R}^{2n}}\opD(\x) \,\op\rho\, \opD(\x)^\dagger \frac{e^{-\frac12\x^T\alpha^{-1}\x}}{\sqrt{(2\pi)^n\det(\alpha)}} \dd{\x}\,.
    \end{equation}
\end{definition}

The following lemma states how the heat semigroup evolution of the subsystems $X_i$ is related to the heat semigroup evolution of the overall system $X$.

\begin{lemma}\label{lem:diag-heatsemi}
    Let $X=X_1\cdots X_K$ be a multipartite bosonic quantum system, where each $X_i$ is a $n$-mode bosonic quantum system and $\alpha=\bigoplus_{i=1}^K\alpha_i$, $\alpha_i\in\rdntdn_{>0}\ \forall i$, a block diagonal matrix, then it holds
    \begin{equation}
       \N_X(\alpha) = \bigotimes_{i=1}^K\N_{X_i}(\alpha_i)\,.
    \end{equation}
\end{lemma}
\begin{proof}
    Let $\z=\mqty(\z_1&\cdots&\z_K)^T$, $\z_i\in\RR^{2n}\ \forall i$. It is easy to notice that $\det(\alpha)=\prod_{i=1}^{K}\det(\alpha_i)$ and $\alpha^{-1}=\bigoplus_{i=1}^K\alpha_i^{-1}$. Furthermore, from the definition of displacement operator, $\opD_X(\z)=\bigotimes_{i=1}^K\opD_{X_i}(\z_i)$ holds. From the definition of the heat semigroup, $\forall\ \rx$ it holds:
    \begin{equation}
	\begin{split}
           &\N_X(\alpha)\rx = \mathbb{E}_\mathds{Z}\qty[\opD_X(\mathds{Z})\rx \opD_X^\dagger(\mathds{Z})]\\
           &= \hspace{-4pt}\int_{\RR^{2Kn}}\hspace{-10pt} \opD_X(\z)\, \rx\, \opD_X^\dagger(\z)\, e^{-\frac12\z^T\alpha^{-1}\z}\!\frac{\dd{\z}}{\sqrt{(2\pi)^{2Kn}\det(\alpha)}}\\
           &= \hspace{-4pt}\int_{\RR^{2Kn}}\hspace{-10pt} \opD_X(\z)\, \rx\, \opD_X^\dagger(\z)\, e^{-\frac12\sum_{i=1}^K\z_i^T\alpha_i^{-1}\z_i}\!\prod_{i=1}^{K}\!\frac{\dd{\z_i}}{\sqrt{(2\pi)^{2n}\det(\alpha_i)}}\\
           &= \prod_{i=1}^{K} \int_{\RR^{2n}}\hspace{-10pt} \opD_{X_i}(\z_i)\, \rxi\, \opD_{X_i}^\dagger(\z_i)\, e^{-\frac12\z_i^T\alpha_i^{-1}\z_i}\frac{\dd{\z_i}}{\sqrt{(2\pi)^{2n}\det(\alpha_i)}}\\
           &= \bigotimes_{i=1}^K\N_{X_i}(\alpha_i)\rxi\,,
	\end{split}
    \end{equation}
    where $\mathds{Z}$ is a classical Gaussian random variable with zero mean and covariance matrix $\alpha$.
\end{proof}

For the map defined in Equation \eqref{eq:heat-semigroup} the following composition rules apply.
\begin{lemma}\label{lem:Phi-N}
    Let $B$ be as in \dref{def:PhiB}, for each $\alpha,\beta\in\rdntdn_{>0}$, it holds
    \begin{equation}
	\N_X(\alpha)\circ\N_X(\beta) = \N_X(\alpha+\beta)\,.
    \end{equation}
\end{lemma}
\begin{proof}
    See \cite[Section 3]{de2021generalized}.
\end{proof}

The following two lemmas describe how the transformation $\Phi^B$ (\dref{def:PhiB}) behaves under the heat semigroup evolution and displacements.
\begin{lemma}[compatibility with the heat semigroup]
    Let $B$ be as in \dref{def:PhiB}, for each $\alpha,\beta\in\rdntdn_{\geq0}$, it holds
    \begin{equation}
        \Phi_B\circ\N_X(\alpha) = \N_Y(B\,\alpha\, B^T)\circ\Phi_B\,.
    \end{equation}
\end{lemma}
\begin{proof}
    See \cite[Lemma 1]{de2021generalized}.
\end{proof}

\begin{lemma}[compatibility with displacements]\label{lem:CompDisp}
    For any $\alpha\in\RR^{2n\times2n}_{\geq0}$ and any $\vb{x}\in\RR^{2n}$,
    \begin{equation}
        \Phi_B\qty(\opD_{\vb{x}}\cdot\opD^\dagger_{\vb{x}}) = \opD_{B\vb{x}}\,\Phi_B(\,\cdot\,)\,\opD^\dagger_{B\vb{x}}\,,
    \end{equation}
    where $\opD_{\vb{x}}$ acts on system $X$ while $\opD_{B\vb{x}}$ acts on system $Y$.
\end{lemma}
\begin{proof}
    See \cite[Lemma 1]{de2021generalized}.
\end{proof}

A fundamental result for the proof of the MCQEPI is the \emph{asymptotic scaling} of the conditional entropy with respect to the time evolution induced by the heat semigroup with a positive definite matrix.

\begin{proposition}[asymptotic scaling of the entropy]\label{prop:AsyScaling}
    Let $X$ be a $n$-mode bosonic quantum system, then, for any quantum state $\op\rho_{XM}$ of $XM$, such that $S(M)$ is finite and $\op\rho_X$ has finite energy, and any $\alpha\in\mathbb{R}_{>0}^{2n\times 2n}$, \modifica{for $t\to+\infty$}
    \begin{equation}
    \modifica{
        S(X|M)(\N_X(t\,\alpha)(\op\rho_{XM})) = \frac12\ln(\det(e\,t\,\alpha)) + \order{\frac1t}\,.
    }
    \end{equation}
\end{proposition}
\begin{proof}
    See \cite[Proposition 10]{de2021generalized}.
\end{proof}

\subsection{Quantum integral de Bruijn identity}\label{sec:QCFI}

Let us start by defining the integral version of the conditional quantum Fisher information.
\begin{definition}[integral conditional quantum Fisher information]\label{def:5.3}
    Let $A$ be a $n$-mode bosonic quantum system, $M$ a generic quantum system and $\op\rho_{AM}$ a quantum state of $AM$. For any $\alpha\in\mathbb{R}_{\geq0}^{2n\times 2n}$, we define the quantum integral Fisher information of $A$ conditioned on $M$ as
    \begin{subequations}
        \begin{align}
            &\Delta_{A|M}(\op\rho_{AM})(\alpha) \coloneqq I(A:\mathds{X}|M)_{\op\sigma_{AM\mathds{X}}}\geq0\,,\\
            &\Delta_{A|M}(\op\rho_{AM})(0) \coloneqq 0\,,
        \end{align}
    \end{subequations}
    where $\mathds{X}$ is a classical Gaussian random variable with zero mean and covariance matrix $\alpha$ and $\op\sigma_{AM\mathds{X}}$ is the joint quantum-classical state of $AM\mathds{X}$ such that for any $\vb{x}\in\mathbb{R}^{2n}$
    \begin{equation}
        \op\sigma_{AM|\mathds{X}=\vb{x}} = \opD_A(\vb{x})\op\rho_{AM}\opD_A(\vb{x})^\dagger\,.
    \end{equation}
\end{definition}

The following \emph{integral de Bruijn identity} proves that the integral Fisher information is equal to the increase in the conditional entropy induced by the heat semigroup.
\begin{proposition}[integral de Bruijn identity]\label{prop:IntDeBurijn}
    For any quantum state $\op\rho_{XM}$ of $XM$ such that $S(M)$ is finite and $\op\rho_X$ has finite energy any $\alpha\in\mathbb{R}_{\geq0}^{n\times n}$, it holds
    \modifica{
    \begin{equation}
        \Delta_{X|M}\!(\op\rho_{XM})\!(\alpha)\! =\! S(X|M)_{\!(\N_X(\alpha)\otimes\1_M)\!(\op\rho_{XM})} - S(X|M)_{\op\rho_{XM}}\,.
    \end{equation}
    }
\end{proposition}
\begin{proof}
    \cite[Proposition 2]{de2021generalized}.
\end{proof}

We can now give the definition of the conditional quantum Fisher information.
\begin{definition}[conditional quantum Fisher information]\label{def:FisherInfo}
    Let $\op\rho_{XM}$ a quantum state of $XM$, such that $S(M)$ is finite and $\op\rho_X$ has finite energy. For any $\alpha\in\mathbb{R}_{\geq 0}^{n\times n}$, the \emph{conditional quantum Fisher information} of $\op\rho_{XM}$ is defined as
    \begin{equation}\label{eq:FisherInformation}
        J_{X|M}(\op\rho_{XM})(\alpha) \coloneqq \lim_{t\to 0}\frac{\Delta_{X|M}(\op\rho_{XM}(t\,\alpha))}{t}\,.
    \end{equation}
\end{definition}
\begin{remark}
    From \cite[Proposition 5]{de2021generalized}, the limit in Equation \eqref{eq:FisherInformation} always exists, finite or infinite.
\end{remark}

\begin{lemma}[linearity of the conditional quantum Fisher information]\label{lem:linearity}
    For any quantum state $\op\rho_{XM}$ of $XM$, such that $S(M)$ is finite and $\op\rho_X$ has finite energy, any $\alpha\in\mathbb{R}_{\geq0}^{n\times n}$ and any $t\geq 0$, it holds
    \begin{equation}
        J_{X|M}(\op\rho_{XM})(t\,\alpha) = t\,J_{X|M}(\op\rho_{XM})(\alpha)\,.
    \end{equation}
\end{lemma}
\begin{proof}
    \cite[Lemma 2]{de2021generalized}.
\end{proof}

The following de Bruijn identity relates the conditional quantum Fisher information to the time derivative of the conditional entropy with respect to the heat semigroup time evolution \cite{de2021generalized}.
\begin{proposition}[de Bruijn identity]\label{pro:gdbi}
    Let $X$ be a $n$-mode bosonic quantum system and $\op\rho_{XM}$ a quantum state of $XM$, such that $S(M)$ is finite and $\op\rho_X$ has finite energy. For each $\alpha\in\rdntdn_{\geq0}$ and $t\in\RR$, let us define $\rxm(t)\ce(\N_X(t\,\alpha)\otimes\1_M)\rxm$. The following relation holds
    \begin{equation}\label{eq:gdbi}
        \dv{t}\sxmt = \jxmt(\alpha)\,,
    \end{equation}
    where $\jxmt(\alpha)$ is defined in \dref{def:FisherInfo}.
\end{proposition}
\begin{proof}
    See \cite[Proposition 7]{de2021generalized}.
\end{proof}

\subsection{Conditional quantum Stam Inequality}\label{sec:QCSI}

The following new \emph{integral Stam inequality} proves that the conditional quantum Fisher information is decreasing with respect to the application of the quantum channel $\Phi^B$ (\dref{def:PhiB}).

\begin{theorem}[conditional quantum Stam inequality]\label{theo:QCondStamIneq}
    Let $X_1, \ldots, X_K$ and $Y$ be $n$-mode bosonic quantum systems, let $X=X_1\cdots X_K$, let $M$ a quantum system and $\Phi_B:X\to Y$ the transformation defined in \dref{def:PhiB}. Let $\rxm$ be a state of $XM$ such that $\hat{\rho}_X$ has finite average energy \modifica{and $S(\oprho_M)<+\infty$},
    and let us suppose that the systems $\{X_i\}_{i=1,\ldots,K}$ are conditionally independent given $M$, \emph{i.e.}, such that
    \begin{equation}\label{eq:CondIndependentS}
        S(X_1\ldots X_K|M) = \sum_{k=1}^K S(X_k|M)\,.
    \end{equation}
    Then, for any set $\lambda=\qty{\lambda_1,\ldots,\lambda_K}$ such that $\lambda_k=0$ if $\det(B_k)=0$ and $0\leq\lambda_i\leq1$ with $\sum_{i=1}^K\lambda_i=1$, the conditional quantum linear Stam inequality holds:
    \begin{equation}\label{eq:LinearStamIneq}
        \boxed{
            \jym\qty\big(I_{2n}) \leq \sum_{i=1}^K\jxim\qty(\lambda_i^2B_i^{-1}B_i^{-T})\,,
        }
    \end{equation}
    where
    \begin{equation}
        \rym \coloneqq (\Phi_X^B\otimes\1_M)\rxm\,.
    \end{equation}
\end{theorem}
\begin{proof}
    We will prove the following inequality for the integral conditional quantum Fisher information:
    \begin{equation}\label{eq:IntegralFisherIneq}
        \dym(t\,I_{2n}) \leq \sum_{i=1}^K \dxim\qty\big(t\,\lambda_i^2B_i^{-1}B_i^{-T})\,.
    \end{equation}
    The linear conditional quantum Stam inequality \eqref{eq:LinearStamIneq} follows taking the derivative of \eqref{eq:IntegralFisherIneq} in $t = 0$. For any $t \geq 0$
    \begin{equation}\label{eq:I(Y:Z|M)}
        \dym(t\,I_{2n}) = I(Y:\mathds{Z}|M)_{\sigma_{YM\mathds{Z}}}\,,
    \end{equation}
    where $\mathds{Z}$ is a Gaussian random variable with zero mean
    and covariance matrix $t\,I_{2n}$ and $\op\sigma_{YM\mathds{Z}}$ is the joint quantum-classical state of $YM\mathds{Z}$ such that for any $\z\in\mathbb{R}^{2n}$
    \begin{equation}
        \op\sigma_{YM|\mathds{Z}=\z} = \op D_Y(\z) \rym \op D_Y^\dagger(\z)\,.
    \end{equation}
    We define the joint quantum-classical state $\op\sigma_{XM\mathds{Z}}$ of $XM\mathds{Z}$ such that for any $\z \in \mathbb{R}^{2n}$
    \begin{equation}\label{eq:sigmaXMZ}
    \begin{split}
        &\op\sigma_{XM|\mathds{Z}=\z}\\
        &=\opD_{X_1}\!(\lambda_1B_1^{-1}\!\!\z) \cdots \opD_{X_K}\!(\lambda_KB_K^{-1}\!\z) \rxm \opD_{X_1}^\dagger\!(\lambda_1B_1^{-1}\!\!\z) \cdots\\
        &\phantom{=} \cdots \opD_{X_K}^\dagger\!(\lambda_KB_K^{-1}\!\z)\,.
    \end{split}
    \end{equation}
    Applying \lref{lem:CompDisp}, with $\vb{x}=\mqty(\lambda_1B_1^{-1}\z&\cdots&\lambda_KB_K^{-1}\z)^T$, $B=\mqty(B_1&\cdots&B_K)$ and exploiting $\sum_{i=1}^K\lambda_i=1$, we then have that for any $t \geq 0$
    \begin{equation}
        \op\sigma_{YM\mathds{Z}}(t\,I_{2n}) = \qty(\Phi_X^B\otimes\1_{M\mathds{Z}})\op\sigma_{XM\mathds{Z}}(t\,I_{2n})\,.
    \end{equation}
    By hypothesis, from Equation \eqref{eq:CondIndependentS} and \lref{lem:IndipSyst}, we have that, for every disjoint sets of initial systems $X_\A=\qty{X_{i_1},\ldots,X_{i_k}}$ and $X_\B=\qty{X_{j_1},\ldots,X_{j_m}}$, it holds
    \begin{equation}\label{eq:XMZ}
        \begin{split}
            I(X_\A:X_\B|M\mathds{Z})_{\op\sigma_{XM\mathds{Z}}} &= \hspace{-3pt} \int_{\mathbb{R}^{2n}} \hspace{-5pt} I(X_\A:X_\B|M)_{\op\sigma_{XM|\mathds{Z}=\vb{z}}}\dd{p_{\mathds{Z}}(\vb{z})}\\
            &= \hspace{-3pt} \int_{\mathbb{R}^{2n}} \hspace{-5pt} I(X_\A:X_\B|M)_{\op\rho_{XM}}\dd{p_{\mathds{Z}}(\vb{z})}\\
            &= 0\,.
        \end{split}
    \end{equation}
    In the second equality we have exploited the fact that $\op\sigma_{XM|\mathds{Z}=\vb{z}}$ is a quantum state of the form $\opU\op\rho_{XM}\opU^\dagger$, with $\opU$ a unitary operator factorised on the two systems, \emph{i.e.}, $\opU=\opU_X\otimes\opU_M$, and thus, since the conditional mutual information can be written in terms of von Neumann entropies and the entropy is invariant under the tensor product of local unitary transformations, it holds $I(X_\A:X_\B|M)_{\op\sigma_{XM|\mathds{Z}=\vb{z}}} = I(X_\A:X_\B|M)_{\opU\op\rho_{XM}\opU^\dagger} = I(X_\A:X_\B|M)_{\op\rho_{XM}}$. From the data-processing inequality for quantum mutual information (\pref{prop:2.8}) and from Equation \eqref{eq:XMZ}, we then have
    \begin{equation}\label{eq:DataProcIneq}
        \begin{split}
            I(Y:\mathds{Z}|M)_{\op\sigma_{YM\mathds{Z}}}
            &\leq I(X:\mathds{Z}|M)_{\op\sigma_{XM\mathds{Z}}}\phantom{\sum^K}\\
            &\leq \sum_{i=1}^KI(X_i:\mathds{Z}|M)_{\op\sigma_{X_iM\mathds{Z}}}\,,
        \end{split}
    \end{equation}
    where we used inequalities \eqref{eq:2.96} and \eqref{eq:2.89}. The last inequality in Equation \eqref{eq:DataProcIneq} is a consequence of the definition of the conditional mutual information. Let us prove it by induction. By expressing the conditional quantum mutual information in terms of the von Neumann entropies of the individual subsystems, it is easy to verify that for $K=2$ the following equality is satisfied:
    \modifica{
    \begin{equation}\label{eq:5.38}
        \begin{split}
            &I(X_1X_2\!:\!\mathds{Z}|M)_{\op\sigma_{XM\mathds{Z}}}\\
            &= I(X_1\!:\!\mathds{Z}|M)_{\op\sigma_{XM\mathds{Z}}} + I(X_2\!:\!\mathds{Z}|M)_{\op\sigma_{XM\mathds{Z}}} + \\
            &\phantom{=} + I(X_1\!:\!X_2|M\mathds{Z})_{\op\sigma_{XM\mathds{Z}}} - I(X_1\!:\!X_2|M)_{\op\sigma_{XM\mathds{Z}}}\\
            &\leq I(X_1\!:\!\mathds{Z}|M)_{\op\sigma_{XM\mathds{Z}}} + I(X_2\!:\!\mathds{Z}|M)_{\op\sigma_{XM\mathds{Z}}}\,,
        \end{split}
    \end{equation}
    }
    where we used Equation \eqref{eq:XMZ} with $X_\A=X_1$ and $X_\B=X_2$. Let us now assume that this equation is valid for $K-1$ systems, which we denote altogether by $J=X_1\,\cdots\,X_{K-1}$, \emph{i.e.}, that we have
    \begin{equation}\label{eq:5.39}
        I(X_1\,\cdots\,X_{K-1} : \mathds{Z} | M)_{\op\sigma_{XM\mathds{Z}}} \leq \sum_{i=1}^{K-1} I(X_i:\mathds{Z}|M)_{\op\sigma_{XM\mathds{Z}}}\,.
    \end{equation}
    We now verify that this relation is also valid for $K$ systems, which we denote altogether by $X = X_1\,\cdots\,X_K = J\,X_K$. From Equations \eqref{eq:5.38} and \eqref{eq:5.39} we have that
    \begin{equation}\label{eq:5.48}
        \begin{split}
            I&(X_1\,\cdots\,X_K : \mathds{Z} | M)_{\op\sigma_{XM\mathds{Z}}}\\
            &\leq I(X_1\,\cdots\,X_{K-1}:\mathds{Z}|M)_{\op\sigma_{XM\mathds{Z}}} + I(X_K:\mathds{Z}|M)_{\op\sigma_{XM\mathds{Z}}}\\
            &\leq \sum_{i=1}^{K-1} I(X_i:\mathds{Z}|M)_{\op\sigma_{XM\mathds{Z}}} + I(X_K:\mathds{Z}|M)_{\op\sigma_{XM\mathds{Z}}}\\
            &= \sum_{i=1}^K I(X_i:\mathds{Z}|M)_{\op\sigma_{XM\mathds{Z}}}\,.
        \end{split}
    \end{equation}	
    Being true for $K=2$, we conclude that inequality \eqref{eq:5.48} is true for every $K\geq2$. First of all, let us note that, from inequality \eqref{eq:sigmaXMZ}, it holds
    \begin{equation}
        \begin{split}
            \op\sigma_{X_iM} &= \int \opD_{X_i}(\lambda_iB_i^{-1}\z)\oprho_{X_iM}\opD_{X_i}^\dagger(\lambda_iB_i^{-1}\z)\dd{p}_{\alpha}\!(\z)\\
            &= \mathbb{E}_\mathds{Z}\qty[\opD_X(\lambda_iB_i^{-1}\mathds{Z})\, \oprho_{X_iM}\, \opD_X^\dagger(\lambda_iB_i^{-1}\mathds{Z})]\\
            &= \mathbb{E}_\mathds{Y}\qty[\opD_X(\mathds{Y})\, \oprho_{X_iM}\, \opD_X^\dagger(\mathds{Y})]\\
            &= \qty(\N_{X_i}\qty(t\,\lambda_i^2B_i^{-1}B_i^{-T})\otimes\1_M)\,\oprho_{X_iM}\,,
        \end{split}
    \end{equation}
    where we defined the Gaussian random variable $\mathds{Y}=\lambda_iB_i^{-1}\mathds{Z}$, which has zero mean and covariance matrix $\lambda_i^2B_i^{-1}B_i^{-T}$.
    Proceeding in the proof we get
    \begin{equation}\label{eq:I(X_i:Z|M)}
        \begin{split}
            I(X_i:\mathds{Z}|M)_{\op\sigma_{X_iM\mathds{Z}}} &= S(X_i|M)_{\op\sigma_{X_iM}} - S(X_i|M)_{\op\rho_{X_iM}}\phantom{\sum}\\
            &= S(X_i|M)_{\qty(\N_{X_i}\qty(t\,\lambda_i^2B_i^{-1}B_i^{-T})\otimes\1_M)\oprho_{X_iM}} +\\
            &\phantom{=}- S(X_i|M)_{\op\rho_{X_iM}}\phantom{\sum}\\
            &= \dxim\qty(t\,\lambda_i^2B_i^{-1}B_i^{-T})\,.
        \end{split}
    \end{equation}
    From Equations \eqref{eq:I(Y:Z|M)} and \eqref{eq:I(X_i:Z|M)} together with inequality \eqref{eq:DataProcIneq}, Equation \eqref{eq:IntegralFisherIneq} is easily obtained and from this we finally obtain Equation \eqref{eq:LinearStamIneq}.
\end{proof}

\begin{example}\label{exam:beamspl_squeez}
    For $K=2$ we have, in the case of the beam splitter with parameter $0\leq\eta\leq1$, the matrices
    \begin{equation}
        B_1=\sqrt{\eta}\,I_{2n}
        \quad\text{and}\quad
        B_2=\sqrt{1-\eta}\,I_{2n}\,,
    \end{equation}
    or, in the case of squeezing with parameter $\kappa>1$, the matrices
    \begin{equation}
        B_1=\sqrt{\kappa}\,I_{2n}
        \quad\text{and}\quad
        B_2=\sqrt{\kappa-1}\,Z_{2n}\,,
    \end{equation}
    where $Z_{2n}$ has been defined in \eref{ex:5.1}. With these matrices, from Equation \eqref{eq:LinearStamIneq}, we obtain the result found in \cite[Equation (66)]{de2018conditional}.
\end{example}

\subsection{Proof of the multimode conditional quantum Entropy Power Inequality}\label{sec:MCQEPI-proof}

\begin{theorem}[multimode conditional quantum Entropy Power Inequality]\label{theo:MCQEPI}
    Let $X_1,\ldots,X_K$ and $Y$ be $n$-mode bosonic quantum systems, let $X$ be the multipartite system $X=X_1\cdots X_K$, let $M$ be a quantum system and let $\Phi_B:X\to Y$ be the linear mixing defined in \dref{def:PhiB}. Let $\op\rho_{XM}$ be a quantum state on $XM$ such that $\hat{\rho}_X$ has finite average energy (as per \dref{def:finite_average_energy}) and \modifica{$S(\oprho_M)<+\infty$}, and let us suppose that the systems $X_1\ldots X_K$ are conditionally independent given $M$, \emph{i.e.}, such that
    \begin{equation}\label{eq:CondIndependent}
        S(X_1\ldots X_K|M)_{\hat{\rho}_{XM}} = \sum_{k=1}^K S(X_k|M)_{\hat{\rho}_{XM}}\,.
    \end{equation}
    Then, defining $b_i\coloneqq\bi$, the \emph{multimode conditional quantum Entropy Power Inequality} holds:
    \begin{tcolorbox}[title = \textbf{\emph{Multimode conditional quantum Entropy Power Inequality}}]
        \begin{equation}\label{eq:MCQEPI}
            \exp(\frac{\sym}{n}) \geq \sum_{i=1}^{K} b_i \exp(\frac{\sxim}{n})\,,
        \end{equation}
    \end{tcolorbox}
    \noindent where $\hat{\rho}_{YM} = (\Phi_B\otimes\im)\hat{\rho}_{XM}$.
\end{theorem}
\begin{proof}
    Let us first recall \dref{def:PhiB} for the matrix $B$:
    \begin{equation}
        B = \mqty(B_1&\cdots&B_K)\in\rdntdkn\,,
    \end{equation}
    where $B_i\in\rdntdn$. As a first step, we observe that if all the blocks $B_i$ of the matrix $B$ are non-invertible, then the inequality \eqref{eq:MCQEPI} we wish to prove is trivial. We therefore assume in the following that there exists at least one invertible block. Let's define
    \begin{equation}
        \Lambda_i \ce \lambda_iI_{2n}\,,
        \qquad
        \Lambda \ce \bigoplus_{i=1}^K\Lambda_i\,,
        \qquad
        \tilde{B} \ce \bigoplus_{i=1}^K B_i\,,
    \end{equation}
    where $0\leq\li\leq1$, such that $\sum_{i=1}^K\li=1$ and $\lambda_k=0$ if $B_k$ is non-invertible.  In the following, with a little abuse of notation, we will define as $0$ all those quantities that are given by the product of a certain $\lambda_k=0$ and a quantity that is not defined, such as $B_k^{-1}$ for a certain non-invertible $B_k$, \emph{i.e.}, $\lambda_k B_k^{-1}\coloneqq0$. We define, for each $t\geq0$, the time evolution:
    \begin{subequations}\label{eq:rmt}
    \begin{align}
        \rxmt &\coloneqq \qty(\bigotimes_{i=1}^K\nxi{t\,\li\bimu\bimt}\otimes\im)\rxm\,, \label{eq:rmt1}\\
        \rymt &\coloneqq \qty\big(\Phi_B\otimes\im)\rxmt\,, \label{eq:rmt2}
    \end{align}
    \end{subequations}
    where $X=X_1\cdots X_K$, $0\leq\li\leq1$ and $\sum_{i=1}^K\li=1$\,.
    The time evolution of the subsystems is then given by:
    \begin{equation}\label{eq:rxim}
        \rximt = \qty(\nxit\otimes\im)\rxim\,.
    \end{equation}
    From Equation \eqref{eq:rmt1} and \lref{lem:diag-heatsemi}, it holds
    \begin{equation}
        \rxmt = \qty(\N_X\qty(t\,\btmu\Lambda\btmt)\otimes\1_M)\rxm\,.
    \end{equation}
    While from Equation \eqref{eq:rmt2} and \lref{def:PhiB}, it holds
    \begin{equation}
        \begin{split}
            \rymt &= \qty\Big(\Phi_B\otimes\1_M)\qty(\N_X\qty(t\,\btmu\Lambda\btmt)\otimes\1_M)\rxm\\
            &= \qty(\N_Y\qty(t\,B\btmu\Lambda\btmt B^T)\otimes\1_M)\rym\\
            &= \qty\Big(\N_Y(t\,I_{2n})\otimes\im)\rym\,,
        \end{split}
    \end{equation}
    where we have exploited the fact that $\sum_{i=1}^K\li=1$. We define the function
    \begin{equation}\label{eq:5.59}
        \phi(t) \ce \symt - \sum_{i=1}^K\lambda_i\sximt\,.
    \end{equation}
    From \pref{pro:gdbi} and \lref{lem:linearity}, the following applies
    \begin{equation}\label{eq:disug}
    \begin{split}
        \phi'(t) &= \jymt(I_{2n}) +\\
        &\phantom{=}- \sum_{i=1}^K\jximt \qty(\lambda_i^2\bimu\bimt)\\
        &\leq 0\,,
    \end{split}
    \end{equation}
    where the inequality was obtained from \tref{theo:QCondStamIneq}, with $\alpha_i=\bimu\Lambda_i\bimt$ (whence $\alpha = \btmu\Lambda\btmt$ and $B\,\alpha\, B^T=I_{2n}$).
    Inequality \eqref{eq:disug}, together with the \emph{asymptotic scaling} of the conditional entropy stated in \pref{prop:AsyScaling}, implies that
    \modifica{
    \begin{equation}\label{eq:phi0}
        \begin{split}
            \phi(0) &\geq \lim_{t\to+\infty}\phi(t)\\
            &= \lim_{t\to+\infty} \qty(\symt - \sum_{i=1}^K\lambda_i\sximt)\\
            &= \frac12\ln\qty\Big(\det\qty\big(e\,t\,I_{2n})) - \sum_{i=1}^K\frac12\lambda_i\ln\qty(\det(e\,t\,\li\bimu\bimt))\\
            &= n\,\qty(\sum_{i=1}^K\lambda_i\ln\frac{b_i}{\lambda_i})\,,
        \end{split}
    \end{equation}
    }
    where $b_i\ce|\det B_i|^\frac{1}{n}$. From inequality \eqref{eq:phi0} immediately follows the \emph{multimode linear conditional quantum Entropy Power Inequality}:
    \modifica{
    \begin{equation}\label{eq:MLCQEPI}
        \frac{\sym}{n} \geq \sum_{i=1}^K\lambda_i\frac{\sxim}{n} + \sum_{i=1}^K\lambda_i\ln(\frac{b_i}{\lambda_i})\,.
    \end{equation}
    }
    Maximising the RHS of Equation \eqref{eq:MLCQEPI} with respect to $\qty{\lambda_i\;:\;i=1,\ldots,K}$ (see \aref{sec:MLCQEPI}), \emph{i.e.}, choosing
    \begin{equation}\label{eq:maximization}
        \lambda_j = \frac{b_j\exp(\frac{\sxjm}{n})}{\sum_{i=1}^Kb_i\exp(\frac{\sxim}{n})}\,,
    \end{equation}
    for all $j$, we obtain precisely the MCQEPI in Equation \eqref{eq:MCQEPI}.
\end{proof}

\begin{observation}
    In the special case of $K=2$ and $S$ the symplectic matrices associated with the attenuator and the amplifier (see \eref{exam:beamspl_squeez}) we find the inequality proved in \cite[Theorem 6]{de2018conditional}.
\end{observation}
\section{The squashed entanglement of bosonic Gaussian states}\label{sec:squashed}

This section is devoted to determine upper and lower bounds to the squashed entanglement of the bosonic Gaussian states whose covariance matrix has half of the symplectic eigenvalues equal to $1/2$\,.

\subsection{The problem}\label{sec:the-prob}

Let $C$ and $D$ be two $n$-mode bosonic quantum systems. Let us define the family $\mathcal{F}$ of bosonic Gaussian states of the system $CD$ characterised by a covariance matrix with at least half of the symplectic eigenvalues equal to $1/2$. From Williamson's theorem (\tref{teo:Williamson}) and properties \eqref{eq:TraslMomPrim} and \eqref{eq:TraslMomSec}, every state $\op\rho_{CD}\in\mathcal{F}$ can be written in the following form:
\begin{equation}\label{eq:rho-state}
    \op\rho_{CD}^{S,\EE} = \dr\, \opU_{\!S}\, \qty\big(\op\omega_A(\EE) \otimes \ket{\0}_{\!B\!\!}\bra{\0})\, \opU_{\!S}^\dagger\, \drd \,,
\end{equation}
where $A$ and $B$ are suitable $n$-mode subsystems of $X$, $\opU_{\!S}: \H_A\otimes\H_B \to\H_C\otimes\H_D$ is the unitary transformation associated with an appropriate symplectic matrix $S\in\text{Sp}(4n,\RR)$, defined in \pref{prop:opUsimpl}, and $\op\omega_A(\EE)$ is defined as the tensor product of $n$ one-mode thermal states with average number of photons $\EE=\qty(\En_1,\ldots,\En_n)$:
\begin{equation}
    \op\omega_A(\EE) \coloneqq \op\omega_{A_1}(\En_1)\otimes\cdots\otimes\op\omega_{A_n}(\En_n) \,.
\end{equation}
The symplectic matrix $S$ defining the isometric transformation $\opU_{\!S}$ can be seen as a real $4n\times4n$ block matrix
\begin{equation}\label{eq:S-blocks}
    S = \mqty(B_1^C & B_2^C\\B_1^D & B_2^D)\,,
\end{equation}
where the blocks are $2n\times2n$ real-valued matrices. The first $2n$ rows and $2n$ columns of matrix $S$ act on system $A$ while the last $2n$ rows and $2n$ columns act on system $B$. The upper left block implements therefore the transformation on the system $A$ while the lower right block implements the transformation on system $B$.

\begin{remark}
    Since the translation operators act independently on each subsystem and the squashed entanglement is invariant under local unitary transformations (\pref{prop:SEinvariance}), then, in what follows, the translation operators $\dr$ in Equation \eqref{eq:rho-state} will be neglected in the calculation of the squashed entanglement.
\end{remark}

\begin{example}[squeezing]
    If $S$ is the squeezing symplectic matrix, then $B^C=\mqty(B_1^C&B_2^C)$ and $B^D=\mqty(B_1^D&B_2^D)$ (see \eref{ex:5.1}) are given by
    \begin{equation}
	\begin{split}
            &B_1^C = B_2^D = \diag{\sqrt{\kappa_1}\,I_2,\,\cdots\,,\sqrt{\kappa_n}\,I_2}\,,\\
		&B_1^D = B_2^C = \diag{\sqrt{\kappa_1-1}\,Z_2,\,\cdots\,,\sqrt{\kappa_n-1}\,Z_2}\,,
	\end{split}
    \end{equation}
    where \modifica{$\kappa_i>1\ \forall i$} and $I_2=\diag{1,1}$, $Z_2=\diag{1,-1}$.
    The output states are given by
    \begin{subequations}
        \begin{align}
            \op\rho_C^{S,\EE} = \Phi_{B^C}\qty\big(\op\omega_A(\EE)\otimes\ket{\0}_{\!B\!\!}\bra{\0})\,,\\
            \op\rho_D^{S,\EE} = \Phi_{B^D}\qty\big(\op\omega_A(\EE)\otimes\ket{\0}_{\!B\!\!}\bra{\0})\,.
        \end{align}
    \end{subequations}
\end{example}

Given the above definitions, the following theorem holds.
\begin{theorem}\label{theo:SqEnt-bounds}
    For any $S\in\text{Sp}(4n,\RR)$ as in Equation \eqref{eq:S-blocks} and any $\EE=\qty(\En_1,\ldots,\En_n)\in\mathbb{R}^n_{+}$, the squashed entanglement of the bosonic Gaussian state $\op\rho_{CD}^{S,\EE}$ defined in Equation \eqref{eq:rho-state} satisfies
    \begin{tcolorbox}[title = \textbf{\emph{Lower bound}}]
        \begin{equation}\label{eq:Esq}
            \frac{n}{2}\ln\!\qty\Big(2\,b_1^Cb_1^D+b_1^Db_2^C+b_1^Cb_2^D) \leq E_\text{sq}\qty(\op\rho_{CD}^{S,\EE})\,,
        \end{equation}
        \end{tcolorbox}
    \noindent where 
    $b_i^Y\coloneqq\abs{\det B_i^Y}^{1/n}$ for $i=1,2$ and $Y=C,D$\,, and
    \begin{tcolorbox}[title = \textbf{\emph{Upper bound}}]
    \modifica{
    \begin{equation}\label{eq:SqEnt-UpperBound}
        \begin{split}
            &E_\text{sq}\qty(\op\rho_{CD}^{S,\EE}) \\
            &\phantom{}\!\leq\!\frac14 \ln\!\qty\bigg[\!\det(\!B_{12\EE}^C\!) \det(\!C_\EE \!-\! B_\EE {B_1^C}^T \!\qty(\!B_{12\EE}^C\!)^{-1}\! {B_1^C} B_\EE)\!]\\
            &\phantom{\leq}\!+\!\frac14 \ln\!\qty\bigg[\!\det(\!B_{12\EE}^D\!) \det(\!C_\EE \!-\! B_\EE {B_1^D}^T \!\qty(\!B_{12\EE}^D\!)^{-1}\! {B_1^D} B_\EE)\!]\\
            &\phantom{\leq}\!+\!2n \!-\! \sum_{i=1}^n\,g\qty\big(\En_i/2)\,,
        \end{split}
    \end{equation}
    }
    \end{tcolorbox}
    \noindent where
    \begin{equation}\label{eq:g-function}
    \begin{split}
	&B_{12\EE}^Y \coloneqq \bigoplus_{i=1}^n\qty(\En_i+\frac12)B_1^Y{B_1^Y}^T+\frac12B_2^Y{B_2^Y}^T\,,\\
	&Z_{2n} \coloneqq \bigoplus_{i=1}^n\mqty(1&0\\0&-1)\,,
    \end{split}
    \end{equation}
    for $Y=C,D$ and the $g$ function as in Equation \eqref{eq:g-funct}\modifica{ and
    \begin{equation}
        B_\EE \coloneqq \bigoplus_{i=1}^n \sqrt{\En_i(\En_i+1)/2}\,Z_2\,,
        \quad
        C_\EE \coloneqq \bigoplus_{i=1}^n \qty(\frac{\En_i}{2} + \frac12)\,I_2\,.
    \end{equation}
    }
\end{theorem}

\begin{observation}
    The symplectic matrix $S$ which brings into normal form the covariance matrix of the state \eqref{eq:rho-state} is not unique. However, if exactly half of the symplectic eigenvalues of this state are equal to $1/2$, thanks to \lref{lem:SimplOrtogMatr}, the only freedom on the choice of this matrix is to multiply the diagonal blocks by orthogonal symplectic matrices. Since such matrices have unit determinant, they do not change the determinants of the blocks of the symplectic matrix and thus the lower bound of \tref{theo:SqEnt-bounds}.
\end{observation}

\subsection{Preliminary results}\label{sec:SE-PrelRes}
In this section we show some important preliminary results for the squashed entanglement.

\begin{proposition}[convexity]\label{prop:7.1}
    The squashed entanglement is a convex function of the state. In particular, given a finite alphabet $X$, an ensemble of quantum states $\{\op\rho^x_{AB}\}_{x\in X}$ and a probability distribution $p(x):X\to[0,1]$, it holds
    \begin{equation}
        E_\text{sq}\qty(\sum_{x\in X} p(x) \, \op\rho^x_{AB}) \leq \sum_{x\in X} p(x) \, E_\text{sq}\qty(\op\rho^x_{AB})
        \,.
    \end{equation}
\end{proposition}
\begin{proof}
    See \cite[Proposition\;3]{christandl2004squashed}.
\end{proof}

The following lemma provides a useful application of the multimode conditional quantum Entropy Power Inequality to a broad class of bosonic Gaussian states. This result will play a key role for the proof of \tref{theo:SqEnt-bounds}.

\begin{lemma}[application of the MCQEPI]\label{lem:appl-MCQEPI}
    Let $A$ be an $n$-mode bosonic quantum system, and let $R$ be a generic quantum system. Let $\op\gamma_{AR}$ be a joint quantum state of $AR$ such that its marginal $\op\gamma_A$ on $A$ has finite average energy and its marginal $\op\gamma_R$ on $R$ has finite entropy. Let
    \begin{equation}
        \begin{split}
            \op\rho_{CDR} &= \opU_{\!S}\,\qty(\op\rho_{ABR})\,\opU_{\!S}^\dagger\\
            &= \opU_{\!S}\,\qty(\op\gamma_{AR}\otimes\ket{\0}_{\!B\!\!}\bra{\0})\,\opU_{\!S}^\dagger\,,
        \end{split}
    \end{equation}
    where $\ket{\0}_B=\ket{0}_B^{\otimes n}$ is the vacuum state of system $B$ and $\opU_{\!S}$ is the $2n$-mode isometric operator acting on $AB$ defined in \pref{prop:opUsimpl}. Then,
    \modifica{
    \begin{subequations}\label{eq:2modeMCQEPI}
        \begin{align}
            S(C|R)_{\op\rho_{CDR}} & \geq n\, \ln(b_1^C \exp(\frac{S(A|R)_{\op\gamma_{AR}}}{n}) + b_2^C)\,, \label{eq:2modeMCQEPIa} \\
            S(D|R)_{\op\rho_{CDR}} & \geq n\, \ln(b_1^D\exp(\frac{S(A|R)_{\op\gamma_{AR}}}{n}) + b_2^D)\,, \label{eq:2modeMCQEPIb}
        \end{align}
    \end{subequations}
    }
    where the constants are defined as in \secref{sec:the-prob}.
\end{lemma}
\begin{proof}
    The multimode conditional quantum Entropy Power Inequality, proven in \tref{theo:MCQEPI}, states that
    \begin{equation}
        \exp(\frac{S(Y|R)_{\op\rho_{YR}}}{n}) \geq \sum_{i=1}^K b_i^Y\exp(\frac{S(X_i|R)_{\op\rho_{X_iR}}}{n})\,,
    \end{equation}
    where, in this case, $K=2$, $Y=C$ or $Y=D$, $X_1=A$, $X_2=B$ and $b_i^Y=\abs{\det B_i^Y}^{1/n}$. With these parameters, it holds:
    \begin{subequations}\label{eq:4.10}
        \begin{align}
            \exp\!\qty(\!\!\frac{S(\!C|R\!)_{\op\rho_{\!C\!R}}}{n}\!\!) &\!\geq\! b_1^C\!\! \exp\!\qty(\!\!\frac{S(\!A|R\!)_{\op\rho_{\!A\!R}}}{n}\!\!) \!+\! b_2^C\!\! \exp\!\qty(\!\!\frac{S(\!B|R\!)_{\op\rho_{\!B\!R}}}{n}\!\!)\,,\\
            \exp\!\qty(\!\!\frac{S(\!D|R\!)_{\op\rho_{\!D\!R}}}{n}\!\!) &\!\geq\! b_1^D\!\! \exp\qty(\!\!\frac{S(\!A|R\!)_{\op\rho_{\!A\!R}}}{n}\!\!) \!+\! b_2^D\!\! \exp\!\qty(\!\!\frac{S(\!B|R\!)_{\op\rho_{\!B\!R}}}{n}\!\!)\,.
        \end{align}
    \end{subequations}
    Let's now observe that
    \begin{equation}
        S(B|R)_{\op\rho_{BR}} = S(B|R)_{\op\gamma_{R}\otimes\ket{\0}_{\!B\!\!}\bra{\0}} = 0\,.
    \end{equation}
        In fact, taking advantage of the property $S(AB)_{\op\rho_A\otimes\op\rho_B}=S(A)_{\op\rho_A} + S(B)_{\op\rho_B}$ and the fact that the entropy of a pure state is zero, in our case $S(B)\coloneqq S(B)_{\ket{\0}_{\!B\!\!}\bra{\0}}=0$, we obtain:
    \begin{equation}
    \begin{split}
        S(B|R) &\coloneqq S(BR) - S(R)\\
        &= S(B) + S(R) - S(R) = S(B) = 0\,.
        \end{split}
    \end{equation}
    Equation \eqref{eq:4.10} is therefore rewritten as
    \begin{subequations}\label{eq:4.17ab}
        \begin{align}
            \exp\qty(\frac{S(C|R)_{\op\rho_{CR}}}{n}) &\geq b_1^C\exp\qty(\frac{S(A|R)_{\op\rho_{AR}}}{n}) + b_2^C\,,\\
            \exp\qty(\frac{S(D|R)_{\op\rho_{DR}}}{n}) &\geq b_1^D\exp\qty(\frac{S(A|R)_{\op\rho_{AR}}}{n}) + b_2^D\,.
        \end{align}
    \end{subequations}
    Hence, taking the logarithm of the LHS and RHS of Equations \eqref{eq:4.17ab} gives precisely the thesis, Equation \eqref{eq:2modeMCQEPI}.
\end{proof}

In the next lemma we state a result that will be useful for the proof of the lower bound of \tref{theo:SqEnt-bounds}.
\begin{lemma}\label{lem:7.5}
    Let $\op\rho_{CD}$ be a quantum state of the form $\op\rho_{CD}=\opU\qty\big(\op\sigma_A\otimes\ket{\psi}_{\!B\!\!}\bra{\psi})\opU^\dagger$ where $\opU$ is an isometric operator acting on system $AB$, then the extensions of this state will be all and only those of the form
    \begin{equation}\label{eq:estensione}
        \op\tau_{CDR}=\opU\qty\big(\op\sigma_{AR}\otimes\ket{\psi}_{\!B\!\!}\bra{\psi})\opU^\dagger\,,
    \end{equation}
    where $\op\sigma_{AR}$ is an extension of $\op\sigma_A$.
\end{lemma}
\begin{proof}
    Let us prove the two implications.
    \begin{enumerate}
        \item The fact that $\op\tau_{CDR}$ in Equation \eqref{eq:estensione} is an extension of $\op\rho_{CD}$ is obvious, indeed:
        \begin{equation}
            \begin{split}
                \tr_R\!\qty\big[\op\tau_{CDR}] &= \tr_R\!\qty[\opU\qty\big(\op\sigma_{AR}\otimes\ket{\psi}_{\!B\!\!}\bra{\psi})\opU^\dagger]\\
                &= \opU\qty\Big(\tr_R\!\qty[\op\sigma_{AR}]\otimes\ket{\psi}_{\!B\!\!}\bra{\psi})\opU^\dagger\\
                &= \opU\qty\big(\op\sigma_A\otimes\ket{\psi}_{\!B\!\!}\bra{\psi})\opU^\dagger\\
                &= \op\rho_{CD}\,.
            \end{split}
        \end{equation}
        \item The fact that all extensions of $\op\rho_{CD}$ are of the form \eqref{eq:estensione} is less obvious. Let $\op\tau_{CDR}$ be a generic extension of $\op\rho_{CD}$, then $\opU^\dagger\,\op\tau_{CDR}\,\opU$ is an extension of $\op\sigma_A\otimes\ket{\psi}_{\!B\!\!}\bra{\psi}$, indeed
        \begin{equation}
            \begin{split}
                \tr_R\!\qty[\opU^\dagger\op\tau_{CDR}\opU] = \op\sigma_A\otimes\ket{\psi}_{\!B\!\!}\bra{\psi}\,.
            \end{split}
        \end{equation}
        Then, $\opU^\dagger\op\tau_{CDR}\opU$ has the form
        \begin{equation}
            \opU^\dagger\op\tau_{CDR}\opU = \op\sigma_{AR}\otimes\ket{\psi}_{\!B\!\!}\bra{\psi}\,,
        \end{equation}
        for some quantum state $\op\sigma_{AR}$, extension of $\op\sigma_A$. From this follows Equation \eqref{eq:estensione}.
    \end{enumerate}
\end{proof}

We can now prove \tref{theo:SqEnt-bounds}.

\subsection{Lower bound}\label{sec:SE-LB}

Applying \lref{lem:appl-MCQEPI}, is now possible to provide a lower bound to the squashed entanglement of the state defined in Equation \eqref{eq:rho-state}. In the following we will distinguish the input systems $A$ and $B$ from the output systems $C$ and $D$ of the operator $\opU_S$. According to \lref{lem:7.5}, the extensions of the state $\op\rho_{CD}^{S,\EE}$ are all and only those of the form
\begin{equation}\label{eq:extension}
	\op\rho_{CDR} = \opU_S \qty(\op\omega_{AR} \otimes \ket{\0}_{\!B\!\!} \bra{\0}) \opU_S^\dagger \,,
	\quad
	\tr_R\!\qty[\op\omega_{AR}]=\op\omega_A(\EE)\,.
\end{equation}
The conditional mutual information of this state is therefore written as:
\begin{equation}\label{eq:4.21}
    \begin{split}
        I(\!C\!:\!D|R\!)_{\op\rho_{\!C\!D\!R}} &\!\coloneqq\! S(\!C|R\!)_{\op\rho_{\!C\!D\!R}} \!\!+\! S(\!D|R\!)_{\op\rho_{\!C\!D\!R}} \!\!-\! S(\!CD|R\!)_{\op\rho_{\!C\!D\!R}}\\
        &\!=\! S(\!C|R\!)_{\op\rho_{\!C\!D\!R}} \!\!+\! S(\!D|R\!)_{\op\rho_{\!C\!D\!R}} \!\!-\! S(\!A|R\!)_{\op\omega_{\!A\!R}}\,,
    \end{split}
\end{equation}
where we used the unitary invariance of the entropy, the additivity with respect to the tensor product between states and the fact that the entropy of a pure state is zero:
\begin{equation}
    \begin{split}
        S(\!C\!D|R\!)_{\op\rho_{\!C\!D\!R}} &\!\coloneqq\! S(\!C\!D\!R\!)_{\op\rho_{\!C\!D\!R}} \!\!-\! S(\!R\!)_{\op\rho_{\!C\!D\!R}}\\
        &\!=\! S(\!C\!D\!R\!)_{\opU_{\!S}\qty(\op\omega_{\!A\!R}\otimes\ket{\0}_{\!B\!}\bra{\0})\opU_S^\dagger} \!\!-\! S(\!R\!)_{\tr_{\!C\!D}\!\!\qty[\op\rho_{\!C\!D\!R}]}\\
        &\!=\! S(\!A\!B\!R\!)_{\op\omega_{\!A\!R}\otimes\ket{\0}_{\!B\!}\bra{\0}} \!\!-\! S(\!R\!)_{\tr_{\!A\!B}\!\!\qty[\op\omega_{\!A\!R}\otimes\ket{\0}_{\!\!B\!\!}\bra{\0}]}\\
        &\!=\! S(\!A\!R\!)_{\op\omega_{\!A\!R}} \!\!-\! S(\!R\!)_{\op\omega_{\!R}}\\
        &\!\coloneqq\! S(\!A|R\!)_{\op\omega_{\!A\!R}}
        \,.
    \end{split}
\end{equation}
For the sake of simplicity, let us now define $x\coloneqq S(A|R)_{\op\omega_{AR}}/n$. From \lref{lem:appl-MCQEPI}, we have
\modifica{
\begin{equation}\label{eq:4.22}
    \begin{split}
        I&(C:D|R)_{\op\rho_{CDR}}\\
        &\geq n\,\ln(b_1^C e^x + b_2^C) + n\,\ln(b_1^D e^x + b_2^D) - n\,\ln(e^x)\phantom{\Big[}\\
        &= n\,\ln\!\qty\Big(b_1^C b_1^D e^{x} + b_2^C b_2^D e^{-x} + \qty(b_2^C b_1^D + b_1^C b_2^D))\,.
    \end{split}
\end{equation}
}
Exploiting \lref{lem:propSimplMatrices}, \emph{i.e.}, $b_1^C b_1^D=b_2^C b_2^D$, inequality \eqref{eq:4.22} reads as
\begin{equation}\label{eq:4.23}
    \begin{split}
        I(C:D|R)_{\op\rho_{CDR}} &\geq n\,\ln\!\qty\Big(b_1^C b_1^D \qty(e^{x} + e^{-x}) + \qty(b_2^C b_1^D + b_1^C b_2^D))\\
        &= n\,\ln\!\qty\Big(2 b_1^C b_1^D \cosh(x) + \qty(b_2^C b_1^D + b_1^C b_2^D))\\
        &\geq n\,\ln\!\qty\Big(2 b_1^C b_1^D + b_2^C b_1^D + b_1^C b_2^D)\,,
    \end{split}
\end{equation}
from which the lower bound of Equation \eqref{eq:Esq} is obtained.

\begin{remark}
The lower bound \eqref{eq:Esq} can be saturated only when the last inequality in Equation \eqref{eq:4.23} is saturated, \emph{i.e.}, when $\cosh(x)=1$ and thus when the following condition holds:
\begin{equation}\label{eq:condizione}
    S(A|R)_{\op\omega_{AR}}=0 \,.
\end{equation}
\end{remark}
This remark tells us that if under certain conditions (e.g. for the number of photons $\EE$ of the thermal state $\op\omega_A(\EE)$ going to infinity) the gap between the lower bound and the upper bound we are going to study closes, then the extension $\op\omega_{AR}$ of the state $\op\omega_A(\EE)$ will necessarily have to satisfy the condition in Equation \eqref{eq:condizione}.

\begin{example}[squeezing transformation]
    If we consider the specific case in which the unitary operator $\opU_S$ is associated with a one-mode squeezing transformation with parameter $\kappa$, it holds
    \begin{equation}
        B_1^C = B_2^D = \sqrt{\kappa}\,I_2\,,
        \qquad
        B_1^D = B_2^C = \sqrt{\kappa-1}\,Z_2\,.
    \end{equation}
    Equation \eqref{eq:Esq}, with $n=1$, thus implies that
    \begin{equation}
        E_\text{sq}\qty(\op\rho_{CD}^{S,\En}) \geq \ln(2k - 1)\,,
    \end{equation}
    which is exactly the result proven in \cite[Theorem 1]{de2019squashed}.
\end{example}

\subsection{Upper bound}\label{sec:SE-UB}

To prove the upper bound in Equation \eqref{eq:SqEnt-UpperBound} we consider a collection of extensions of the state $\op\rho_{CD}^{S,\EE}$. First of all, given an $n$-mode bosonic quantum system, we define the noisless bosonic Gaussian attenuator on such a system as the tensor product of $n$ one-mode noisless bosonic Gaussian attenuators with attenuation parameters $\eta_i$ on each individual mode, \emph{i.e.}, $\mathcal{E}_{\vb*{\eta}} \coloneqq \mathcal{E}_{\eta_1}\otimes \cdots \otimes \mathcal{E}_{\eta_n}$. Specifically, let us consider the $n$-parameter family of extensions $\qty{\op\rho_{CDR}(\ee)}_{\ee\in[0,1]^n}$ of the form \eqref{eq:extension} where $R$ is a $n$-mode bosonic quantum system and
\begin{equation}\label{eq:estensioni}
    \op\omega_{AR}(\vb*{\eta}) = (\1_A \otimes \mathcal{E}_{\vb*{\eta}})(\ket{\phi_\EE}_{\!AR\!\!}\bra{\phi_\EE})
\end{equation}
is the bosonic Gaussian state obtained applying the noiseless bosonic Gaussian attenuator with attenuation parameter $\vb*{\eta}$ to half of the $2n$-mode squeezed vacuum state $\ket{\phi_\EE}_{AR}$ with average number of photons per mode $\En_i$. The state $\ket{\phi_\EE}_{\!AR\!\!}\bra{\phi_\EE}$ is defined as the tensor product of $n$ two-mode squeezed vacuum states.  We leave $\vb*{\eta}$ as a free parameter over which we will optimize in the end. The covariance matrix of $\op\omega_{AR}(\vb*{\eta})$ is\footnote{For the calculation of this covariance matrix, please refer to \aref{sec:CacoloDisigmaAReta}.}
\begin{equation}\label{eq:sigmaAReta}
    \sigma(\op\omega_{AR}(\vb*{\eta})) =
    \bigoplus_{i=1}^n
    \mqty(\qty(\En_i+\dfrac12)I_2 & \sqrt{\eta_i \En_i(\En_i+1)}\,Z_2\\
    \sqrt{\eta_i \En_i(\En_i+1)}\,Z_2 & \qty(\eta_i \En_i + \dfrac12)I_2)\,,
\end{equation}
where for convenience we rearranged the modes as $A_1R_1$ $\cdots$ $A_nR_n$, with $A_i$ and $R_i$ are one-mode bosonic quantum systems and $I_2=\diag{1,1}$ and $Z_2=\diag{1,-1}$.

\begin{lemma}\label{lem:7.6}
    The symplectic eigenvalues of $\sigma(\op\omega_{AR}(\vb*{\eta}))$ defined in Equation \eqref{eq:sigmaAReta} are
    \begin{equation}
        \nu_{2k-1}(\op\omega_{AR}(\vb*{\eta})) = (1-\eta_k)\En_k+\frac12\,,
        \qquad
        \nu_{2k}(\op\omega_{AR}(\vb*{\eta})) = \frac12\,,
    \end{equation}
    for $k=1,\ldots,n\,$.
\end{lemma}
\begin{proof}
    In the case $n=1$, covariance matrix \eqref{eq:sigmaAReta} is of the form
    \begin{equation}
        \sigma(\op\omega_{AR}(\eta)) = \mqty(\qty(\En+\dfrac12)I_2 & \sqrt{\eta \En(\En+1)}\,Z_2\\
        \sqrt{\eta \En(\En+1)}\,Z_2 & \qty(\eta \En+\dfrac12)I_2)\,,
    \end{equation}
    Its symplectic eigenvalues are
    \begin{equation}\label{eq:onemodeautovalsimpl}
        \nu_{+}(\op\omega_{AR}(\eta)) = (1-\eta)\En+\frac12\,,
        \qquad
        \nu_{-}(\op\omega_{AR}(\eta)) = \frac12\,.
    \end{equation}
    For the quantum state $\op\omega_{AR}(\vb*{\eta})=\bigotimes_{i=1}^n\op\omega_{A_iR_i}(\eta_i)$, the covariance matrix is
    \begin{equation}\label{eq:block-matrix}
        \sigma(\op\omega_{AR}(\vb*{\eta})) = \bigoplus_{i=1}^n\mqty(\qty(\En_i+\dfrac12)I_{2} & \sqrt{\eta_i \En_i(\En_i+1)}\,\sigma_z\\
        \sqrt{\eta_i \En_i(\En_i+1)}\,\sigma_z & \qty(\eta_i \En_i+\dfrac12)I_{2})\,.
    \end{equation}
    The matrices \eqref{eq:block-matrix} and $\Delta$ are block diagonal, then so will be the matrix $\Delta\,\sigma(\op\omega_{AR}(\vb*{\eta}))$. Since the eigenvalues of a diagonal block matrix are the eigenvalues of the individual blocks, it is clear that the symplectic eigenvalues of matrix \eqref{eq:block-matrix} are
    \begin{equation}\label{eq:autovalsimpl}
        \nu_{2k-1}(\op\omega_{AR}(\vb*{\eta})) = (1-\eta_k)\En_k+\frac12\,,
        \qquad
        \nu_{2k}(\op\omega_{AR}(\vb*{\eta})) = \frac12\,,
    \end{equation}
    with $k=1,\ldots,n$.
\end{proof}

From \pref{prop:7.4} and \lref{lem:7.6} it holds
\modifica{
\begin{equation}\label{eq:4.26}
    \begin{split}
        S(CDR)_{\op\rho_{CDR}(\ee)} &= S(AR)_{\op\omega_{AR}(\ee)} \\
        &= \sum_{k=1}^n\biggl[g\qty(\nu_{2k-1}(\op\omega_{A_kR_k}(\eta_k))-\frac12) +\\
        &\phantom{=}+g\qty(\nu_{2k}(\op\omega_{A_kR_k}(\eta_k))-\frac12)\biggr]\\
        &= \sum_{k=1}^n g\qty\big((1-\eta_k)\En_k)\,.
    \end{split}
\end{equation}
}
As far as entropy $S(R)$ is concerned, we must consider the blocks associated with the systems $R_1,\ldots,R_n$ of the matrix in Equation \eqref{eq:sigmaAReta}:
\modifica{
\begin{equation}\label{eq:4.28}
    \begin{split}
	S(R)_{\op\rho_{CDR}(\ee)} &= S(R)_{\op\omega_{AR}(\ee)} \\
	&= \sum_{k=1}^n g\qty(\nu_{k}(\op\omega_{R_k}(\eta_k)) -\frac12)\\
	&= \sum_{k=1}^n g\qty(\eta_k\En_k)\,.
    \end{split}
\end{equation}
}
Let $\op\rho_{CR}(\ee)$ and $\op\rho_{DR}(\ee)$ be the marginals of $\op\rho_{CDR}(\ee)$ on systems $CR$ and $DR$, respectively. They are bosonic Gaussian states with covariance matrices (see \aref{sec:A.2})
\begin{subequations}\label{eq:4.27}
    \begin{align}
        &\sigma(\op\rho_{CR}(\ee)) = \mqty(A_\EE\,{B_1^C}\,{B_1^C}^T + \dfrac12\,{B_2^C}\,{B_2^C}^T & {B_1^C}\,B_\EE\\B_\EE\,{B_1^C}^T & C_\EE)\,, \label{eq:4.27a} \\
        &\sigma(\op\rho_{DR}(\ee)) = \mqty(A_\EE\,{B_1^D}\,{B_1^D}^T + \dfrac12\,{B_2^D}\,{B_2^D}^T & {B_1^D}\,B_\EE\\B_\EE\,{B_1^D}^T & C_\EE)\,, \label{eq:4.27b}
    \end{align}
\end{subequations}
where we have defined the following quantities
\begin{equation}
\begin{split}
    &A_\EE\coloneqq\bigoplus_{i=1}^n\qty(\En_i+\frac12)\,I_2\,,\\
    \qquad
    &B_\EE\coloneqq\bigoplus_{i=1}^n\sqrt{\eta_i\,\En_i(\En_i+1)}\,Z_2\,,\\
    \qquad
    &C_\EE\coloneqq\bigoplus_{i=1}^n\qty(\eta_i\,\En_i+\frac12)\,I_2\,.
\end{split}
\end{equation}

We can now derive the upper bound of the squashed entanglement for the state $\op\rho_{CD}^{S,\EE}$. To do so we exploit the fact that the von Neumann entropy $S_Q(\gamma)$ of a bosonic Gaussian state with covariance matrix $\gamma$ satisfies the inequality given by the following
\begin{lemma}\label{lem:7.7}
    For any $\gamma\in\mathbb{R}^{2n\times2n}_{>0}$ with $\nu_\text{min}(\gamma)\geq1/2$ it holds
    \modifica{
    \begin{equation}\label{eq:SQ<=SG}
	S_G(\gamma) - \frac{n}{4 \nu_\text{min}(\gamma)^2}\ln(\frac{e}{2}) \leq S_Q(\gamma) \leq S_G(\gamma)\,,
    \end{equation}
    }
    where $\nu_\text{min}(\gamma)$ is the minimum symplectic eigenvalue of $\gamma$ and $S_G(\gamma)$ is the Shannon entropy of a Gaussian random variable with covariance matrix $\gamma$,
    \modifica{
    \begin{equation}\label{eq:disuguaglianza_G-Q}
	S_G(\gamma)=\frac12\ln(\det(e\,\gamma))\,.
    \end{equation}
    }
\end{lemma}
\begin{proof}
    See \cite[Lemma 9]{de2021generalized}.
\end{proof}
 
From Equations \eqref{eq:disuguaglianza_G-Q} and \eqref{eq:SQ<=SG}, it is therefore true that
\modifica{
\begin{equation}\label{eq:disGQ}
    \begin{split}
        S(CR)_{\op\rho_{CDR}(\ee)} &\leq S_G(\sigma(\op\rho_{CR}(\ee))) \\
        &= \frac12\ln(\det\!\qty\big(e\,\sigma(\op\rho_{CR}(\ee))))\\
        &= 2n + \frac12\ln(\det\!\qty\big(\sigma(\op\rho_{CR}(\ee))))\,,
    \end{split}
\end{equation}
}
where $n$ is the number of modes of the individual subsystems. Using the formula for the determinant of a block matrix\footnote{If $A$ is invertible, it holds
\begin{equation*}
    \det\mqty(A&B\\C&D)=\det(A)\det(D-CA^{-1}B)\,.
\end{equation*}}, Equation \eqref{eq:disGQ} becomes
\begin{equation}\label{eq:7.65}
    \begin{split}
        &S(CR)_{\op\rho_{CDR}(\ee)}\\
        &\phantom{i} \leq 2n \!+\! \frac12 \ln\biggl[\!\det(\!B_{12\EE}^C\!) \det(\!C_\EE \!-\! B_\EE {B_1^C}^T\! \qty(B_{12\EE}^C\!)^{-1}\! {B_1^C}\! B_\EE\!)\!\biggr]\,,
    \end{split}
\end{equation}
where, for typographical reasons, we have defined the matrix
\begin{equation}
    B_{12\EE}^Y = A_\EE\,B_1^Y\,{B_1^Y}^T + \frac12\,B_2^Y\,{B_2^Y}^T\,,
\end{equation}
where $Y=C,D$. Similarly (with the replacement of blocks $B_1^C\to B_1^D$ and $B_2^C\to B_2^D$) it holds
\begin{equation}
    \begin{split}
        &S(DR)_{\op\rho_{CDR}(\ee)}\\
        &\phantom{i} \leq 2n \!+\! \frac12 \ln\biggl[\!\det(\!B_{12\EE}^D\!) \det(\!C_\EE \!-\! B_\EE {B_1^D}^T\! \qty(B_{12\EE}^D)^{-1}\! {B_1^D}\! B_\EE\!)\!\biggr]\,.
    \end{split}
\end{equation}
Therefore, for any $0\leq\eta_i\leq1$ it holds
\modifica{
\begin{equation}
    \begin{split}
        E_\text{sq}&(\op\rho_{CD}^{S,\EE})\\
        &\leq \frac12 I(C:D|R)_{\op\rho_{CDR}(\ee)}\\
        &= \frac12 \Bigl(S(CR)_{\op\rho_{CDR}(\ee)} + S(DR)_{\op\rho_{CDR}(\ee)} +\\
        &\phantom{=} - S(R)_{\op\rho_{CDR}(\ee)} - S(CDR)_{\op\rho_{CDR}(\ee)}\Bigr)\\
        &= \frac14 \ln\biggl[\det(\!B_{12\EE}^C\!)\det(\!C_\EE \!-\! B_\EE {B_1^C}^T \qty(B_{12\EE}^C)^{-1} {B_1^C} B_\EE\!)\biggr]\\
        &\phantom{=}+ \frac14 \ln\biggl[\det(\!B_{12\EE}^D\!)\det(\!C_\EE \!-\! B_\EE {B_1^D}^T \qty(B_{12\EE}^D)^{-1} {B_1^D} B_\EE\!)\biggr]\\
        &\phantom{=}+ 2n - \frac12\sum_{i=1}^n g\qty\big(\eta_i \En_i) - \frac12\sum_{i=1}^n g\qty\big((1-\eta_i)\En_i)\,.
    \end{split}
\end{equation}
}
Now, from an explicit form of the individual blocks of the symplectic matrix, we could optimise the expression above to obtain the value of $\ee$ that minimises this upper bound. However, we know from Equation \eqref{eq:4.23} that the lower bound can be saturated only when $S(A|R)_{\op\omega_{AR}}=0$, which, with the notation of Equation \eqref{eq:estensioni} and from Equations \eqref{eq:4.26} and \eqref{eq:4.28} can happen for example for\footnote{Actually this condition can also be fulfilled for $E=0$, but we are mostly interested in the case $E>0$.} $\eta_i=1/2\ \forall i$. Indeed
\begin{equation}
    \begin{split}
	S(A|R)_{\op\omega_{AR}(\ee)} &\coloneqq S(AR)_{\op\omega_{AR}(\ee)} - S(R)_{\op\omega_{AR}(\ee)}\\
	&= \sum_{i=1}^n g((1-\eta_i)\En_i) - \sum_{i=1}^ng(\eta_i \En_i)\\
	&= 0\,,
    \end{split}
\end{equation}
which is fulfilled for $\eta_i=1/2\ \forall i$.
We can give an expression for the upper bound by setting $\eta_i=1/2\ \forall i$:
\modifica{
\begin{equation}\label{eq:RawUpperBound}
    \begin{split}
        E_\text{sq}&\qty(\op\rho_{CD}^{S,\EE})\\
        &\leq \frac14 \ln\biggl[\det(\!B_{12\EE}^C\!) \det(\!C_\EE \!-\! B_\EE {B_1^C}^T \!\qty(B_{12\EE}^C)^{-1}\! {B_1^C} B_\EE\!)\biggr]\\
        &\phantom{=}+ \frac14 \ln\biggl[\det(\!B_{12\EE}^D\!) \det(\!C_\EE \!-\! B_\EE {B_1^D}^T \!\qty(B_{12\EE}^D)^{-1}\! {B_1^D} B_\EE\!)\biggr]\\
        &\phantom{=}+ 2n - \sum_{i=1}^n\,g\qty\big(\En_i/2)\,,
    \end{split}
\end{equation}
}
which is the thesis, Equation \eqref{eq:SqEnt-UpperBound}.
\section{The squashed entanglement of multimode extreme bosonic Gaussian channels}\label{sec:channels}

In this section we determine the lower bound to the squashed entanglement of all the \emph{multimode extreme bosonic Gaussian channels}, \emph{i.e.}, the bosonic Gaussian channels that cannot be decomposed as a non-trivial convex combination of quantum channels. The proof will be based on the results obtained in \secref{sec:squashed}.

Every bosonic Gaussian channel $\mathcal{N}$ can be written in the physical representation as
\begin{equation}\label{eq:8.2}
    \mathcal{N}(\op\rho_S) = \tr_E\!\qty\Big[\opU_{SE}\,\qty\big(\op\rho_S\otimes\op\phi_E)\,\opU^\dagger_{SE}]\,,
\end{equation}
where $\op\phi_E$ is a Gaussian environment state with which the system interacts via the unitary transformation $\opU_{SE}$ associated to a symplectic matrix as in \pref{prop:opUsimpl}.

\subsection{Preliminary results}

This section is devoted to some preliminary results that are needed to prove the lower bound to the squashed entanglement of the multimode extreme bosonic Gaussian channels.

\begin{lemma}\label{lem:ChoiG}
    Let $A$ and $A'$ be $m$-mode bosonic quantum systems and let
    \begin{equation}\label{eq:cov-matrix}
        \Sigma = \left(
        \begin{array}{cc}
            \sigma_{AA} & \sigma_{AA'} \\
            \sigma_{A'A} & \sigma_{A'A'} \\
        \end{array}
        \right) \ge \pm\frac{i}{2}  \left(
        \begin{array}{cc}
            \Delta & 0 \\
            0 & \Delta \\
        \end{array}
        \right)
    \end{equation}
    be the covariance matrix of a pure bosonic Gaussian state such that the matrices $\sigma_{A'A'} \pm \frac{i}{2} \Delta$ are both invertible. Let $K,\,\alpha\in\mathbb{R}^{2n\times2n}$. Then, $\alpha \ge \pm \frac{i}{2} (\Delta - K\,\Delta\,K^T)$ iff
    \begin{equation}\label{eq:channel-cov-matrix}
        \Sigma' = \left(
        \begin{array}{cc}
            K\,\sigma_{AA}\,K^T + \alpha & K\,\sigma_{AA'} \\
            \sigma_{A'A}\,K^T & \sigma_{A'A'} \\
        \end{array}
        \right) \ge \pm\frac{i}{2}  \left(
        \begin{array}{cc}
            \Delta & 0 \\
            0 & \Delta \\
        \end{array}
        \right)\,.
    \end{equation}
\end{lemma}
\begin{proof}
    Let us show the two implications separately.
    
    \paragraph{``$\implies$''}
    Inequality \eqref{eq:channel-cov-matrix} is equivalent to the following
    \begin{equation}
        \mqty(K\,\sigma_{AA}\,K^T+\alpha\mp\frac{i}{2}\Delta & K\,\sigma_{AA'} \\ \sigma_{A'A}\,K^T & \sigma_{A'A'}\mp\frac{i}{2}\Delta) \geq 0\,.
    \end{equation}
    By hypothesis, conditions \eqref{eq:cov-matrix} and $\alpha\geq\pm\frac{i}{2}\qty(\Delta - K\,\Delta\,K^T)$ imply that
    \begin{equation}
        \begin{split}
            &\mqty(K\,\sigma_{AA}\,K^T+\alpha\mp\frac{i}{2}\Delta & K\,\sigma_{AA'} \\ \sigma_{A'A}\,K^T & \sigma_{A'A'}\mp\frac{i}{2}\Delta)\\
            &\phantom{mm}\geq
            \mqty(K\,\qty(\sigma_{AA}\mp\frac{i}{2}\Delta)\,K^T & K\,\sigma_{AA'} \\ \sigma_{A'A}\,K^T & \sigma_{A'A'}\mp\frac{i}{2}\Delta)\\
            &\phantom{mm}=
            \mqty(K&0\\0&I)\mqty(\sigma_{AA}\mp\frac{i}{2}\Delta & \sigma_{AA'} \\ \sigma_{A'A} & \sigma_{A'A'}\mp\frac{i}{2}\Delta)\mqty(K^T&0\\0&I)\\
            &\phantom{mm}\geq0\,.
        \end{split}
    \end{equation}
    
    \paragraph{``$\Longleftarrow$''}
    Given condition \eqref{eq:channel-cov-matrix} and being $\sigma_{A'A'}\pm\frac{i}{2}\Delta$ invertible by hypothesis, \lref{prop:positività-blocchi} ensures that
    \begin{equation}\label{eq:eq-positiv}
        K\,\sigma_{\!AA}\,K^T + \alpha \mp \frac{i}{2}\Delta - K\,\sigma_{\!AA'}\qty(\!\sigma_{\!A'\!A'}\mp\frac{i}{2}\Delta\!)^{-1}\!\!\!\sigma_{\!A'\!A'}\,K^T \geq 0
        \,.
    \end{equation}
    Being, by hypothesis, $\Sigma$ the covariance matrix of a pure bosonic Gaussian state, \lref{lem:autoval-simpl} and \lref{lem:autoval-simpl-vs-autoval} imply $\rank(\Sigma\pm\frac{i}{2}\Delta)=2n$. By hypothesis $\sigma_{A'A'}\pm\frac{i}{2}\Delta$ is invertible, so $\rank(\sigma_{A'A'}\pm\frac{i}{2}\Delta) = 2n = \rank(\Sigma\pm\frac{i}{2}\Delta)$. Since $\rank(\Sigma\pm\frac{i}{2}\Delta) = \rank(\sigma_{A'A'}\pm\frac{i}{2}\Delta)$, \pref{prop:Schur} implies that
    \begin{equation}\label{eq:GRankFormula}
        \sigma_{AA} \mp \frac{i}{2}\Delta = \sigma_{AA'}\qty(\sigma_{AA}\mp\frac{i}{2}\Delta)^{-1}\!\!\sigma_{A'A}\,.
    \end{equation}
    Substituting Equation \eqref{eq:GRankFormula} into Equation \eqref{eq:eq-positiv}, it holds
    \begin{equation}
        \alpha \geq \pm \frac{i}{2}\qty(\Delta - K\Delta K^T)\,.
    \end{equation}
\end{proof}

\begin{lemma}\label{lem:sympl}
    Let $\Sigma'\in\mathbb{R}^{4n\times4n}$ satisfy
    \begin{equation}
        \Sigma' \ge \pm\frac{i}{2}\left(
        \begin{array}{cc}
            \Delta & 0 \\
            0 & \Delta \\
        \end{array}
        \right)\,.
    \end{equation}
    Let us assume that the only $\delta\in\mathbb{R}^{2n\times2n}$ such that $\delta\ge0$ and
    \begin{equation}\label{eq:11}
        \Sigma' - \left(
        \begin{array}{cc}
            \delta & 0 \\
            0 & 0 \\
        \end{array}
        \right)  \ge \pm\frac{i}{2}\left(
        \begin{array}{cc}
            \Delta & 0 \\
            0 & \Delta \\
        \end{array}
        \right)
    \end{equation}
    is $\delta=0$.
    Then, $\Sigma'$ has at least $n$ symplectic eigenvalues equal to $1/2$\,.
\end{lemma}
\begin{proof}
    Let us assume that $\Sigma'$ has at least $n+1$ symplectic eigenvalues strictly larger than $1/2$\,. Therefore, \lref{lem:indet} and \lref{lem:autoval-simpl-vs-autoval} ensure that $\Sigma'+\frac{i}{2}\Delta$ has rank at least $2n + n + 1 = 3n+1$. Then, the intersection between the support of $\Sigma'+\frac{i}{2}\Delta$ and the span of the first $2n$ vectors of the canonical basis of $\mathbb{C}^{4n}$ has dimension at least $3n+1 + 2n - 4n = n+1$. The statement follows if we prove that such an intersection contains at least one unit real vector $\vb{v}\in\mathbb{R}^{4n}$ by choosing $\delta = \varepsilon\, \vb{v}\,\vb{v}^T$ with $\varepsilon$ strictly smaller than the smallest eigenvector of the matrix $\Sigma'+\frac{i}{2}\Delta\,$.
	
    Since $\mathbb{C}$ is isomorphic to $\mathbb{R}^2$, we can identify the vector space $\mathbb{C}^{2n}$ given by the span of the first $2n$ vectors of the canonical basis of $\mathbb{C}^{4n}$ with $W=\mathbb{R}^{4n}$. That is, we can identify $\mathbb{C}^{2n}$ with the set of ordered pairs $\qty(\vb{v},\vb{w})$, where $\vb{v},\vb{w}\in\mathbb{R}^{2n}$ are the real and the imaginary part, respectively, of the complex vector. By this analogy, the space given by the intersection of the support of $\Sigma'+\frac{i}{2}\Delta$ and the span of the first $2n$ vectors of the canonical basis of $\mathbb{C}^{4n}$ will be $V=\mathbb{R}^{2n+2}$. We therefore want to show that $V\subset W$ contains a vector with only real components, that is, a vector of the form $\qty(\vb{r},0)$. This must be necessarily true because the orthogonal space of a subspace of dimension $2n+2$ has dimension $2n-2$ and if in $\mathbb{R}^{2n}$ we impose the orthogonality condition to $2n-2$ vectors the solutions have at least dimension $2$.
	
    The proof is concluded by the general property which states that subtracting from a positive semidefinite matrix $\varepsilon\, \vb{v}\,\vb{v}^T$, where $\vb{v}$ is a unit vector in the matrix support, and $\varepsilon$ is smaller than the smallest eigenvector of the matrix, then the matrix is still positive semidefinite.
\end{proof}

\begin{lemma}\label{lem:AutovalUnMezzo}
    With the same notation of \lref{lem:ChoiG}, let the $(K,\alpha)$ Gaussian channel be extreme. Then, $\Sigma'$ has at least $n$ symplectic eigenvalues equal to $1/2$.
\end{lemma}
\begin{proof}
    Let $\delta\in\mathbb{R}^{2n\times2n}$, $\delta\ge0$ and let us assume that
    \begin{equation}
	\Sigma' - \left(
	\begin{array}{cc}
		\delta & 0 \\
		0 & 0 \\
	\end{array}
	\right) \ge \pm\frac{i}{2}\left(
	\begin{array}{cc}
		\Delta & 0 \\
		0 & \Delta \\
	\end{array}
	\right)\,.
    \end{equation}
    \lref{lem:ChoiG} (with $\alpha\to\alpha-\delta$) implies that
    \begin{equation}
	\alpha \ge \alpha-\delta \ge \pm\frac{i}{2}(\Delta - K\,\Delta\,K^T)\,,
    \end{equation}
    which means $\delta=0$, being the $(K,\alpha)$ Gaussian channel extreme. The claim follows from \lref{lem:sympl}.
\end{proof}

Thanks to these observations, we can now state the following proposition, which will play an important role throughout the rest of the work.
\begin{proposition}\label{prop:extreme_autoval_un_mezzo}
    Let $\op\rho$ be a $2n$-mode pure bosonic Gaussian state and $\Phi$ an extreme bosonic quantum channel acting on $n$ modes, then the covariance matrix of the state $(\1~\otimes~\Phi)(\op\rho)$ has at least half of its symplectic eigenvalues equal to $1/2$.
\end{proposition}
\begin{proof}
    This is a direct consequence of \lsref{lem:ChoiG}, \ref{lem:sympl} and \ref{lem:AutovalUnMezzo}.
\end{proof}

\subsection[Extremality condition]{Extremality condition}\label{sec:ExtremalityCond}

In this subsection we provide an algorithm to check that a given quantum channel fulfils the condition of being extreme.
\begin{algorithm}[htb!]
    \caption{\centering \emph{Test for the extremality condition}}
    \begin{algorithmic}
        \Input Two matrices $K,\alpha\in\mathbb{R}^{2n\times2n}$ describing a bosonic Gaussian channel.
        \Output \texttt{True} if the channel is extreme, \texttt{False} otherwise.
        \Proced
        \State 1. Output \texttt{True} if the following equality is satisfied, \texttt{False} otherwise:
        \begin{equation*}
            -\qty(\alpha\qty(\Delta-K\,\Delta\,K^T)^{-1})^2=\frac{I}{4}\,.
        \end{equation*}
    \end{algorithmic}
    \label{alg:ExtremalityCondition}
\end{algorithm}

\begin{proposition}
    \alref{alg:ExtremalityCondition} provides a method for verifying that a bosonic Gaussian channel is extreme.
\end{proposition}
\begin{proof}
    As stated in \pref{prop:extreme}, a bosonic Gaussian channel defined by the matrices $(K,\alpha)$ is extreme if and only if it does not exist a matrix $\gamma\neq\alpha$ such that $\alpha\geq\gamma\geq\pm\frac{i}{2}\qty(\Delta - K\, \Delta\, K^T)\coloneqq\pm\frac{i}{2}\Delta'$. To verify that a $(K,\alpha)$ Gaussian channel is extreme we have to check that $\alpha$ is minimal among all matrices satisfying $\alpha\geq\pm\frac{i}{2}\Delta'$. The condition $\alpha\geq\pm\frac{i}{2}\Delta'$ tells us that $\alpha$ is a covariance matrix with respect to the symplectic form $\Delta'$. A covariance matrix of a quantum state is minimal if and only if its symplectic eigenvalues are all equal to $1/2$. Recalling \dref{def:symplectic_eigenvalues}, we are therefore interested in the moduli of the eigenvalues of $\alpha\,\Delta'^{-1}$. Requiring $\alpha$ to be minimal is then equivalent to require that it has all symplectic eigenvalues equal to $1/2$, or alternatively that the matrix $\alpha\,\Delta'^{-1}$ has all eigenvalues equal to $\pm i/2$, since these are pure imaginary and pairwise opposite, $\alpha$ being a real matrix. This is then equivalent to requiring that the eigenvalues of $-\qty(-\alpha\,\Delta'^{-1})^2$ are all equal to $-(\pm i/2)^2=1/4$. Being a matrix proportional to the identity, \emph{i.e.}, with all eigenvalues identical, equal in all bases, this implies that for $\alpha$ to be minimal $-\qty(-\alpha\,\Delta'^{-1})^2$ must equal to $I/4$.
\end{proof}

\subsection{Upper bound}\label{subs:upperbound}

In this section we provide an algorithm to compute an upper bound to the squashed entanglement of any multimode (not necessarily extreme) bosonic Gaussian channel. The squashed entanglement of a quantum channel $\mathcal{N}_{A\to B}$ can also be written in the alternative form \cite[Lemma 4]{takeoka2014squashed}:
\begin{equation}\label{eq:SqEntChannel}
    E_\text{sq}(\mathcal{N}) = \sup_{\oprho_{A}}\frac12 \inf_{\mathcal{S}_{E'\to E''}} \qty\Big(S(B|E'')_{\op\psi} + S(B|F')_{\op\psi})\,,
\end{equation}
where $\op\psi=\op\psi_{BE''F'}$ is the quantum state given by
\begin{equation}\label{eq:SqEntState}
    \op\psi_{BE''F'} = \mathcal{U}^\mathcal{S}_{E'F\to E''F'}\,\mathcal{U}^\N_{AE\to BE'}\,\qty\big(\oprho_{A}\otimes\ket{\0}_{\!E\!\!}\bra{\0}\otimes\ket{\0}_{\!F\!\!}\bra{\0})\,,
\end{equation}
where $\mathcal{U}^\mathcal{S}_{E'F\to E''F'}$ and $\mathcal{U}^\N_{AE\to BE'}$ are the unitary quantum channels associated to the unitary dilations of the \emph{squashing channel} $\mathcal{S}_{E'\to E''}$ and the considered channel $\mathcal{N}_{A\to B}$ respectively.

\begin{theorem}\label{theo:SqUpperBound}
    The squashed entanglement of a bosonic Gaussian channel $\mathcal{N}$ satisfies
    \begin{tcolorbox}[title = \textbf{\emph{Upper bound}}]
        \begin{equation}\label{eq:EsqChannelLimit}
            E_\text{sq}(\mathcal{N}) \leq \lim_{t\to+\infty} \frac12 \qty(S(B|E'')_{\op\psi(t)} + S(B|F')_{\op\psi(t)})\,,
	\end{equation}
    \end{tcolorbox}
    \noindent where $\op\psi(t)$ is the quantum state given by
    \begin{equation}\label{eq:oppsi}
        \op\psi_{BE''F'}(t) \coloneqq \mathcal{U}^{\mathcal{L}}_{E'F\to E''F'}\, \mathcal{U}^{\mathcal{N}}_{AE\to BE'}\, \qty\big(\oprho_{A}\otimes\ket{\0}_{\!E\!\!}\bra{\0}\otimes\ket{\0}_{\!F\!\!}\bra{\0})\,.
    \end{equation}
    $\mathcal{U}^{\mathcal{L}}_{E'F\to E''F'}$ is a unitary quantum channels associated to the unitary dilation of a multimode extreme Gaussian attenuator $\mathcal{L}:E'\to E''$ with transmissivity $1/2$ on all modes and the environment state a generic pure Gaussian state. $\mathcal{U}^{\mathcal{N}}_{AE\to BE'}$ is a unitary quantum channels associated to the unitary dilation of the channel $\mathcal{N}:A\to B$. $\oprho_{A}(t)$ is a bosonic Gaussian state with covariance matrix $t\,\gamma_{A}$, where $\gamma_{A}$ is an arbitrary positive-definite fixed matrix.
\end{theorem}

\begin{proof}
    Our proof is divided into two parts. In the first part we show that
    \begin{equation}\label{eq:EsqChannelSup}
        E_\text{sq}(\mathcal{N}) \leq \sup_{\gamma_{A}} \frac12 \qty(S(B|E'')_{\op\psi(t)}+S(B|F')_{\op\psi(t)})\,,
    \end{equation}
    where $\op\psi(t)$ is defined in Equation \eqref{eq:oppsi}. In the second part, we show that the $\sup$ in Equation \eqref{eq:EsqChannelSup} can be restricted to matrices proportional to a given matrix and that Equation \eqref{eq:EsqChannelSup} is equivalent to Equation \eqref{eq:EsqChannelLimit}.
	
    It is convenient to use the equivalent definition \eqref{eq:SqEntChannel} of the squashed entanglement of a quantum channel. In what follows, we will consider the channel in Stinespring representation and use the notation of Equation \eqref{eq:8.2}. It is not an easy task to optimize over the squashing channel $\mathcal{S}$. Instead, we consider a specific squashing channel: a multimode extreme Gaussian attenuator $\mathcal{S} = \mathcal{L}$ with transmissivity $1/2$ on all modes and the environment state a generic pure bosonic Gaussian state. Since the overall channel from $A$ to $BE''$ is Gaussian and the overall channel from $A$ to $BF'$ is also Gaussian, it follows from \pref{prop:GaussStMax} that the maximization in Equation \eqref{eq:SqEntChannel} can be restricted to the set of Gaussian states. The covariance matrix of the initial state in the system $AEF$ is given by $\gamma_{A}\oplus \sigma_E\oplus \sigma_F\,$, where $\sigma_E$ is the covariance matrix of the state of the environment when we consider the physical representation of the channel $\mathcal{N}$ as in Equation \eqref{eq:8.2}. The beamsplitting operation and the symplectic operation act on the covariance matrices as
    \begin{equation}
        \begin{split}
        &\gamma_{A}\oplus\sigma_E\oplus\sigma_F\\
        &\longrightarrow\!
        \qty(\!I_{2n} \!\oplus\! S_{1/2}\!) \qty(\!S \!\oplus\! I_{2n}\!) \qty(\!\gamma_{\!A} \!\oplus\! \sigma_{\!E} \!\oplus\! \sigma_{\!F}\!) \qty(\!S \!\oplus\! I_{2n}\!)^{T} \qty(\!I_{2n} \!\oplus\! S_{1/2}\!)^{T},
        \end{split}
    \end{equation}
    where
    \begin{equation}\label{eq:S1/2}
        S_{1/2} = \frac{1}{\sqrt{2}}\mqty(I_{2n} & I_{2n} \\ -I_{2n} & I_{2n})
        \quad\text{ and }\quad
        S = \mqty(B_1^C&B_2^C\\B_1^D&B_2^D)\,.
    \end{equation}
    We obtain
    \begin{equation}\label{eq:8.9}
        \begin{split}
            E_\text{sq}(\mathcal{N})
            &\leq \sup_{\oprho_{A}}\frac12 \qty\Big(S(B|E'')_{\op\psi(t)} + S(B|F')_{\op\psi(t)})\\
            &= \sup_{\gamma_{A}} \frac12 \qty(S(B|E'')_{\op\psi(t)}+S(B|F')_{\op\psi(t)})\,,
        \end{split}
    \end{equation}
    where we have replaced the $\sup$ on Gaussian states with the sup on covariance matrices, the expression in brackets being dependent only on the latter, according to \pref{prop:7.4}, and where $\op\psi(t)$ is the quantum state given by Equation \eqref{eq:oppsi}. With the same notation as in \lref{lem:Hbar-monotonicity}, given a Gaussian state $\oprho^\gamma_{A}$ with covariance matrix $\gamma$ and a bosonic Gaussian channel $\Phi:A\to BE''F'$, we define the function
    \begin{equation}\label{eq:Sbar}
        \bar{S}(\gamma) \coloneqq S(B|E'')_{\Phi(\oprho^\gamma_{A})} + S(B|F')_{\Phi(\oprho^\gamma_{A})}\,.
    \end{equation}
    
    Given a generic positive definite matrix $\alpha$ and a parameter $t$\footnote{Every positive definite matrix multiplied by a sufficiently large constant $t$ is a covariance matrix, \emph{i.e.}, has all symplectic eigenvalues greater than or equal to $1/2$ (the latter being linear in $t$).}, let us now prove that $\sup$ over all possible covariance matrices of the function $\bar{S}$ defined in Equation \eqref{eq:Sbar} is equivalent to evaluating that function in the matrix $t\,\alpha$ and then taking the limit for $t$ going to infinity. First of all, the $\sup$ of $\bar{S}$ over all possible covariance matrices is surely greater than or equal to the limit for $t$ going to infinity of $\bar{S}(t\,\alpha)$, since the latter case is limited to a particular class of covariance matrices: $\sup_{\gamma_{A}} \bar{S}(\gamma_{A}) \geq \lim_{t\to+\infty} \bar{S}(t\,\alpha)$. On the other hand, since the quantity in Equation \eqref{eq:Sbar} is increasing in the covariance matrix of the input state $\oprho_{A}$ of the channel, from \lref{lem:Hbar-monotonicity}, if we consider any covariance matrix $\gamma_{A}$, for $t$ sufficiently large, $t\,\alpha$ will be greater than or equal to the matrix $\gamma_{A}$, we therefore have $\sup_{\gamma_{A}} \bar{S}(\gamma_{A}) \leq \lim_{t\to+\infty} \bar{S}(t\,\alpha)$. These two inequalities imply that
    \begin{equation}
        \sup_{\gamma_{A}}\bar{S}(\gamma_{A}) = \lim_{t\to+\infty}\bar{S}(t\,\alpha)\,.
    \end{equation}
\end{proof}

\begin{remark}
    The particular choice we have made for the squashing channel is due to the fact that this choice is known to be optimal for the one-mode extreme bosonic Gaussian attenuator and amplifier.
\end{remark}

\begin{remark}
    This result greatly simplifies the calculation of the upper bound of the squashed entanglement of a quantum channel. We have in fact reduced the sup over an infinite number of parameters to a limit in a single variable, thus avoiding multiple variable optimisations through, for example, gradient descent.
\end{remark}

From \tref{theo:SqUpperBound}, it is easy to see that the following algorithm provides a method for calculating an upper bound of the form \eqref{eq:EsqChannelLimit} to the squashed entanglement of a generic multimode bosonic Gaussian channel.

\begin{algorithm}[H]
    \caption{\centering \emph{Upper bound to $E_\text{sq}(\mathcal{N})$}}
    \begin{algorithmic}
        \Input Two matrices $K,\alpha\in\mathbb{R}^{2n\times2n}$ describing a bosonic Gaussian channel $\mathcal{N}$.
        \Output A constant $C_{UB}$ such that $E_\text{sq}(\mathcal{N})\leq C_{UB}\,.$
        \Proced
        \State Calculate numerically the \modifica{limit $C_{UB}=\lim_{t\to+\infty}C_{UB}(t)$}, where the quantity $C_{UB}(t)$ is computed as follows.
        \State \indent 1. From matrices $K,\alpha$, derive an environment-state covariance matrix $\sigma_E\in\mathbb{R}^{2n\times2n}_{>0}$ and a coupling symplectic matrix $S\in\text{Sp}(4n,\mathbb{R})$ describing the bosonic Gaussian channel in Stinespring representation as in \cite[Section 5.3]{serafini2017quantum}.
        \State \indent 2. Evaluate the matrix $\sigma_{BE''F'}(t)$:
            \begin{equation}\label{eq:sigmaBEsFp}
                \begin{split}
                &(\!I_{2n} \!\oplus\! S_{1/2}\!)
                (\!S \!\oplus\! I_{2n}\!)
                (\!t\,I_{2n} \!\oplus\! \sigma_{\!E} \!\oplus\! \sigma_{\!F}\!)
                (\!S \!\oplus\! I_{2n}\!)^{\!T}
                (\!I_{2n} \!\oplus\! S_{1/2}\!)^{\!T}\!,
                \end{split}
            \end{equation}
            where $S_{1/2}$ is defined in Equation \eqref{eq:S1/2} and $\sigma_{F}$ the environment-state covariance matrix of a multimode bosonic Gaussian channel.
        \State \indent 3. Calculate the symplectic eigenvalues $\qty{\nu_\sigma^j(t)}_j$ of the matrices: $\sigma_{E''}(t)\,,\sigma_{F'}(t)\,,\sigma_{BE''}(t)\,,\sigma_{BF'}(t)$\,;
        \State \indent 4. Compute the quantity:
            \modifica{
            \begin{equation}\label{eq:C_UB}
                \begin{split}
                    C_{UB}(t) &= \frac12\, \sum_{j=1}^{2n} \biggl[g\qty(\nu^j_{\sigma_{BE''}}(t) - \frac12) + g\qty(\nu^j_{\sigma_{BF'}}(t)-\frac12)\biggr] +\\
                    &\phantom{=}- \frac12\, \sum_{j=1}^{n} \biggl[g\qty(\nu^j_{\sigma_{E''}}(t)-\frac12) + \sum_{j=1}^{n} g\qty(\nu^j_{\sigma_{F'}}(t) -\frac12)\biggr]\,.
                \end{split}
            \end{equation}
            }
    \end{algorithmic}
    \label{alg:UpperBound}
\end{algorithm}

\subsection{Lower bound}\label{sec:SE-lower-bound}

In this section, we provide an algorithm for calculating a lower bound to the squashed entanglement of any multimode extreme bosonic Gaussian channel. The proof of this algorithm is based on \tref{theo:SqEnt-bounds} and on the definition \eqref{eq:SqEntChannel} of the squashed entanglement of a quantum channel. Regarding this last definition, the results for attenuator and amplifier \cite{de2019squashed} lead us to formulate the following

\begin{conjecture}\label{congj:OptimalGaussianChannel}
    For any bosonic Gaussian channel $\N$, the optimal squashing channel in Equation \eqref{eq:SqEntState} is always a bosonic Gaussian channel.
\end{conjecture}

\begin{algorithm}[H]
    \caption{\centering \emph{Lower bound to $E_\text{sq}(\mathcal{N})$}}
    \begin{algorithmic}
        \Input Two matrices $K,\alpha\in\mathbb{R}^{2n\times2n}$ describing an extreme bosonic Gaussian channel $\mathcal{N}$, an average number of photons $N$.
        \Output A constant $C_{LB}$ such that $C_{LB}\leq E_\text{sq}(\mathcal{N})\,.$
        \Proced
        \State Calculate numerically the \modifica{limit $C_{LB}=\lim_{N\to+\infty}C_{LB}(N)$}, where the quantity $C_{LB}(N)$ is computed as follows.
        \State \qquad 1. Build the matrix:
        \begin{equation}\label{eq:sigmaCDN}
            \sigma_{CD}(N) = (I_{2n} \oplus K)\, \sigma_{AB}(N)\, (I_{2n} \oplus K)^T + (0_{2n} \oplus \alpha)\,,
        \end{equation}
        where $\sigma_{AB}(N) \in \mathbb{R}^{4n\times4n}_{>0}$ is the covariance matrix of the tensor product of $n$ two-mode squeezed vacuum states with average number of photons per mode $N$.
        \State \qquad 2. Find a symplectic matrix (which exists according to Williamson's theorem, \tref{teo:Williamson}) $S\in\text{Sp}(4n,\mathbb{R})$ such that the matrix
        \begin{equation*}
            S\, \sigma_{CD}(N)\, S^T
        \end{equation*}
        is diagonal.
        \State \qquad 3. With the same notation of Equation \eqref{eq:S-blocks}, compute the quantity:
        \begin{equation}\label{eq:C_LB}
            C_{LB}(N) = \dfrac{n}{2} \ln\!\qty\Big(2\, b_1^C b_1^D + b_1^D b_2^C + b_1^C b_2^D)\,.
        \end{equation}
    \end{algorithmic}
    \label{alg:LowerBound1}
\end{algorithm}

\begin{theorem}
    Every constant $C_{LB}(N)$ computed by the \alref{alg:LowerBound1} is a lower bound to the squashed entanglement of a generic multimode extreme bosonic Gaussian channel.
\end{theorem}
\begin{proof}
    Any choice of the input state will provide a lower bound to the squashed entanglement of the quantum channel. We choose as input state the tensor product $\ket{\phi_\EE}_{\!AR\!\!}\bra{\phi_\EE}$ of $n$ two-mode squeezed vacuum states with average number of photons per mode $N$. From \pref{prop:extreme_autoval_un_mezzo}, when an extreme bosonic Gaussian channel acts on half of a pure Gaussian state the covariance matrix of the resulting output state has at least half of its symplectic eigenvalues equal to $1/2$. This insight enables us to represent the output state in the form given by Equation \eqref{eq:rho-state}. In particular, the proof of Williamson's theorem (\tref{teo:Williamson}) states that a symplectic matrix that diagonalises the covariance matrix $\sigma_{CD}$ is given by $S = M \, \sigma_{CD}^{-1/2}$, where $M$ is a non-singular matrix that brings the antisymmetric real matrix $\Delta' = \sigma_{CD}^{-1/2} \, \Delta \, \sigma_{CD}^{-1/2}$ into canonical form, as discussed in \lref{lem:6.2}. Subsequently, having the explicit form for the matrix $S$, thanks to \tref{theo:SqEnt-bounds}, we can determine a lower bound to the squashed entanglement of the output state $\qty(\1_A \otimes \N_B) \qty(\ket{\phi_\EE}_{\!AR\!\!}\bra{\phi_\EE})$, using the expression in Equation \eqref{eq:Esq}. Finally, sending the average number of photons per mode $N$ of the tensor product of $n$ two-mode squeezed vacuum states to infinity, makes the \modifica{constant $C_{LB} = \lim_{N\to+\infty} C_{LB}(N)$} still a lower bound to the squashed entanglement of the channel. This last step is justified by the fact that in the one-mode case this limit provides the optimal lower bound to the squashed entanglement for the extreme Gaussian attenuator and amplifier.
\end{proof}

In the case of the one-mode extreme Gaussian attenuator and amplifier, the lower bound to the squashed entanglement obtained with the above method is sharp and coincides with the actual value of the squashed entanglement \cite{de2019squashed}. In the next section we will explicitly show a numerical example of the calculation of this quantity.

\subsection{Some relevant examples}
In this subsection, we show how our results (\alref{alg:UpperBound} and \alref{alg:LowerBound1}) for the lower and upper bound of the squashed entanglement for multimode extreme bosonic Gaussian channels are perfectly compatible with known results for some very important examples of one-mode extreme bosonic Gaussian channels, \emph{i.e.}, the one-mode noiseless bosonic Gaussian attenuator and the one-mode noiseless bosonic Gaussian amplifier \cite{de2019squashed}.

\subsubsection{One-mode noiseless bosonic Gaussian attenuator}
\paragraph{Lower bound.}
Let us begin by computing the lower bound of a one-mode pure-loss bosonic Gaussian attenuator. To do this we apply step by step \alref{alg:LowerBound1} with $n=1$.
\begin{enumerate}
    \item With the same notation as in \alref{alg:LowerBound1} and thanks to Equations \eqref{eq:A4}, \eqref{eq:S_beam_splitter} and \eqref{eq:K_alpha_from_S} we have:
    \begin{equation}
    \begin{split}
        \sigma_{AB}(N) &= 
        \mqty(
            \qty(\En+\dfrac12)\,I_2 & \sqrt{\En(\En+1)}\,Z_2 \\
            \sqrt{\En(\En+1)}\,Z_2 & \qty(\En+\dfrac12)\,I_2
            )\,,\\
        K &= \sqrt{\eta}\,I_2
        \qquad\text{and}\qquad
        \alpha = \frac{1-\eta}{2}\,I_2\,.
    \end{split}
    \end{equation}
    From Equation \eqref{eq:sigmaCDN} therefore follows
    \begin{equation}
        \sigma_{CD}(N) = 
        \mqty(
            \qty(\En+\dfrac12)\,I_2 & \sqrt{\eta\,\En(\En+1)}\,Z_2 \\
            \sqrt{\eta\,\En(\En+1)}\,Z_2 & \qty(\eta\,\En+\dfrac12)\,I_2
            )\,.
    \end{equation}
    \item It is easy to verify that a symplectic matrix that diagonalises $\sigma_{CD}(N)$ is given by
    \modifica{
    \begin{equation}\label{eq:Skpkp}
        \begin{split}
        &S_{\kappa'} = 
        \mqty(
            \sqrt{\kappa'}\,I_2 & \sqrt{\kappa'-1}\,Z_2 \\
            \sqrt{\kappa'-1}\,Z_2 & \sqrt{\kappa'}\,I_2
            )\,,\\
        &\kappa'=\frac{N+1}{(1-\eta)N+1}\,.
        \end{split}
    \end{equation}
    }
    \item From Equation \eqref{eq:Skpkp}, it is easy to see that the quantities defined in \tref{theo:SqEnt-bounds} are $b_1^C=b_2^D=k'$ and $b_1^D=b_2^C=k'-1$, from which it follows that Equation \eqref{eq:C_LB} takes the form
    \modifica{
    \begin{equation}
        \begin{split}
        C_{LB}(N) &= \ln(2\kappa'-1)\\ 
        &\xrightarrow{N\to+\infty} \ln(\frac{1-\eta}{1+\eta})\,.
        \end{split}
    \end{equation}
    }
\end{enumerate}

\paragraph{Upper bound}
Although the upper bound for this channel has already been proved in \cite{takeoka2014squashed,de2019entropy,takeoka2014fundamental}, we compare this result with our new algorithm \alref{alg:UpperBound} for calculating the upper bound.
\begin{enumerate}
    \item Thanks to Equations \eqref{eq:S1/2} and \eqref{eq:S_beam_splitter} and fixing the state of the system $F$ to be the vacuum state, it holds
    \begin{equation}
    \begin{split}
        &S_{1/2} = \frac{1}{\sqrt{2}} \mqty(I_2 & I_2 \\ -I_2 & I_2)\,,\\
        &S_{\eta} = \mqty(\sqrt{\eta}\,I_2 & \sqrt{1-\eta}\,I_2 \\ -\sqrt{1-\eta}\,I_2 & \sqrt{\eta}\,I_2)\,,\\
        &\sigma_E = \sigma_F = \frac12\,I_2\,,
    \end{split}
    \end{equation}
    \item Thanks to Equation \eqref{eq:sigmaBEsFp}, $\sigma_{BE''F'}(t)$ then turns to be
    \begin{strip}
    \modifica{
    \begin{equation}
        \sigma_{BE''F'}(t) =
        \mqty(
            \qty(\eta\qty(t-\dfrac12)+\dfrac12)\,I_2 & \dfrac{1}{2 \sqrt{2}}\sqrt{\bar{\eta}\,\eta} (1-2t)\,I_2 & \dfrac{1}{2 \sqrt{2}}\sqrt{\bar{\eta}\,\eta} (2t-1)\,I_2 \\
            \dfrac{1}{2 \sqrt{2}}\sqrt{\bar{\eta}\,\eta} (1-2t)\,I_2 & \dfrac14\,(\eta +2 \bar{\eta}\,t+1)\,I_2 & \dfrac14\,\bar{\eta}\,(1-2t)\,I_2 \\
            \dfrac{1}{2 \sqrt{2}}\sqrt{\bar{\eta}\,\eta} (1-2t)\,I_2 & \dfrac14\,\bar{\eta}\,(1-2t)\,I_2 & \dfrac14\,(\eta +2 \bar{\eta} t+1)\,I_2
        )\,.
    \end{equation}
    }
    \end{strip}
    \item We must now compute the symplectic eigenvalues of the matrices $\sigma_{BE''}$, $\sigma_{BF'}$, $\sigma_{E'}$ and $\sigma_{F'}$, which are respectively
    \begin{equation}
    \begin{split}
        &\nu_{\sigma_{BE''}} = \nu_{\sigma_{BF'}} = \qty{\frac12, \frac14\qty\Big(1-\eta+2t(1+\eta))}\,,\\
        &\nu_{\sigma_{E''}} = \nu_{\sigma_{F'}} = \frac14\qty\Big(1 + 2t (1 - \eta) + \eta)\,.
    \end{split}
    \end{equation}
    \item The upper bound in Equation \eqref{eq:C_UB} then becomes
    \modifica{
    \begin{equation}
        \begin{split}
            C_{UB}(t) &= \biggl[g(0) + g\qty(\frac14\qty\Big(1-\eta+2t(1+\eta))) +\\
            &\phantom{=} - g\qty(\frac14\qty\Big(1+2t(1-\eta)+\eta))\biggr]\\
            &\xrightarrow{t\to+\infty} \ln(\frac{1+\eta}{1-\eta})\,.
        \end{split}
    \end{equation}
    }
\end{enumerate}
Let us observe that the values for the lower and upper bound computed via \alref{alg:LowerBound1} and \alref{alg:UpperBound} coincide, thus providing the exact value for the squashed entanglement of a one-mode noisless bosonic Gaussian attenuator, a result already proved in \cite{de2019squashed}.

\subsubsection{One-mode noisless bosonic Gaussian amplifier}
In this example, we are going to compute the squashed entanglement of a one-mode noisless bosonic Gaussian amplifier and verify that the results provided by \alref{alg:UpperBound} and \alref{alg:LowerBound1} are compatible with the results shown in \cite{de2019squashed}.

\paragraph{Lower bound}
Let us begin compute the lower bound of a one-mode noisless bosonic Gaussian amplifier. Let us apply step by step \alref{alg:LowerBound1} with $n=1$.
\begin{enumerate}
    \item With the same notation as in \alref{alg:LowerBound1} and thanks to Equations \eqref{eq:A4}, \eqref{eq:S_squeezing} and \eqref{eq:K_alpha_from_S} we have:
    \begin{equation}
        \begin{split}
        \sigma_{AB}(N) &= 
        \mqty(
            \qty(\En+\dfrac12)\,I_2 & \sqrt{\En(\En+1)}\,Z_2 \\
            \sqrt{\En(\En+1)}\,Z_2 & \qty(\En+\dfrac12)\,I_2
            )\,,\\
        K &= \sqrt{\kappa}\,I_2
        \qquad\text{and}\qquad
        \alpha = \frac{\kappa-1}{2}\,I_2\,.
        \end{split}
    \end{equation}
    From Equation \eqref{eq:sigmaCDN} it holds
    \begin{equation}
        \sigma_{CD}(N) = 
        \mqty(
            \qty(\En+\dfrac12)\,I_2 & \sqrt{\eta\,\En(\En+1)}\,Z_2 \\
            \sqrt{\eta\,\En(\En+1)}\,Z_2 & \qty(\eta\,\En+\dfrac12)\,I_2
            )\,.
    \end{equation}    
    \item It is easy to verify that a symplectic matrix that diagonalises $\sigma_{CD}(N)$ is given by
    \begin{equation}
        \begin{split}
        &S_{\kappa'} = 
        \mqty(
            \sqrt{\kappa'}\,I_2 & \sqrt{\kappa'-1}\,Z_2 \\
            \sqrt{\kappa'-1}\,Z_2 & \sqrt{\kappa'}\,I_2
            )\,,\\
        &\kappa'=\frac{\kappa(N+1)}{(\kappa-1)N+1}\,.
        \end{split}
    \end{equation}    
    \item From Equation \eqref{eq:Skpkp}, it is easy to see that the quantities defined in \tref{theo:SqEnt-bounds} are $b_1^C=b_2^D=\kappa'$ and $b_1^D=b_2^C=\kappa'-1$, from which it follows that Equation \eqref{eq:C_LB} takes the form
    \begin{equation}
        \begin{split}
        C_{LB}(N) &= \ln(2\kappa'-1)\\ &\xrightarrow{N\to+\infty} \ln(\frac{\kappa+1}{\kappa-1})
        \,.
        \end{split}
    \end{equation}
\end{enumerate}

\paragraph{Upper bound.}
As mentioned for the one-mode noisless bosonic Gaussian attenuator, also the upper bound for this channel has already been proved in \cite{takeoka2014squashed,de2019entropy,takeoka2014fundamental}. We compare this result with our new algorithm \alref{alg:UpperBound} for calculating the upper bound.
\begin{enumerate}
    \item Thanks to Equations \eqref{eq:S1/2} and \eqref{eq:S_beam_splitter} and fixing the state of the system $F$ to be the vacuum state, we have:
    \begin{equation}
    \begin{split}
        &S_{1/2} = \frac{1}{\sqrt{2}}\mqty(I_2&I_2\\-I_2&I_2)\,,\\
        &S_{\kappa} = \mqty(\sqrt{\kappa}\,I_2&\sqrt{\kappa-1}\,Z_2\\\sqrt{1-\kappa}\,Z_2&\sqrt{\kappa}\,I_2)\,,\\
        &\sigma_E = \sigma_F = \frac12\,I_2\,,
    \end{split}
    \end{equation}
    \item Thanks to Equation \eqref{eq:sigmaBEsFp}, $\sigma_{BE''F'}(t)$ then turns to be
    \vspace{60pt}
    \begin{strip}
    \modifica{
    \begin{equation}
        \sigma_{BE''F'}(t) =
        \mqty(
            \dfrac12(2\kappa t+\bar{\kappa})\,I_2 & \dfrac{1}{2 \sqrt{2}}\sqrt{\bar{\kappa}\,\kappa} (2 t+1)\,Z_2 & -\dfrac{1}{2 \sqrt{2}}\sqrt{\bar{\kappa}\,\kappa} (2 t+1)\,Z_2 \\
            \dfrac{1}{2 \sqrt{2}}\sqrt{\bar{\kappa}\,\kappa} (2 t+1)\,Z_2 & \dfrac14 (2\bar{\kappa} t+\kappa+1)\,I_2 & -\dfrac14 \bar{\kappa} (2 t+1)\,I_2 \\
            -\dfrac{1}{2 \sqrt{2}}\sqrt{\bar{\kappa}\,\kappa} (2 t+1)\,Z_2 & -\dfrac14 \bar{\kappa} (2 t+1)\,I_2 & \dfrac14 (2\bar{\kappa} t+\kappa+1)\,I_2
        )\,.
    \end{equation}
    }
    \end{strip}
    \item At this point, the algorithm \alref{alg:UpperBound} requires computing the symplectic eigenvalues of the matrices, $\sigma_{BE''}$, $\sigma_{BF'}$, $\sigma_{E''}$ and $\sigma_{F'}$, which are respectively:
    \begin{equation}
    \begin{split}
        &\nu_{\sigma_{BE''}} = \nu_{\sigma_{BF'}} = \qty{\frac12, \frac14\qty(\kappa-1+2t(1+\kappa))}\,,\\
        &\nu_{\sigma_{E''}} = \nu_{\sigma_{F'}} = \frac14\qty\Big(1 + 2t (\kappa-1) + \kappa)\,.
    \end{split}
    \end{equation}
    \item The upper bound in Equation \eqref{eq:C_UB} then becomes
    \modifica{
    \begin{equation}
        \begin{split}
            C_{UB}(t) &= \biggl[g(0) + g\qty(\frac14\qty(\kappa-1+2t(1+\kappa))) +\\
            &\phantom{=} - g\qty(\frac14\qty(1+2t(\kappa-1)+\kappa))\biggr]\\
            &\xrightarrow{t\to+\infty}
            \ln(\frac{1+\kappa}{\kappa-1})\,.
        \end{split}
    \end{equation}
    }
\end{enumerate}
Also in the example of a one-mode noisless bosonic Gaussian amplifier, the calculation of the lower bound via \alref{alg:LowerBound1} coincides with the calculation of the upper bound found using \alref{alg:UpperBound}. We have therefore determined the exact value for the squashed entanglement of a one-mode noisless bosonic Gaussian amplifier, a result already proved in \cite{de2019squashed}.
\section{Numerical simulations}\label{sec:simulations}

In this section we present some numerical results regarding the lower and upper bounds to the squashed entanglement of a multimode extreme bosonic Gaussian channel. Without loss of generality, for this numerical simulation we set the number of modes of each system equal to $2$\,. We will consider a Gaussian attenuator with different attenuation parameters on each mode, and the state of the environment $E$ interacting with the state of the input system $A$ is assumed to be a two-mode squeezed vacuum state. This particular choice of the channel is clearly inspired by reality, however the strategy we are going to follow is completely generic and can be applied to any extreme bosonic Gaussian channel on an arbitrary number of modes. In formulae, the channel we are going to consider is defined as
\begin{equation}\label{eq:canale_eta1eta2}
    \N_{A\to B}^{(\eta_1,\eta_2,\kappa)}(\oprho_A) = \tr_E\!\qty[\opU_{AE}^{(\eta_1,\eta_2)}\,\qty(\oprho_A\otimes\op\phi_E^\kappa)\,{\opU_{AE}^{(\eta_1,\eta_2)}}^{\hspace{-0pt}\dagger}\hspace{0pt}]\,,
\end{equation}
where $\opU_{AE}^{(\eta_1,\eta_2)}$ is the unitary operator implementing the symplectic matrix
\begin{equation}
    S^{(\eta_1,\eta_2)}_{AE} \!=\! 
    \begin{pNiceMatrix}[first-row,first-col]
        & \gray{A_1}& \gray{A_2} & \gray{E_1} & \gray{E_2} \\ 
        \gray{B_1}\phantom{i} & \sqrt{\eta_1}\,I_2 & 0 & \sqrt{\bar{\eta}_1}\,I_2 & 0 \\
        \gray{B_2}\phantom{i} & 0 & \sqrt{\eta_2}\,I_2 & 0 & \sqrt{\bar{\eta}_2}\,I_2 \\
        \gray{E_1}\phantom{i} & -\sqrt{\bar{\eta}_1}\,I_2 & 0 & \sqrt{\eta_1}\,I_2 & 0 \\
        \gray{E_2}\phantom{i} & 0 & -\sqrt{\bar{\eta}_1}\,I_2 & 0 & \sqrt{\eta_1}\,I_2
    \end{pNiceMatrix}\,,
\end{equation}
where $\bar{\eta}_i=1-\eta_i\,,i=1,2$, whereas the state of the environment $\op\phi_E^\kappa=\ket{\phi_\kappa}_E\bra{\phi_\kappa} = \opU^\kappa_E\ket{00}_E\bra{00}{\opU^\kappa_E}^\dagger$ is the two-mode squeezed vacuum state with squeezing parameter \modifica{$\kappa>0$}. Systems $A$, $B$ and $E$ are two-mode bosonic quantum systems respectively made up by the one-mode bosonic quantum systems $A_1$, $A_2$, $B_1$, $B_2$ and $E_1$, $E_2$.

First of all, we observe that, as proved in \secref{sec:TestExtremality}, the bosonic Gaussian channel we are considering is indeed extreme. Regarding the explicit lower and upper bounds calculation, instead, please refer to \secref{sec:ExampleLowerBound} and \secref{sec:ExampleUpperBound}. Here we present some plots showing the behaviour of the lower and upper bounds of the squashed entanglement of the channel taken into account by varying certain parameters of interest.

\begin{figure}[H]
    \centering
    \includegraphics[width=0.95\linewidth]{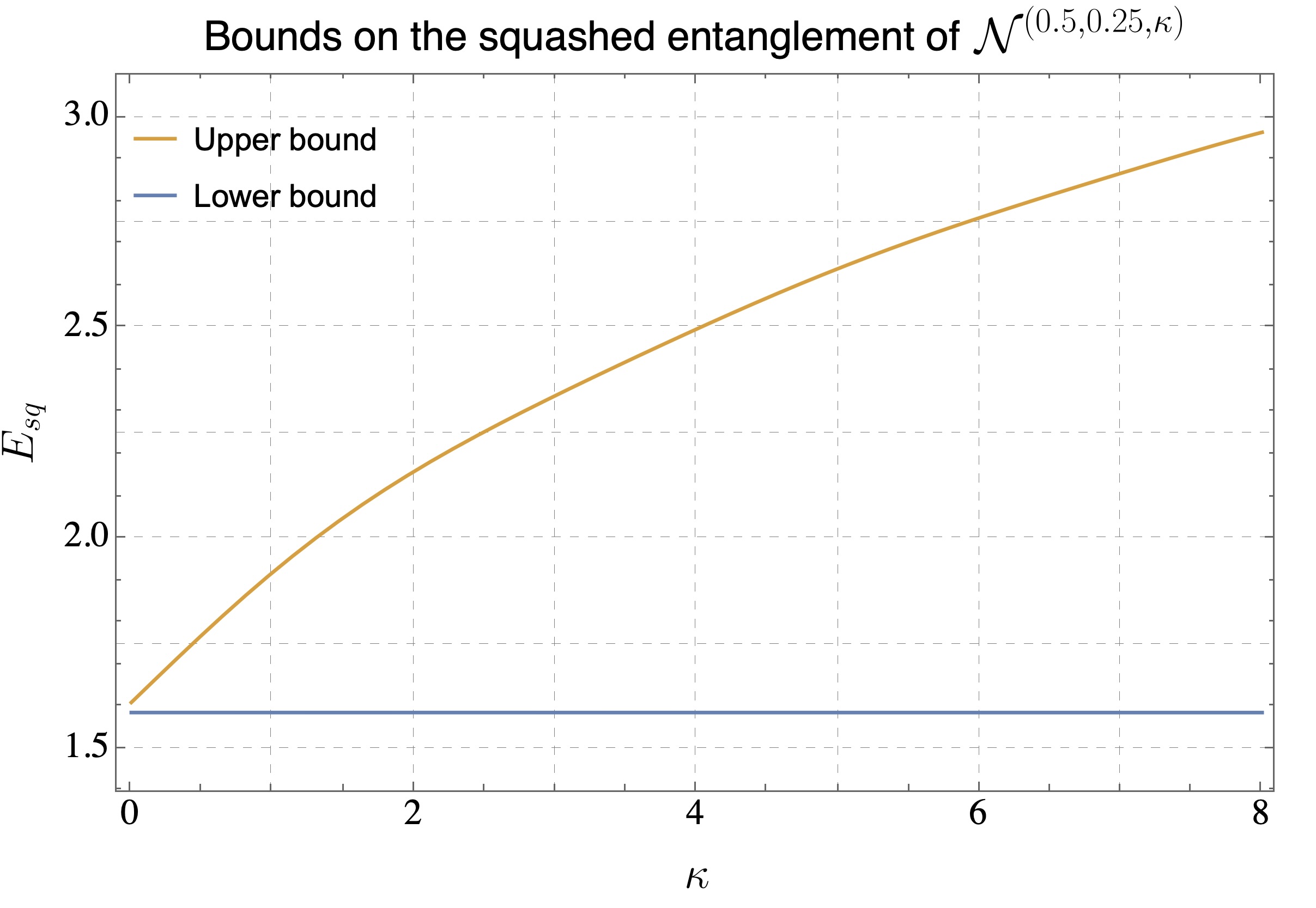}
    \caption{
        Lower bound (blue line) and upper bound (orange line) to the squashed entanglement of the channel $\N_{A\to B}^{(\eta_1,\eta_2,\kappa)}$ as a function of the squeezing parameter $\kappa$ of the environment state, fixed $\eta_1=0.5$ and $\eta_2=0.25$.
        The upper bound has been obtained by considering as a squashing channel a bosonic Gaussian attenuator with transmissivity $1/2$ and the environment state a two-mode squeezed vacuum state and optimising on this squeezing parameter for each value of $\kappa$.
    }
    \label{fig:SqEnt_vs_k}
\end{figure}

\begin{figure}[H]
    \centering
    \includegraphics[width=0.95\linewidth]{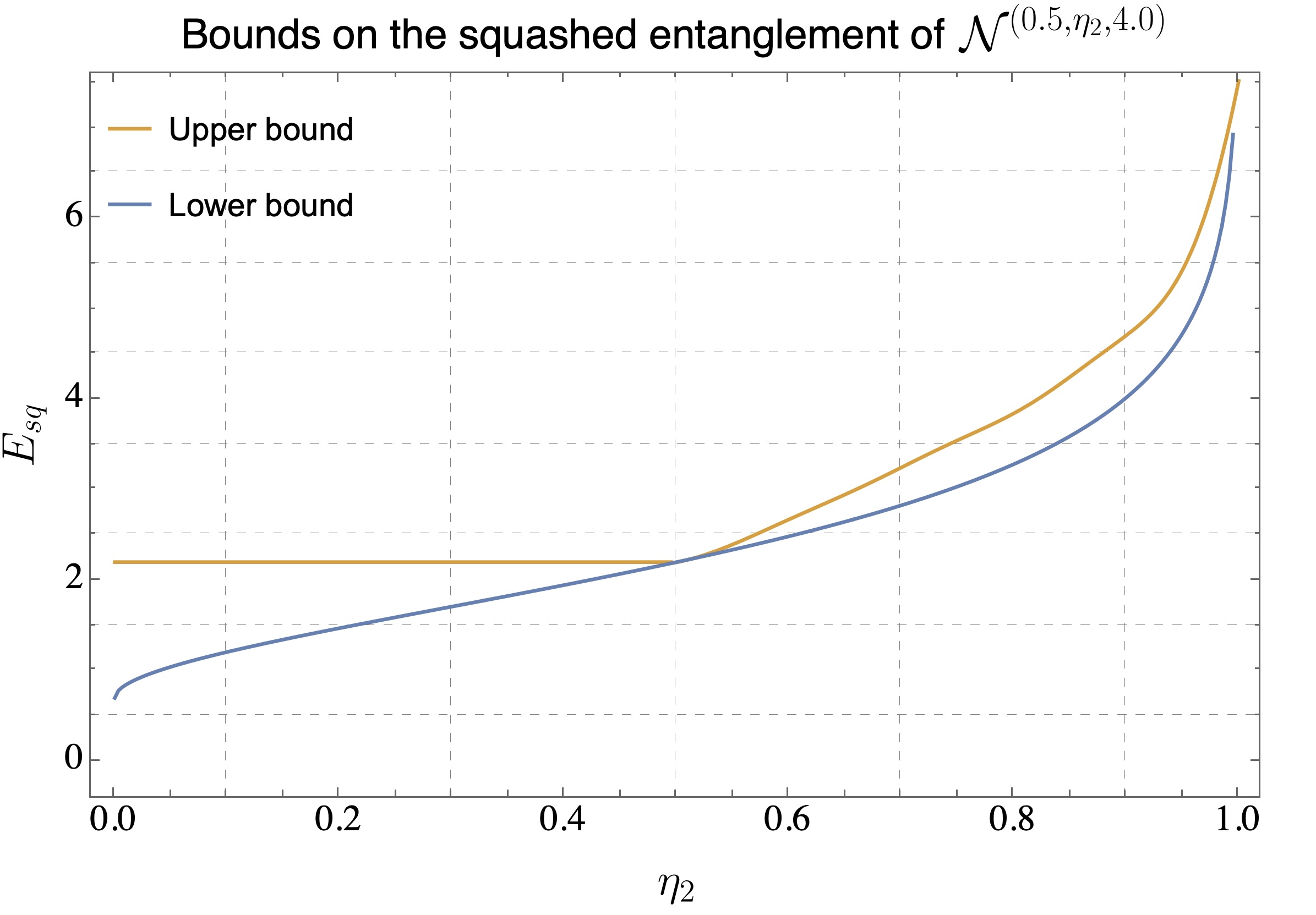}
    \caption{
        Lower bound (blue line) and upper bound (orange line) of the squashed entanglement of the channel $\N_{A\to B}^{(\eta_1,\eta_2,\kappa)}$ as a function of the attenuation parameter $\eta_2$, fixed $\eta_1=0.5$ and $\kappa=4.0$.
        The upper bound has been obtained by considering as a squashing channel a bosonic Gaussian attenuator with transmissivity $1/2$ and the environment state a two-mode squeezed vacuum state and optimising on this squeezing parameter for each value of $\eta_2$.
    }
    \label{fig:SqEnt_vs_eta}
\end{figure}

\begin{remark}
    We observe that, in the graph in \autoref{fig:SqEnt_vs_k}, for $\kappa=0$ we get back to the case discussed in \cite[Theorem 2]{de2019squashed} and thus exactly determine the squashed entanglement of this channel, $E_\text{sq}(\N_{A\to B}^{(\eta_1,\eta_2,0)}) = \ln(\frac{1+\eta_1}{1-\eta_1})+\ln(\frac{1+\eta_2}{1-\eta_2})$, where we used the additivity of the squashed entanglement with respect to the tensor product of quantum channels \cite[Corollary 8]{takeoka2014squashed}.
\end{remark}

\begin{remark}
    Knowing that applying a quantum channel to a state can only decrease the squashed entanglement of the state, \emph{i.e.},
    \begin{equation}\label{eq:Esq_channel_subadditivity}
        E_\text{sq}\qty(\Phi(\oprho)) \leq E_\text{sq}\qty(\oprho)\,,
    \end{equation}
    it is easy to show that the squashed entanglement of a composition of quantum channels is smaller or equal to the squashed entanglement of the individual channels, \emph{i.e.},
    \begin{equation}\label{eq:Esq_channel_composition_subadditivity}
        E_\text{sq}\qty(\Phi_2\circ\Phi_1) \leq E_\text{sq}\qty(\Phi_1),E_\text{sq}\qty(\Phi_2)\,.
    \end{equation}
    The proof is straightforward. For the channel $\Phi_2$ it holds:
    \begin{equation}
        \begin{split}
            E_\text{sq}\qty(\Phi_2\circ\Phi_1) &= \sup_{\oprho_{AB}} E_\text{sq}\qty(\qty(\1\otimes\Phi_2\circ\Phi_1)\qty(\oprho_{AB})) \\
            &\leq \sup_{\op\tau_{AB}} E_\text{sq}\qty(\qty(\1\otimes\Phi_2)\qty(\op\tau_{AB}))\\
            & \coloneqq E_\text{sq}\qty(\Phi_2)\,,
        \end{split}
    \end{equation}
    where we used the fact that optimisation for the compositions of the channels $\Phi_1$ and $\Phi_2$ is suboptimal compared to optimisation for the channel $\Phi_2$ alone. For the channel $\Phi_1$ it holds:
    \begin{equation}
        \begin{split}
            E_\text{sq}\qty(\Phi_2\circ\Phi_1) &= \sup_{\oprho_{AB}} E_\text{sq}\qty(\qty(\1\otimes\Phi_2\circ\Phi_1)\qty(\oprho_{AB})) \\
            &\leq \sup_{\op\rho_{AB}} E_\text{sq}\qty(\qty(\1\otimes\Phi_1)\qty(\op\rho_{AB}))\\
            & \coloneqq E_\text{sq}\qty(\Phi_1)\,,
        \end{split}
    \end{equation}
    where we used Equation \eqref{eq:Esq_channel_subadditivity} applied to the channel $\Phi_2$.
    It is therefore easy now to understand why the bounds in \autoref{fig:SqEnt_vs_eta} decrease as $\eta_2$ decreases. Indeed, from the well-known composition rule for bosonic Gaussian attenuators ($\mathcal{E}_{\eta_2}\circ\mathcal{E}_{\eta_1}=\mathcal{E}_{\eta_2\eta_1}$), decreasing $\eta_2$ is equivalent to applying a further attenuator to the system $A_2$ and thus the squashed entanglement is expected to decrease.
\end{remark}

\begin{remark}
    An important observation on the graph in \autoref{fig:SqEnt_vs_eta} is that for $\eta_1=\eta_2$ the lower bound coincides with the upper bound and thus we we are able to exactly determine the squashed entanglement of the channel $\N_{A\to B}^{(\eta_1,\eta_2,\kappa)}$ taken into consideration. This follows from the fact that the squashed entanglement is additive for the tensor product of channels and if $\eta_1=\eta_2$ the channel we are considering is the two-mode extension of the extreme bosonic Gaussian attenuator studied in \cite[Theorem 2]{de2019squashed}. This result follows from the observation that applying a symplectic trasformation to one input of a beam splitter with attenuation parameters $\eta_i=\eta\ \forall i$ is equivalent to applying that symplectic to the output and its inverse to the other input, as proven in \lref{lem:BeamSplitter}. With the same notation of \lref{lem:BeamSplitter}, if the symplectic transformation we are going to consider is the squeezing operation, then the channel whose squashed entanglement we are going to calculate is given by
    \begin{equation}
        \Phi^{\opU_\kappa\,\ketbra{\0}\,\opU_\kappa^\dagger}_\eta(\cdot) = \opU_\kappa\,\Phi^{\ketbra{\0}}_\eta(\opU_\kappa^\dagger\,\cdot\,\opU_\kappa)\,\opU_\kappa^\dagger\,,
    \end{equation}
    where $\ket{\0}\coloneqq\ket{00}$. This means that applying the squeezing to one input of the beam splitter is equivalent to applying it to the output of the channel and applying its inverse to the other input. Now, being the squashed entanglement an entanglement measure \cite{christandl2004squashed}, and thus invariant under local unitaries \cite{vedral1997quantifying,plenio2005introduction}, it holds
    \begin{equation}
    \begin{split}
        &E_\text{sq} \qty(\qty(\I\otimes\opU_\kappa) \qty\big[\1\otimes\Phi^{\ketbra{\0}}_\eta\qty(\opU_\kappa^\dagger\,\cdot\,\opU_\kappa)]\qty(\I\otimes\opU_\kappa^\dagger))\\
        &= E_\text{sq} \qty(\qty(\1\otimes\Phi^{\ketbra{\0}}_\eta)\qty(\opU_\kappa^\dagger\,\cdot\,\opU_\kappa))\,.
    \end{split}
    \end{equation}
    Finally, the optimisation on the input states of the channel implies that
    \begin{equation}
        \begin{split}
            E_\text{sq}\qty(\Phi^{\ketbra{\0}}_\eta\qty(\opU_\kappa^\dagger\,\cdot\,\opU_\kappa)) &= \sup_{\oprho_{AB}} E_\text{sq} \qty(\qty(\1\otimes\Phi^{\ketbra{\0}}_\eta)\qty(\opU_\kappa^\dagger\,\oprho_{AB}\,\opU_\kappa))\\
            &= \sup_{\oprho_{AB}} E_\text{sq} \qty(\qty(\1\otimes\Phi^{\ketbra{\0}}_\eta)\qty(\oprho_{AB}))\\
            &= E_\text{sq}\qty(\Phi^{\ketbra{\0}}_\eta)\,.
        \end{split}
    \end{equation}
    We have thus shown that the case $\eta_1=\eta_2$ is equivalent to the trivial two-mode extension of the squashed entanglement calculation for a one-mode extreme Gaussian attenuator, as discussed in \cite[Theorem 2]{de2019squashed}.
\end{remark}

\begin{remark}
    Another observation that can be made is the following. If the channel $\N_{A\to B}^{(\eta_1,\eta_2,\kappa)}$ defined in Equation \eqref{eq:canale_eta1eta2} is characterised by having $\eta_1=\eta_2$ then the optimal squashing channel for the squashed entanglement calculation, according to the definition \eqref{eq:SqEntChannel}, is a two-mode Gaussian attenuator, with the two attenuation parameters both equal to $1/2$, and with the environment state a two-mode squeezed vacuum state with squeezing parameter $\kappa$, the same squeezing parameter that appears in $\N_{A\to B}^{(\eta_1,\eta_2,\kappa)}$. Let $\mathcal{N_\eta^\kappa}$ be a two-mode Gaussian attenuator with both attenuation parameters equal to $\eta$ and environment state a two-mode squeezed vacuum state with squeezing parameter $\kappa$. Let $\mathcal{U}^\kappa_X(\cdot)\coloneqq (\opU_\kappa)_X\,\cdot\,(\opU_\kappa^\dagger)_X$ be the squeezing unitary channel on system $X$ with attenuation parameter $\kappa$ defined in \secref{sec:qGchannels}. Given a quantum state $\oprho_{BE''F'}$ of the system $BE''F'$, let's define the function $\bar{S}(\oprho_{BE''F'}) \coloneqq S(B|E'')_{\oprho_{BE''F'}} + S(B|F')_{\oprho_{BE''F'}}$. It holds
    \begin{strip}
    \begin{equation}
        \begin{split}
            E_\text{sq}(\mathcal{N}) &= \sup_{\oprho_{A}}\frac12 \inf_{\mathcal{S}_{E'\to E''}} \qty(\bar{S}\qty(\mathcal{U}^\mathcal{S}_{E'F\to E''F'}\,\mathcal{U}^{\N^\kappa_\eta}_{AE\to BE'}\,\qty\big(\oprho_{A}\otimes\ket{\0}_{\!E\!\!}\bra{\0}\otimes\ket{\0}_{\!F\!\!}\bra{\0})))\\
            &= \sup_{\oprho_{A}}\frac12 \inf_{\mathcal{S}_{E'\to E''}} \qty(\bar{S}\qty(\mathcal{U}^\mathcal{S}_{E'F\to E''F'}\,\mathcal{U}^\kappa_{BE'}\,\mathcal{U}^{\N^0_\eta}_{AE\to BE'}\,{\mathcal{U}^{\kappa\;\dagger}_{A}}\,\qty\big(\oprho_{A}\otimes\ket{\0}_{\!E\!\!}\bra{\0}\otimes\ket{\0}_{\!F\!\!}\bra{\0})))\\
            &= \sup_{\oprho_{A}}\frac12 \inf_{\mathcal{S}_{E'\to E''}} \qty(\bar{S}\qty(\mathcal{U}^\mathcal{S}_{E'F\to E''F'}\,\mathcal{U}^\kappa_{BE'}\,\mathcal{U}^{\N^0_\eta}_{AE\to BE'}\,\qty\big(\oprho_{A}\otimes\ket{\0}_{\!E\!\!}\bra{\0}\otimes\ket{\0}_{\!F\!\!}\bra{\0})))\\
            &= \sup_{\oprho_{A}}\frac12 \inf_{\mathcal{S}_{E'\to E''}} \qty(\bar{S}\qty(\mathcal{U}^\mathcal{S}_{E'F\to E''F'}\,\mathcal{U}^\kappa_{E'}\,\mathcal{U}^{\N^0_\eta}_{AE\to BE'}\,\qty\big(\oprho_{A}\otimes\ket{\0}_{\!E\!\!}\bra{\0}\otimes\ket{\0}_{\!F\!\!}\bra{\0})))\,.
        \end{split}
    \end{equation}
    \end{strip}
    In the second equality we used the fact that applying a symplectic unitary (in our case the squeezing operator) to the two outputs of a beam splitter and its inverse to the two inputs leaves the overall transformation unchanged (\lref{lem:BeamSplitter}). In the third equality, we simply incorporated the squeezing unitary channel $\mathcal{U}^{\kappa\;\dagger}_{A}$ when we optimise over all possible input states $\op\rho_{A}$. Finally, in the fourth equality we used the fact that $\mathcal{U}^\kappa_{BE'}$ is factorised on the systems on which it acts, \emph{i.e.}, $\mathcal{U}^\kappa_{BE'}=\mathcal{U}^\kappa_{B} \otimes \mathcal{U}^\kappa_{E'}$, and that $\bar{S}$ is invariant under local unitaries. Now, for the additivity of squashed entanglement under the tensor product of channels and from Ref.s \cite{takeoka2014squashed,de2019squashed}, we know that the optimal squashing channel $\mathcal{S}$ is such that it holds $\mathcal{U}^\mathcal{S}_{E'F\to E''F'}\,\mathcal{U}^\kappa_{E'} = \mathcal{U}^{\N_{1/2}^0}_{E'F\to E''F'}$. Whence, in our case, the unitary extension of the squashing channel satisfies
    \begin{equation}
        \mathcal{U}^\mathcal{S}_{E'F\to E''F'} = \mathcal{U}^{\N_{1/2}^0}_{E'F\to E''F'}\,{\mathcal{U}^\kappa_{E'}}^\dagger = {\mathcal{U}^{\kappa\;\dagger}_{E''F'}}\, \mathcal{U}^{\N_{1/2}^\kappa}_{E'F\to E''F'}\,.
    \end{equation}
    Since ${\mathcal{U}^{\kappa\;\dagger}_{E''F'}} = {\mathcal{U}^\kappa_{E''}}^\dagger \otimes {\mathcal{U}^\kappa_{F'}}^\dagger$, and $\bar{S}$ being invariant under local unitaries, it is sufficient that $\mathcal{U}^\mathcal{S}_{E'F\to E''F'} = \mathcal{U}^{\N_{1/2}^\kappa}_{E'F\to E''F'}$. Thus, apart from local unitaries, the optimal squashing channel is ${\N_{1/2}^\kappa}$, from which the thesis follows.
\end{remark}
\section{Conclusions}\label{sec:disc}

We have proved the multimode conditional quantum Entropy Power Inequality for bosonic quantum systems (\tref{theo:MCQEPI}), which determines the minimum von Neumann conditional entropy of the output of any linear mixing of bosonic modes among all the input states with given conditional entropies. Moreover, we have determined new lower bounds to the squashed entanglement of a large family of bosonic Gaussian states (\tref{theo:SqEnt-bounds}), and we have applied such bounds to determine new lower bounds to the squashed entanglement of all the multimode extreme bosonic Gaussian channels (\secref{sec:channels}), \emph{i.e.}, the bosonic Gaussian channels that cannot be decomposed as a convex combination of quantum channels.
Our lower bound is known to be optimal in the case of the extreme bosonic Gaussian attenuator and amplifier \cite{de2019squashed}.

\vspace{\baselineskip}

It is well known that proving the lower bounds to the squashed entanglement of a quantum state or a quantum channel is a challenging task since the optimizations in Equations \eqref{eq:defEsq} and \eqref{eq:ChannelSqEnt} over all the possible extensions of the quantum state are almost never analytically treatable. We overcome this problem by employing the new multimode conditional quantum Entropy Power Inequality \eqref{eq:MCQEPI}, which holds for any conditioning quantum system. We significantly extended the results of Ref. \cite{de2019squashed}, where the squashed entanglement of the noiseless bosonic Gaussian attenuator and amplifier have been determined. In particular, we extend the set of states for which it is possible to provide a lower bound to the squashed entanglement to all those quantum states whose covariance matrix has at least half of the symplectic eigenvalues equal to $1/2$ and we provide the first method for calculating a non-trivial lower bound to the squashed entanglement of any multimode extreme bosonic Gaussian channel.

\vspace{\baselineskip}

Our results can have applications in quantum key distribution, establishing the first upper bounds to the distillable key of such a large family of bosonic Gaussian states and to the secret-key capacity and the two-way quantum capacity that is achievable over any multimode extreme bosonic Gaussian channel. Furthermore, having a proven lower bound for the squashed entanglement of a quantum channel helps in establishing an optimal upper bound. In fact, if during the optimisation process for the upper bound the lower bound is reached then the squashed entanglement is exactly determined.
\modifica{The upper bounds on the squashed entanglement of quantum channels establish corresponding upper bounds on both the secret-key capacity and the two-way quantum capacity. Conversely, the lower bounds on the squashed entanglement imply that the squashed entanglement cannot yield tighter upper bounds on the aforementioned capacities than those provided by the lower bounds.}

\appendices
\section{Mathematical notes}\label{app:MathNotes}

This section is dedicated to showing and deriving some results used in the main text.

\begin{lemma}[covariance matrix of a product state]
    The covariance matrix of a product state can be written in terms of the covariance matrices of the states of the individual subsystems as
    \begin{equation}
        \sigma(\oprho_1\otimes\oprho_2) = \sigma(\oprho_1)\oplus\sigma(\oprho_2)\,.
    \end{equation}
\end{lemma}

\begin{lemma}[covariance matrix of a two-mode squeezed vacuum state]
    The covariance matrix of a two-mode squeezed vacuum state $\ketbra{\phi_\En}$ as a function of the average number of photons $\En$ per single mode is given by
    \begin{equation}\label{eq:A4}
        \sigma(\ketbra{\phi_\En}) = \mqty(\qty(\En+\dfrac12)\,I_2&\sqrt{\En(\En+1)}\,Z_2\\\sqrt{\En(\En+1)}\,Z_2&\qty(\En+\dfrac12)\,I_2)\,,
    \end{equation}
    where $I_2=\diag{1,1}$ and $Z_2=\diag{1,-1}$\,.
\end{lemma}
\begin{proof}
    Since $\En$ is the average number of photons per single mode, the two-mode squeezed vacuum state $\ketbra{\phi_\En}$ has average number of photons $2\En$. Therefore, recalling Equation \eqref{eq:number-of-photons}, it holds
    \modifica{
    \begin{equation}
        2\En = \frac12\tr[\sigma\qty\big(\ketbra{\phi_\En})-\frac{I_4}{2}]\,,
    \end{equation}
    where $I_4=\diag{1,1}^{\oplus2}$. Whence, from \cite[Equation (5.21)]{serafini2017quantum}, it holds
    \begin{equation}
        \cosh(2r) = 2\En + 1\,.
    \end{equation}
    }
    By substituting this expression into \cite[Equation (5.21)]{serafini2017quantum} Equation \eqref{eq:A4} is obtained.
\end{proof}

\begin{lemma}[covariance matrix of a $2n$-mode squeezed vacuum state]
    The covariance matrix of a $2n$-mode squeezed vacuum state as a function of the average number of photons $\En_i$ per single mode is given by
    \begin{equation}\label{eq:A9}
        \sigma(\ket{\phi_\EE}_{\!AR\!\!} \bra{\phi_\EE}) \!=\! 
        {\setlength{\arraycolsep}{-1pt}
        \mqty(\!\bigoplus_{i=1}^n\! \qty(\!\En_i\!+\!\dfrac12\!) I_2 & \bigoplus_{i=1}^n\! \sqrt{\En_i(\!\En_i\!+\!1\!)} Z_2\! \\ \!\bigoplus_{i=1}^n\! \sqrt{\En_i(\!\En_i\!+\!1\!)} Z_2 & \bigoplus_{i=1}^n\! \qty(\!\En_i\!+\!\dfrac12\!) I_2\!)
        }\,,
    \end{equation}
    where $A$ and $R$ are $n$-mode bosonic quantum systems.
\end{lemma}
\begin{proof}
    Let us think of systems $A$ and $R$ as the union of $n$ one-mode quantum systems $\{A_i\}_{i=1,\ldots,n}$ and $\{R_j\}_{j=1,\ldots,n}$. The $2n$-mode squeezed vacuum state is defined as the tensor product of $n$ two-mode squeezed vacuum state, as follows
    \begin{equation}
        \ket{\phi_\EE}_{\!AR\!\!}\bra{\phi_\EE} = \bigotimes_{i=1}^n\ket{\phi_{\En_i}}_{\!A_iR_i\!\!}\bra{\phi_{\En_i}}\,,
    \end{equation}
    with covariance matrix
    \begin{equation}\label{eq:A12}
        \sigma(\ket{\phi_\EE}_{\!AR\!\!}\bra{\phi_\EE}) = 
        \bigoplus_{i=1}^n \mqty(\qty(\En_i+\dfrac12)\,I_2 & \sqrt{\En_i(\En_i+1)}\,Z_2 \\ \sqrt{\En_i(\En_i+1)}\,Z_2 & \qty(\En_i+\dfrac12)\,I_2)\,.
    \end{equation}
    Matrix \eqref{eq:A12} is a block diagonal matrix which is related to systems, in order, $A_1$, $R_1$, $\ldots$, $A_n$, $R_n$\,. Since we are interested in the ordering $AR=A_1\cdots A_nR_1\cdots R_n$, we can rearrange the rows and columns of matrix \eqref{eq:A12} to obtain Equation \eqref{eq:A9}.
\end{proof}

\subsection{Covariance matrix of the output of a Gaussian attenuator}\label{sec:CacoloDisigmaAReta}

Let us consider the quantum state
\begin{equation}
    \begin{split}
        &\op\omega_{AR}(\ee)\\
        &= (\1_A \otimes \mathcal{E}_\ee)(\ket{\phi_\EE}_{\!AR\!\!}\bra{\phi_\EE})\\
        &= \tr_B\!\qty\bigg[\qty(\I_A\otimes\opU_{RB}^\ee)\qty\Big(\ket{\phi_\EE}_{\!AR\!\!}\bra{\phi_\EE}\otimes\ket{\0}_{\!B\!\!}\bra{\0})\qty(\I_A\otimes{\opU_{RB}^\ee}^{\!\!\!\!\dagger})]
        \,.
    \end{split}
\end{equation}
The covariance matrix $\sigma(\op\omega_{AR}(\ee))$ is derived by only taking the rows and columns relative to the systems $A$ and $R$ of the following covariance matrix:
\begin{equation}
    \begin{split}
        &\sigma(\op\omega_{ARB}(\ee))\\
        &= \sigma\qty(\qty(\I_A\otimes\opU^\ee_{RB})\qty\Big(\ket{\phi_\EE}_{\!AR\!\!}\bra{\phi_\EE}\otimes\ket{\0}_{\!B\!\!}\bra{\0})\qty(\I_A\otimes{\opU^\ee_{RB}}^{\!\!\!\!\dagger}))\\
        &= \qty\Big(I_{2n}\oplus S_\ee)\qty\Big(\sigma\qty\big(\ket{\phi_\EE}_{\!AR\!\!}\bra{\phi_\EE})\oplus\sigma\qty\big(\ket{\0}_{\!B\!\!}\bra{\0}))\qty\Big(I_{2n}\oplus S_\ee)^T
        .
    \end{split}
\end{equation}
If we now explicit this expression in terms of matrices, and, for convenience, we reorder the systems as $A_1R_1B_1\cdots A_nR_nB_n$ we have
\begin{equation}
    I_{2n}\oplus S_\ee = \bigoplus_{i=1}^n \mqty(I_2&0&0\\0&\sqrt{\eta_i}\,I_2&\sqrt{1-\eta_i}\,I_2\\0&-\sqrt{1-\eta_i}\,I_2&\sqrt{\eta_i}\,I_2)\,,
\end{equation}
and
\begin{equation}
    \begin{split}
        &\sigma \qty\big(\ket{\phi_\EE}_{\!AR\!\!} \bra{\phi_\EE}) \oplus \sigma \qty\big(\ket{\0}_{\!B\!\!}\bra{\0})\\
        &= \bigoplus_{i=1}^n \mqty(\qty(\En_i+\frac12)\,I_2&\sqrt{\En_i(\En_i+1)}\,Z_2&0\\\sqrt{\En_i(\En_i+1)}\,Z_2&\qty(\En_i+\frac12)\,I_2&0\\0&0&\frac{I_2}{2})
        \,.
    \end{split}
\end{equation}
whence
\begin{equation}
    \sigma(\op\omega_{\!A\!R\!B}(\ee)) \!=\! \bigoplus_{i=1}^n 
    {\setlength{\arraycolsep}{2pt}
    \mqty(\qty(\En_i+\dfrac12) I_2 & \sqrt{\eta_i \En_i(\En_i\!+\!1)} Z_2 & * \\ \sqrt{\eta_i \En_i(\En_i\!+\!1)} Z_2 & \qty(\eta_i \En_i\!+\!\dfrac12) I_2 & * \\ * & * & *)
    }
    \,,
\end{equation}
from which, considering only the rows and columns relative to the systems $A$ and $R$, matrix $\sigma(\op\omega_{AR}(\ee))$ in Equation \eqref{eq:sigmaAReta} is obtained.

\subsection{Covariance matrix of the output of a generic symplectic transformation}\label{sec:A.2}

In this section, we derive Equations \eqref{eq:4.27a} and \eqref{eq:4.27b}.
The covariance matrices of the marginal states $\op\rho_{CR}(\ee)$ and $\op\rho_{DR}(\ee)$ are derived from $\sigma\qty(\op\rho_{CDR}(\ee))$. This covariance matrix is given by
\begin{equation}\label{eq:A.12CRD}
    \begin{split}
        &\sigma(\op\rho_{CRD}(\ee))\\
        &= \sigma\qty(\qty(\opU^S_{AB}\otimes\I_R)\qty\big(\op\omega_{AR}(\ee)\otimes\ket{\0}_{\!B\!\!}\bra{\0})\qty({\opU^S_{AB}}^{\!\!\!\!\dagger}\otimes\I_R))\\
        &= \qty(S_{AB}\oplus I_R) \sigma\qty\big(\op\omega_{AR}(\ee)\otimes\ket{\0}_{\!B\!\!}\bra{\0}) \qty(S_{AB}\oplus I_R)^T,
    \end{split}
\end{equation}
where $\opU^S_{AB}$ is an isometric transformation acting on the quadratures via a symplectic matrix of the form \eqref{eq:S-blocks}. Now, defining the variables

\begin{equation}
    \begin{split}
        &A_\EE \coloneqq \bigoplus_{i=1}^n\qty(\En_i+\frac12)\,I_2\,,\\
        &B_\EE \coloneqq \bigoplus_{i=1}^n\sqrt{\eta_i\, \En_i(\En_i+1)}\,Z_2\,,\\
        &C_\EE \coloneqq \bigoplus_{i=1}^n\qty(\eta_i\, \En_i+\frac12)\,I_2\,,
    \end{split}
\end{equation}
and recalling that the rows and columns are, in order, associated with the systems $A$, $R$ and $B$, the matrix form of Equation \eqref{eq:A.12CRD} is
\modifica{
\begin{equation}\label{eq:A11}
    \begin{split}
        &\sigma(\op\rho_{CRD}(\ee))\\
        &= {\setlength{\arraycolsep}{2pt}\mqty(B_1^C&0&B_2^C\\0&I_{2n}&0\\B_1^D&0&B_2^D)} {\setlength{\arraycolsep}{2pt}\mqty(A_\EE&B_\EE&0\\ B_\EE&C_\EE&0\\0&0&I_{2n}/2)} {\setlength{\arraycolsep}{2pt}\mqty({B_1^C}^T&0&{B_1^D}^T\\0&I_{2n}&0\\{B_2^C}^T&0&{B_2^D}^T)}\\
        &\!=\! {\setlength{\arraycolsep}{2pt}
        \mqty(\!A_\EE\!{B_1^C}\!{B_1^C}^T\!\!+\!\dfrac12\!{B_2^C}\!{B_2^C}^T&{B_1^C}\!B_\EE&A_\EE\!{B_1^C}\!{B_1^D}^T\!\!+\!\dfrac12\!{B_2^C}\!{B_2^D}^T\!\\B_\EE\!{B_1^C}^T&C_\EE&B_\EE\!{B_1^D}^T\\\!A_\EE{B_1^D}\!{B_1^C}^T\!\!+\!\dfrac12\!{B_2^D}\!{B_2^C}^T&{B_1^D}\!B_\EE&A_\EE\!{B_1^D}\!{B_1^D}^T\!\!+\!\dfrac12\!{B_2^D}\!{B_2^D}^T\!)
        }
        \,.
    \end{split}
\end{equation}
}

\subsection{Covariance matrix of the output of a two-mode Gaussian attenuator}\label{app:cov_mat_CD}

In this section, we calculate the covariance matrix $\sigma_{CD}$ of \secref{sec:ExampleLowerBound}.
Given 3 two-mode squeezed vacuum states $\phi^{\En_1}_{A_1B_1}$, $\phi^{\En_2}_{A_2B_2}$, $\phi^{\en_3}_{E}$ with average number of photons $\En_1$, $\En_2$ and $\en_3$ respectively, let us calculate the covariance matrix of the state
\begin{equation}
    \begin{split}
        &\op\gamma^{S}_{CD}\\
        &= (\1_A \otimes \mathcal{N}_B)\qty(\op\phi^{\En_1}\otimes\op\phi^{\En_2})_{AB}\\
        &= \tr_E\!\qty\bigg[\qty(\I_A\otimes\opU_{BE}^S)\qty\Big(\qty(\op\phi^{\En_1}\otimes\op\phi^{\En_2})_{AB}\otimes\op\phi^{\en_3}_E)\qty(\I_A\otimes{\opU_{BE}^S}^{\!\!\!\!\dagger})]
        \,,
    \end{split}
\end{equation}
where $\opU^S_{BE}$ is an isometric transformation acting on the quadratures of the systems $B$ and $E$ via a $8\times8$ symplectic matrix which can be seen in the form \eqref{eq:S-blocks}. Let us calculate the covariance matrix $\sigma\qty(\op\gamma^{S}_{CDE})$:
\begin{equation}
    \begin{split}
        &\sigma\qty(\op\gamma^{S}_{CDE})\\
        &= \sigma\qty(\qty(\I_A\otimes\opU^S_{BE})\qty\Big(\qty(\op\phi^{\En_1}\otimes\op\phi^{\En_2})_{AB}\otimes\op\phi^{\en_3}_E)\qty(\I_A\otimes{\opU^S_{BE}}^{\!\!\!\!\dagger}))\\
        &= \qty\Big(I_A \!\oplus\! S_{BE}) \qty\Big(\sigma\qty\big(\qty(\op\phi^{\En_1}\!\otimes\!\op\phi^{\En_2})_{AB}) \!\oplus\! \sigma\qty(\op\phi^{\en_3}_E)) \qty\Big(I_A \!\oplus\! S_{BE})^T
        \!.
    \end{split}
\end{equation}
In a more explicit form, it holds

\begin{strip}
\modifica{
\vspace{-25pt}
\begin{equation}\label{eq:A16}
    \begin{split}
        \sigma\bigl(\op\gamma^{S}_{CDE}\bigr) &=
        \mqty(
            I_4 & 0 & 0 \\
            0 & B_1^C & B_2^C \\
            0 & B_1^D & B_2^D
            )
        \mqty(
            \bigoplus_{i=1}^2 \qty(\En_i+\frac12)\,I_2 & \bigoplus_{i=1}^2 \sqrt{\En_i(\En_i+1)}\,Z_2 & 0 \\ \bigoplus_{i=1}^2 \sqrt{\En_i(\En_i+1)}\,Z_2 & \bigoplus_{i=1}^2\qty(\En_i+\frac12)\,I_2 & 0 \\
            0 & 0 & \sigma_E
            )
        \mqty(
            I_4 & 0 & 0 \\
            0 & {B_1^C}^T & {B_1^D}^T \\
            0 & {B_2^C}^T & {B_2^D}^T
            )\\
        &=
        \mqty(
            \bigoplus_{i=1}^2 \qty(\En_i+\frac12)\,I_2 & \bigoplus_{i=1}^2 \sqrt{\En_i(\En_i+1)}\,Z_2\,{B_1^C}^T & * \\ \bigoplus_{i=1}^2\sqrt{\En_i(\En_i+1)}\,B_1^C\,Z_2 & \bigoplus_{i=1}^2\qty(\En_i+\frac12)\,{B_1^C}{B_1^C}^T+{B_2^C}\,\sigma_E\,{B_2^C}^T & * \\
            * & * & *
            )\,.
    \end{split}
\end{equation}
}
\end{strip}
where $\sigma_E$ is defined in Equation \eqref{eq:sigmaE}. Let us now consider the special case in which the symplectic matrix is given by Equation \eqref{eq:SSEeta12}. In this case, the following relations apply:
\begin{equation}
\begin{split}
    &B_1^C = B_2^D = \bigoplus_{i=1}^2 \sqrt{\eta_i}\, I_2\,,\\
    &B_2^C = -B_1^D = \bigoplus_{i=1}^2 \sqrt{1-\eta_i}\, I_2\,.
\end{split}
\end{equation}
By replacing these matrix blocks in Equation \eqref{eq:A16} and considering only the first eight rows and columns, those related to the spaces $C$ and $D$, the explicit form for the matrix $\sigma\qty(\op\gamma^{S}_{CD})$ is obtained:
\begin{strip}
\modifica{
\begin{equation}\label{eq:sigma_CD_explicit}
    \sigma\qty(\op\gamma^{S}_{CD}) =
    \mqty(
    \qty(\En_1+\dfrac12)\,I_2 & 0 & \sqrt{\eta_1\,\En_1(\En_1+1)}\,Z_2 & 0 \\
    0 & \qty(\En_2+\dfrac12)\,I_2 & 0 & \sqrt{\eta_2\,\En_2(\En_2+1)}\,Z_2 \\
    \sqrt{\eta_1\,\En_1(\En_1+1)}\,Z_2 & 0 & \qty(N_3+\dfrac12)\qty(\eta_1+\sqrt{1-\eta_1})\,I_2 & \sqrt{N(N+1)(1-\eta_2)}\,Z_2 \\
    0 & \sqrt{\eta_2\,\En_2(\En_2+1)}\,Z_2 & \sqrt{N(N+1)(1-\eta_1)}\,Z_2 & \qty(N_3+\dfrac12)\qty(\eta_1+\sqrt{1-\eta_2})\,I_2
    )\,.
\end{equation}
}
\end{strip}

\subsection{Maximisation of the multimode linear conditional quantum EPI}\label{sec:MLCQEPI}
The aim of this section is to maximise the RHS of Equation \eqref{eq:MLCQEPI} with respect to $\lambda_i$, $i=1,\ldots,K$ and subject to the constraint $\sum_{i=1}^K\lambda_i=1$. It is natural to treat this problem using the method of Lagrange multipliers. In particular, let us the objective and constraint function as
\modifica{
\begin{equation}
\begin{split}
    &f(\vb*{\lambda}) = \sum_{i=1}^K\lambda_i\frac{x_i}{n} + \sum_{i=1}^K\lambda_i\ln(\frac{b_i}{\lambda_i})\,,\\
    &g(\vb*{\lambda}) = \sum_{i=1}^K\lambda_i - 1
    \,,
\end{split}
\end{equation}
}
where $x_i\coloneqq\sxim$ and $\vb*{\lambda}=\qty(\lambda_1\,\ldots\,\lambda_K)$. We can now build the Lagrangian function
\begin{equation}
    \Lambda(\vb*{\lambda},\delta) = f(\vb*{\lambda}) + \delta\,g(\vb*{\lambda})
\end{equation}
and impose the system
\begin{equation}\label{eq:A20Lagrange}
    \begin{cases}
        \displaystyle\pdv{\lambda_i} \Lambda(\vb*{\lambda},\delta) = 0\\
        \displaystyle\pdv{\delta} \Lambda(\vb*{\lambda},\delta) = 0
    \end{cases}
    \implies
    \begin{cases}
        \displaystyle\lambda_j = \frac{b_j\exp(\frac{x_j}{n})}{e^{1-\delta}}\\
        \displaystyle\sum_{i=1}^K \lambda_i= 1\,,
    \end{cases}
\end{equation}
where it is implicitly understood that the first equation in the system \eqref{eq:A20Lagrange} is valid $\forall i$. By substituting the first $K$ equations of the system \eqref{eq:A20Lagrange} into the last equation, we obtain that $e^{1-\delta}=\sum_{i}^Kb_i\exp(\frac{x_i}{n})$ and thus, $\forall j$, it holds
\begin{equation}
    \lambda_j = \frac{b_j\exp(\frac{x_j}{n})}{\sum_{i}^Kb_i\exp(\frac{x_i}{n})}\,,
\end{equation}
which is exactly Equation \eqref{eq:maximization}. Moreover, since $f$ is a concave function in $\vb*{\lambda}$ with a unique stationary point, that point is necessarily the global maximum.
\section{Auxiliary lemmas}\label{app:lem}

In this appendix, we provide the statements and the proofs of some lemmas and results used in the main text.

\subsection{Some linear algebra lemmas}

\begin{lemma}\label{lem:propSimplMatrices}
    Given a symplectic matrix $S\in\text{Sp}(2n,\mathbb{R})$ that we can write in a block form as
    \begin{equation}
        S = \mqty(A&B\\C&D)\,,
    \end{equation}
    then the following relationship for the blocks applies
    \begin{equation}\label{eq:4.15}
        a\,c = b\,d\,,
    \end{equation}
    where $a=\abs{\det A}$, $b=\abs{\det B}$, $c=\abs{\det C}$ and $d=\abs{\det D}$\,.
    \begin{proof}
        By definition of symplectic matrix, the following relation holds
        \begin{equation}\label{eq:2modeSimpl}
            S\, \Delta\, S^T = \Delta\,,
        \end{equation}
        where
        \begin{equation}
        \begin{split}
            &\phantom{mmmmmmm} S = \mqty(A&B\\C&D)\,,\\
            &\Delta = \mqty(\Delta_{2n}&0\\0&\Delta_{2n})\,,
            \qquad
            \Delta_{2n} = \bigoplus_{i=1}^n\mqty(0&1\\-1&0)\,.
        \end{split}
        \end{equation}
        Explicitly, Equation \eqref{eq:2modeSimpl} can be written as
        \begin{equation}
            {\setlength{\arraycolsep}{2pt}
            \mqty(
            A \Delta_{2n} A^T \!+\! B \Delta_{2n} B^T & A \Delta_{2n} C^T \!+\! B \Delta_{2n} D^T \\ C \Delta_{2n} A^T \!+\! D \Delta_{2n} B^T & C \Delta_{2n} C^T \!+\! D \Delta_{2n} D^T
            )
            }
            \!=\!
            {\setlength{\arraycolsep}{1pt}
            \mqty(\Delta_{2n} & 0 \\ 0 & \Delta_{2n})
            }
            \,.
        \end{equation}
        In particular, looking at the block in the top right corner, this implies that
        \begin{equation}
            A\, \Delta_{2n}\, C^T = - B\, \Delta_{2n}\, D^T\,,
        \end{equation}
        from which, it holds
        \begin{equation}\label{eq:4.20}
            \det(A)\det(C) = -\det(B)\det(D)\,,
        \end{equation}
        from which Equation \eqref{eq:4.15} directly follows.
    \end{proof}
\end{lemma}

\begin{theorem}[Williamson]\label{teo:Williamson}
    Given $h\in\RR^{2n\times2n}_{>0}$ symmetric, there exists a symplectic matrix $S$ such that
    \begin{equation}\label{eq:Williamson}
        ShS^T = \bigoplus_{k=1}^n \nu_k I_2 =: D\,,
    \end{equation}
    where $D$ is a diagonal matrix with pairwise equal eigenvalues. The $\{\nu_k\}_{k=1,\ldots,n}$ are the symplectic eigenvalues of the matrix $h$\,.
\end{theorem}
\begin{proof}
    See \cite{serafini2017quantum}.
\end{proof}

\begin{lemma}\label{prop:ordered_symplectic_eigenvalues}
    Let $A$ and $B$ be two $2n\times2n$ real positive definite matrices such that $A>B$\,. Suppose that $A$ and $B$ have symplectic eigenvalues $d_1(A) \leq \cdots \leq d_n(A)$\,, $d_1(B) \leq \cdots \leq d_n(B)$\,. Then for all $i=1,\ldots,n$\,,
    \begin{equation}
        d_i(A) \geq d_i(B) \,.
    \end{equation}
    This means that the ordered symplectic eigenvalues of matrix $A$ are one-to-one greater than the ordered symplectic eigenvalues of matrix $B$\,. 
\end{lemma}
\begin{proof}
    This is a direct consequence of \cite[Corollary 2]{jain2021sums}. In particular, if $A>B$ then $A=B+\delta$\,, where $\delta$ is a positive definite real matrix. Fixing $k=1$, Corollary 2 of \cite{jain2021sums} states that for every $1\leq i_j\leq n$\,, with $1\leq j\leq n$\,, it holds
    \begin{equation}\label{eq:symplectic_eigenvalues_inequality}
        d_{i_j}(A) = d_{i_j}(B+\delta) \geq d_{i_j}(B) + d_{j}(\delta) \geq d_{i_j}(B)\,.
    \end{equation}
    From Equation \eqref{eq:symplectic_eigenvalues_inequality} it therefore follows that for every $1\leq j\leq n$ it holds $d_i(A) \geq d_i(B)$\,, whence the thesis.
\end{proof}
\lref{prop:ordered_symplectic_eigenvalues} thus states that given two matrices $A$ and $B$ such that $A>B$ the ordered symplectic eigenvalues of matrix $A$ are one-to-one greater than the ordered symplectic eigenvalues of matrix $B$\,. 

\begin{proposition}\label{prop:g_function_symplectic_relation}
    Let $U$ be a $2n\times2n$ positive definite real matrix, $V$ a $2n\times2n$ constant symmetric real matrix, $t$ a real variable and $g(x) \coloneqq(x+1) \ln(x+1) - x \ln(x)$\,. Let $W = t\,U + V$ and let $\{\nu^i_W\}_{i=1,\ldots,n}$ and $\{\nu^i_U\}_{i=1,\ldots,n}$ be the ordered symplectic eigenvalues of $W$ and $U$ respectively. In the limit of $t$ going to infinity for every $i=1,\ldots,n$ holds $g\qty(\nu^i_W-\frac12) \sim g\qty(t\,\nu^i_U-\frac12)$, \emph{i.e.},
    \begin{equation}
        \lim_{t\to\infty} \frac{g\qty(\nu^i_W-\dfrac12)}{g\qty(\nu^i_{t\,U}-\dfrac12)} = 1\,.
    \end{equation}
\end{proposition}
\begin{proof}
    There exists a constant $r>0$ such that $-r\,U\leq V\leq r\,U$, and a possible choice is $r = \norm{U^{-1/2}\,V\,U^{-1/2}}_\infty$, where $\norm{\cdot}_\infty$ denotes the infinity norm\footnote{Multiplying the inequalities $-r\,U\leq V\leq r\,U$ by $U^{-1/2}$ will result in $-r\,I\leq U^{-1/2}\,V\,U^{-1/2} \leq r\,I$ which basically means that $\norm{U^{-1/2}\,V\,U^{-1/2}}_\infty\leq r$, so $r$ exists if $U$ is invertible.}. This means that $(t-r)U\leq t\,U+V\leq (t+r)U$. It also holds that the symplectic eigenvalues respect the inequalities between matrices, \emph{i.e.}, that the ordered symplectic eigenvalues of $t\,U+V$ lie between the ordered symplectic eigenvalues of $(t-r)\,U$ and $(t+r)\,U$ (see \lref{prop:ordered_symplectic_eigenvalues}). This means that for every $i=1,\ldots,n$ it holds:
    \begin{equation}
        \nu_{(t-r)\,U}^i \leq \nu_W^i \leq \nu_{(t+r)\,U}^i\,.
    \end{equation}
    Moreover, since the function $g$ is monotonically increasing in its argument, \emph{i.e.}, for every $x,y$ such that $x>y>0$ $g(x)>g(y)$ holds, for every $i=1,\ldots,n$ it holds:
    \begin{equation}\label{eq:nuW_inequality}
        g\qty(\nu_{(t-r)\,U}^i-\frac12) \leq g\qty(\nu_W^i-\frac12) \leq g\qty(\nu_{(t+r)\,U}^i-\frac12)
        \,.
    \end{equation}
    Simple calculations show that, for $t\gg1$, the following applies
    \modifica{
    \begin{equation}\label{eq:nutU_estimation}
        g\qty(\nu_{t\,U}^i-\frac12) = \ln(t\,\nu_U^i)+1+\order{\frac{1}{t^2}}
    \end{equation}
    }
    and
    \modifica{
    \begin{equation}\label{eq:nutpmrU_estimation}
        g\qty(\nu_{(t\pm r)\,U}^i-\frac12) = \ln(t\,\nu_U^i) + 1 \pm \frac{r}{t} + \order{t^{-2}}\,,
    \end{equation}
    }
    where we used the fact that, for every real constant $c$ and every positive definite matrix $A$, every symplectic eigenvalue of $A$ fulfils the condition $\nu_{c\,A}=c\,\nu_A$.    
    From Equations \eqref{eq:nuW_inequality} and \eqref{eq:nutpmrU_estimation} it follows that
    \modifica{
    \begin{equation}
        \ln(t\,\nu_U^i)\!+\!1\!-\!\frac{r}{t}\!+\!\order{t^{-2}}
        \!\leq\! g\qty(\nu_W^i\!-\!\frac12)
        \!\leq\! \ln(t\,\nu_U^i)\!+\!1\!+\!\frac{r}{t}+\order{t^{-2}}\,.
    \end{equation}
    }
    In the limit of $t$ tending to infinity, it holds
    \modifica{
    \begin{equation}\label{eq:nuW_estimation}
        g\qty(\nu_W^i-\frac12) \sim \ln(t\,\nu_U^i)+1\,,
        \qquad\text{for}\qquad
        t\to\infty\,.
    \end{equation}
    }
    A comparison of Equation \eqref{eq:nuW_estimation} with Equation \eqref{eq:nutU_estimation} yields the thesis, \emph{i.e.},
    \begin{equation}
        g\qty(\nu_W^i-\frac12) \sim g\qty(\nu_{t\,U}^i-\frac12)\,,
        \qquad\text{for}\qquad
        t\to\infty\,.
    \end{equation}
\end{proof}

\begin{lemma}\label{lem:autoval-simpl}
    The symplectic eigenvalues of the covariance matrix of any pure bosonic Gaussian state are all equal to $1/2$\,.
\end{lemma}
\begin{proof}
    It is well known that any pure bosonic Gaussian state can be written in the form $\opomega=\opD_\r\opU_S\ketbra{\0}\opU_S^\dagger\opD_\r^\dagger$, for appropriate symplectic unitaries $\opU_S$ and translation operators $\opD_\r$. Since the covariance matrix of a quantum state is invariant under translation operators, and since the symplectic eigenvalues are invariant under symplectic transformations, every pure bosonic Gaussian state has the same symplectic eigenvalues of the vacuum state, \emph{i.e.}, all equal to $1/2$. See also \cite{serafini2017quantum}.
\end{proof}

\begin{lemma}\label{lem:indet}
    A positive-definite real matrix $\sigma$ satisfies the uncertainty principle $\sigma\geq\pm\frac{i}{2}\Delta$ if and only if all its symplectic eigenvalues $\nu_k$ are greater than or equal to $1/2$\,.
\end{lemma}
\begin{proof}
    See \cite[Section 3.4]{serafini2017quantum}.
\end{proof}

\begin{lemma}\label{lem:autoval-simpl-vs-autoval}
    Let's consider a matrix $\sigma\in\mathbb{R}^{2n\times2n}_{>0}$ satisfying $\sigma\geq\pm\frac{i}{2}\Delta$ and with symplectic eigenvalues $\qty{\nu_k}_{k=1,\ldots,n}$. The rank of $\sigma\pm\frac{i}{2}\Delta$
    is equal to $n$ plus the number of $\nu_k>1/2$.
\end{lemma}
\begin{proof}
    Since rank is an invariant under matrix equivalence, we have that $\rank(\sigma\pm\frac{i}{2}\Delta)=\rank(S\sigma S^T\pm\frac{i}{2}\Delta)$, where $S$ is the symplectic matrix provided by Williamson's theorem that brings the matrix $\sigma$ into its normal form. Being $S\sigma S^T\pm\frac{i}{2}\Delta$ a Hermitian matrix, and therefore a diagonalizable matrix, its rank is equal to the number of its non-zero eigenvalues. Thus the rank of the matrix $\sigma\pm\frac{i}{2}\Delta$ is equal to the non-zero eigenvalues of the matrix
    \begin{equation}
        \bigoplus_{k=1}^n\mqty(\nu_k&\pm i/2\\ \mp i/2&\nu_k)\,,
    \end{equation}
    which are $\qty{\nu_k\pm\frac12}_{k=1,\ldots,n}$. \lref{lem:autoval-simpl} ensures that $\nu_k\geq1/2\ \forall k$, then it holds
    \begin{equation}
        \begin{split}
            \rank(\sigma\pm\tfrac{i}{2}\Delta) &= \rank(S\sigma S^T\pm\tfrac{i}{2}\Delta)\\
            &= \#\qty{\nu_k\,:\,\nu_k+\tfrac12\neq 0}_k + \#\qty{\nu_k\,:\,\nu_k-\tfrac12\neq 0}_k\\
            &= n + \#\qty{\nu_k\,:\,\nu_k>\tfrac12}_k\,,
        \end{split}
    \end{equation}
    with $k=1,\ldots,n$.
\end{proof}

\begin{lemma}\label{prop:positività-blocchi}
    For any symmetric matrix $M$ of the form
    \begin{equation}
        M = \mqty(A&B\\B^T&C)
    \end{equation}
    if $C$ is invertible then the following property hold: $M > 0$ iff $C > 0$ and $A - BC^{-1}B^T > 0$.
\end{lemma}
\begin{proof}
    See \cite[Proposition 2.1]{gallier2019the}.
\end{proof}

\begin{lemma}[Guttman rank additivity formula]\label{prop:Schur}
    Let
    \begin{equation}
        M = \left(
        \begin{array}{cc}
            A & B \\
            C & D \\
        \end{array}
        \right)
    \end{equation}
    be a square matrix such that $D$ is invertible and $\mathrm{rank}\, M = \mathrm{rank}\, D$.
    Then,
    \begin{equation}
        A = B\,D^{-1}\,C\,.
    \end{equation}
\end{lemma}
\begin{proof}
    See \cite{zhang2011matrix}.
\end{proof}

\begin{lemma}[canonical decomposition]\label{lem:decomp-canon}
    Given an anti-symmetric matrix $\Delta'\in\RR^{2n\times2n}$ , there exists an orthogonal matrix $O'\in\RR(2n)$ and a set of values $d_k\in\RR$ such that
    \begin{equation}\label{eq:decomp-canon}
        O'\Delta'O'^T=\bigoplus_{k=1}^n d_k^{-1} \Delta_2\,,
        \quad\text{with}\quad
        \Delta_2=\mqty(0&1\\-1&0)\,.
    \end{equation}
\end{lemma}
\begin{proof}
    See \cite[Section 3.2.3]{serafini2017quantum}.
\end{proof}

\begin{lemma}\label{lem:SimplOrtogMatr}
    The symplectic matrix $S$ of \tref{teo:Williamson} is unique up to left multiplication by a block diagonal orthogonal symplectic matrix where each block acts on an eigenspace of the matrix $D$, \emph{i.e.}, $S\to O\,S$ with $O=\bigoplus_{j=1}^{k\leq n}O_j$.
\end{lemma}
\begin{proof}
    Let $h\in\RR^{2n\times2n}_{>0}$ and let $S$ and $S'$ be two symplectic matrices such that $D = S\,h\,S^T = S'\,h\,S'^T$, where the matrix $D$ is defined as $D\coloneqq\text{diag}\qty(\nu_1 I_2,\ldots,\nu_n I_2)$\,. It is easy to verify that
    \begin{equation}
        D = X\,D\,X^T
        ,
    \end{equation}
    where we defined the symplectic matrix $X \coloneqq S'S^{-1}$.
    Since $D$ is invertible, the equation above is equivalent to
    \begin{equation}
        \qty(D^{-1/2}\,X\,D^{1/2}) \qty(D^{-1/2}\,X\,D^{1/2})^T = I_{2n}\,,
    \end{equation}
    $O \coloneqq D^{-1/2}\,X\,D^{1/2}$ is thus an orthogonal matrix and $X$ can be written as
    \begin{equation}\label{eq:XDOX}
        X = D^{1/2}\,O\,D^{-1/2}\,.
    \end{equation}
    Substituting this expression into the condition for $X$ to be symplectic, \emph{i.e.}, $X\,\Delta\,X^T=\Delta$\,, we have
    \begin{equation}\label{eq:B19}
        D^{-1/2}\,\Delta\,D^{-1/2} = O\,D^{-1/2}\,\Delta\,D^{-1/2}\,O^T\,,
    \end{equation}
    from which it holds
    \begin{equation}
    \begin{split}
        D^{-1/2}\,\Delta\,D^{-1/2} 
        &= \bigoplus_{j=1}^n \frac{1}{\nu_j} \Delta_2\\
        &= \qty(\bigoplus_{i=1}^n u) \qty(\bigoplus_{i=1}^n \frac{1}{\nu_j} d) \qty(\bigoplus_{i=1}^n u)^{\!\dagger}\\
        &=: U\,\tilde{D}\,U^\dagger
        ,
    \end{split}
    \end{equation}
    where we used that being $\Delta_2\coloneqq\mqty(0&1\\-1&0)$ an antisymmetric matrix, and therefore normal, by the spectral theorem is diagonalisable by a unitary matrix $u$ and, defined $U\coloneqq\qty(\bigoplus_{i=1}^n u)$ and $\tilde{D}\coloneqq\qty(\bigoplus_{i=1}^n \frac{1}{\nu_j} d)$ \footnote{Explicitly: $d=\frac{i}{2}\,Z_2$ and $u = S^\dagger H$, where $S = \mqty(1&0\\0&i)$ and $H=\mqty(1&1\\1&-1)$\,.}, Equation \eqref{eq:B19} then becomes
    \begin{equation}
        D^{-1/2}\,\Delta\,D^{-1/2} 
        = U\,\tilde{D}\,U^\dagger = (O\,U)\,\tilde{D}\,(O\,U)^\dagger\,.
    \end{equation}
    This equation tells us that we have two spectral decompositions of matrix $D^{-1/2}\,\Delta\,D^{-1/2}$. The eigenvectors associated with distinct eigenvalues are orthogonal to each other and two orthogonal sets of eigenvectors associated with the same eigenvalue are mapped into each other via a unitary matrix. Then there exists a block diagonal matrix $V=\oplus_{j=1}^{k\leq n}V_j$, where the blocks $V_j$ are unitary matrices of the same size as the blocks of $\tilde{D}$ proportional to the identity, such that $O\,U = V\,U$. This implies that $V = O$ and thus, from Equation \eqref{eq:XDOX},
    \begin{equation}
        X = O = \oplus_{j=1}^{k\leq n}V_j\,,
    \end{equation}
    implying
    \begin{equation}
        S' = O\,S\,.
    \end{equation}
\end{proof}

\begin{lemma}\label{lem:lemB2}
    Given a $2n\times2n$ anti-symmetric non-singular real matrix $B$ and let $\pm b_1i,\ldots,\pm b_ni$ be its eigenvalues and $\vb{u}_1+i\vb{v}_1,\ldots,\vb{u}_n+i\vb{v}_n$ the $n$ orthogonal eigenvectors corresponding to the eigenvalues $b_1i,\ldots,b_ni$. Then the orthogonal matrix
    \begin{equation}
        O = \begin{pmatrix}\begin{array}{c|c|c|c|c}\dfrac{\vb{u}_1}{\norm{\vb{u}_1}} & 	\dfrac{\vb{v}_1}{\norm{\vb{v}_1}} & \cdots & \dfrac{\vb{u}_n}{\norm{\vb{u}_n}} & \dfrac{\vb{v}_n}{\norm{\vb{v}_n}}\end{array}\end{pmatrix}
    \end{equation}
    is such that
    \begin{equation}
        O^T\,B\,O = \bigoplus_{j=1}^n \mqty(0&b_j\\-b_j&0)\,.
    \end{equation}
\end{lemma}
\begin{proof}
    See \cite{zumino1962normal,becker1973transformation,greub2012linear}.
\end{proof}

\begin{remark}\label{rem:rem6.1}
    It is easy to see that for each non-singular matrix $A$ with $\det A=b^{-1}$ it is true that
    \begin{equation}
        A^T\,\mqty(0&b\\-b&0)\,A = \mqty(0&1\\-1&0)\,.
    \end{equation}
\end{remark}
\begin{lemma}\label{lem:6.2}
    Let the matrices $B$, $O$ and $A_i$ be as in \lref{lem:lemB2} and \remref{rem:rem6.1}, defining $A=\bigoplus_{i=1}^n A_i$ and $M=O\,A$\,, then
    \begin{equation}
        M^T\,B\,M = \bigoplus_{j=1}^n \mqty(0&1\\-1&0)\,.
    \end{equation}
\end{lemma}
\begin{proof}
    This is a direct consequence of \lref{lem:lemB2} and \remref{rem:rem6.1}.
\end{proof}

\subsection[Entropic functionals]{Entropic functionals in quantum communication}

In this subsection we show some fundamental entropic inequalities. In the following, for the sake of simplicity, the Von Neumann entropy of a state $\op\rho_A$ will be denoted as $S(A)\coloneqq S(\op\rho_A)$. By analogy with the classical case, we state the following definitions \cite{nielsen2000quantum}.
\begin{definition}[quantum conditional entropy]
    The \emph{quantum conditional entropy} is defined as:
    \begin{equation}
        S(A|B) = S(AB) - S(B)\,.
    \end{equation}
\end{definition}
\begin{definition}[quantum mutual information]
    The \emph{quantum mutual information} is defined as:
    \begin{equation}
        I(A:B) = S(A) + S(B) - S(AB)\,,
    \end{equation}
    which is a non-negative quantity.
\end{definition}

Exploiting the strong subadditivity for the Von Neumann entropy, it is possible to prove the following proposition \cite[Chapter 4]{khatri2020principles}.
\begin{proposition}\label{prop:5.1}
    The following relationships hold.
    \begin{enumerate}[label=\roman*.]
        \item Conditioning always decreases entropy,
        \begin{equation}
            S(A|BC)\leq S(A|C),S(A|B)\,.
        \end{equation}
        \item Discarding systems always decreases the quantum mutual information,
        \begin{equation}\label{eq:2.89}
            I(A:BC) \geq I(A:B)\,.
        \end{equation}
    \end{enumerate}
\end{proposition}

The independence condition \eqref{eq:3.2} for $K$ quantum systems is fully equivalent to the condition \eqref{eq:1.10}. This can be formally expressed by the following
\begin{lemma}\label{lem:IndipSyst}
    Let us consider $K$ quantum systems $\{X_i\}_{i=1,\ldots,K}$, a quantum system $M$ and the set of indices  $\mathcal{K}=\qty{1,\ldots,K}$. Then
    \begin{equation}
        \begin{split}
        &S(A_1\ldots A_K|M) = \sum_{k=1}^K S(A_k|M)\\
        &\iff
        I(X_\mathcal{A}:X_\mathcal{B}|M)=0\ \forall \mathcal{A},\mathcal{B}\subseteq\mathcal{K}\text{ s.t. }\mathcal{A}\cap\mathcal{B}=\emptyset\,,
        \end{split}
    \end{equation}
    where, given a subset of indices $\mathcal{D}=\qty{j_1,\ldots,j_r}\subset\mathcal{K}$, $X_\mathcal{D}$ is defined as $X_\mathcal{D}\coloneq X_{j_1}\ldots X_{j_r}$.
\end{lemma}
\begin{proof}
    Let us prove the two implications separately. First of all, let's notice that
    \begin{equation}\label{eq:B16}
        \begin{split}
        &S(A_1\ldots A_K|M) = \sum_{k=1}^K S(A_k|M)\\
        &\iff
        S(X_\mathcal{C}|M) = \sum_{k=1}^{\abs{\mathcal{C}}} S(X_{l_k}|M)\ \forall \mathcal{C}\subset\mathcal{K}
        \,.
        \end{split}
    \end{equation}
    The implication ``$\Longleftarrow$'' is obvious by considering $\mathcal{C}=\mathcal{K}$\,. The implication ``$\implies$'' is a consequence of the following.
    \begin{equation}
    \begin{split}
        \sum_{k\in\mathcal{C}} S(A_k|M) + \sum_{k\in\mathcal{K}\setminus\mathcal{C}} S(A_k|M)
        &= S(A_1\ldots A_K|M)\\
        &\leq S(X_\mathcal{C}|M) + S(X_{\mathcal{K}\setminus\mathcal{C}}|M)
        \,,
    \end{split}
    \end{equation}
    whence
    \begin{equation}
    \begin{split}
        0&\leq\qty(\sum_{k\in\mathcal{C}} S(A_k|M) - S(X_\mathcal{C}|M)) +\\
        &+\qty(\sum_{k\in\mathcal{K}\setminus\mathcal{C}} S(A_k|M) - S(X_{\mathcal{K}\setminus\mathcal{C}}|M)) \leq 0\,,
    \end{split}
    \end{equation}
    from which the second equation in \eqref{eq:B16}.
    
    \paragraph{``$\implies$''} Let's consider two disjoint subsets $\mathcal{A}=\qty{i_1,\ldots,i_m},\mathcal{B}=\qty{j_1,\ldots,j_r}$ of the set of indexes $\mathcal{K}$, then
    \begin{equation}
        \begin{split}
            &I(X_\mathcal{A}:X_\mathcal{B}|M) \\
            &= S(X_\mathcal{A}|M) + S(X_\mathcal{B}|M) - S(X_\mathcal{A}X_\mathcal{B}|M) \\
            &= \sum_{q=1}^{m} S(X_{i_q}|M) + \sum_{s=1}^{r} S(X_{j_s}|M) +\\
            &\phantom{=}- \qty(\sum_{q=1}^{m} S(X_{i_q}|M) + \sum_{s=1}^{r} S(X_{j_s}|M)) \\
            &= 0\,.
        \end{split}
    \end{equation}
    
    \paragraph{``$\Longleftarrow$''} Given $\mathcal{C}=\qty{j_1,\ldots,j_r}\subset\mathcal{K}$, let's define $\mathcal{A}=\qty{j_1,\ldots,j_{r-1}}$ and $\mathcal{B}=\qty{j_r}$. Then
    \begin{equation}
        \begin{split}
            S(X_\mathcal{A}:X_{j_r}|M) &= S(X_\mathcal{A}|M) + S(X_{j_r}|M) - I(X_\mathcal{A}:X_{j_r}|M) \\
            &= S(X_\mathcal{A}|M) + S(X_{j_r}|M)\,.
        \end{split}
    \end{equation}
    By applying this decomposition recursively, we obtain exactly what we want to prove.
\end{proof}

Under the action of quantum channels, the following proposition holds true \cite[Chapter 4]{khatri2020principles}.
\begin{proposition}[data-processing inequality]\label{prop:2.8}
    Given the quantum channel $\Phi_A\otimes\1_B:AB\to A'B$, the following inequality holds
    \begin{equation}\label{eq:2.96}
        I(A':B) \leq I(A:B)\,.
    \end{equation}
\end{proposition}
\begin{remark}
    In this case it is sufficient that $\op\rho_{AB}$ is an extension of $\op\rho_A$.
\end{remark}
This therefore says that the quantum mutual information between two systems $A$ and $B$ decreases under the action of a quantum channel $\Phi$ on one of the two systems.

\begin{lemma}\label{lem:heat-evol}
    Given two bosonic Gaussian states $\oprho^\gamma$ and $\oprho^\beta$, with the same first moments and covariance matrices $\gamma$ and $\beta$ respectively, with $\gamma\geq\beta$, it follows that the state $\oprho^\gamma$ can be written as
    \begin{equation}\label{eq:CombLin}
        \oprho^\gamma = \int_{\mathbb{R}^n} \opD_\x\,\oprho^\beta\,\opD_\x^\dagger\, \dd{\mu_{\mathds{X}}(\x)}\,,
    \end{equation}
    where $\mathds{X}$ is a centered Gaussian random variable with covariance matrix equal to $\gamma-\beta$.
\end{lemma}
\begin{proof}
    Knowing that $\oprho^\beta$ is a bosonic Gaussian state and that a linear combination of Gaussian states is still a Gaussian state, we can conclude that the RHS of Equation \eqref{eq:CombLin} is still a Gaussian state. Moreover, knowing that Gaussian states are uniquely determined by their first and second moments (\pref{prop:RSigmaGaussianStates}), in order to verify equality \eqref{eq:CombLin} it is sufficient to verify that the states at the LHS and RHS have the same first moments and covariance matrices.
    Assuming $\vb{r}(\oprho^\gamma) = \vb{r}(\oprho^\beta) = \bar{\vb{r}}$, for the first moments we have:
    \begin{equation}\label{eq:gamma-first-moments}
        \begin{split}
            \vb{r}\qty(\int_{\mathbb{R}^n} \opD_\x\,\oprho^\beta\,\opD_\x^\dagger\, \dd{\mu_{\mathds{X}}(\x)})
            &= \int_{\mathbb{R}^n} \vb{r}\qty(\opD_\x\,\oprho^\beta\,\opD_\x^\dagger) \dd{\mu_{\mathds{X}}(\x)}\\
            &= \int_{\mathbb{R}^n} \qty(\bar{\vb{r}}+\vb{x}) \dd{\mu_{\mathds{X}}(\x)}\\
            &= \bar{\vb{r}}\,,
        \end{split}
    \end{equation}
    where we used the linearity of the trace, the property described in \secref{sec:prel-SimplGroup} and the normalisation and symmetry of the Gaussian integral. Equation \eqref{eq:gamma-first-moments} therefore tells us that the states at the RHS and LHS of Equation \eqref{eq:CombLin} have the same first moments. Exploiting the same properties, for the second moments we have
    \modifica{
    \begin{equation}
        \begin{split}
            \int_{\mathbb{R}^n}&\tr\!\qty\Big[\qty(\opR_i-r_i) \,\opD_\x\,\oprho^\beta\,\opD_\x^\dagger\, \qty(\opR_j-r_j)]\dd{\mu_{\mathds{X}}(\x)} \\
            &\phantom{m} = \int_{\mathbb{R}^n}\tr\!\qty\Big[\qty(\opR_i-r_i+x_i) \,\oprho^\beta\, \qty(\opR_j-r_j+x_j)]\dd{\mu_{\mathds{X}}(\x)}\\
            &\phantom{m} = \int_{\mathbb{R}^n}\tr\!\qty\Big[\qty(\opR_i-r_i) \,\oprho^\beta\, \qty(\opR_j-r_j) + x_i\,x_j\,\oprho^\beta]\dd{\mu_{\mathds{X}}(\x)}\\
            &\phantom{m} = \int_{\mathbb{R}^n}\tr\!\qty\Big[\qty(\opR_i-r_i) \,\oprho^\beta\, \qty(\opR_j-r_j)]\dd{\mu_{\mathds{X}}(\x)} +\\
            &\phantom{m=} + \int_{\mathbb{R}^n} x_ix_j \dd{\mu_{\mathds{X}}(\x)}\\
            &\phantom{m} = \int_{\mathbb{R}^n}\tr\!\qty\Big[\qty(\opR_i-r_i) \,\oprho^\beta\, \qty(\opR_j-r_j)]\dd{\mu_{\mathds{X}}(\x)} + \gamma_{ij}-\beta_{ij}\,,
        \end{split}
    \end{equation}
    }
    where we used that the fact that the covariance matrix of the random variable $\mathds{X}$, with zero mean, is written as $\gamma_{ij}-\beta_{ij} = \mathbb{E}\qty[x_i\,x_j] = \int_{\mathbb{R}^n} x_ix_j \dd{\mu_{\mathds{X}}(\x)}$. Finally then, the state covariance matrix of the state to the RHS of Equation \eqref{eq:CombLin} is
    \modifica{
    \begin{equation}\label{eq:gamma-second-moments}
        \begin{split}
            \sigma&\qty(\int_{\mathbb{R}^n} \opD_\x\,\oprho^\beta\,\opD_\x^\dagger\, \dd{\mu_{\mathds{X}}(\x)}) \\
            &\phantom{m} = \frac12\int_{\mathbb{R}^n}\tr\Bigl[\qty(\opR_i-r_i) \,\opD_\x\,\oprho^\beta\,\opD_\x^\dagger\, \qty(\opR_j-r_j) +\\
            &\phantom{m=} + \qty(\opR_j-r_j) \,\opD_\x\,\oprho^\beta\,\opD_\x^\dagger\, \qty(\opR_i-r_i)\Bigr]\dd{\mu_{\mathds{X}}(\x)}\\
            &\phantom{m} = \frac12\Biggl(\int_{\mathbb{R}^n}\tr\Bigl[\qty(\opR_i-r_i) \,\oprho^\beta\, \qty(\opR_j-r_j) +\\
            &\phantom{m=} + \qty(\opR_j-r_j) \,\oprho^\beta\, \qty(\opR_i-r_i)\Bigr]\dd{\mu_{\mathds{X}}(\x)}\Biggr) + \gamma_{ij}-\beta_{ij}\\
            &\phantom{m} = \gamma_{ij}\,.
        \end{split}
    \end{equation}
    }
    Equation \eqref{eq:gamma-second-moments} thus proves that the states at the RHS and LHS of Equation \eqref{eq:CombLin} also have the same second moments.
\end{proof}

\begin{lemma}\label{lem:Hbar-monotonicity}
    Given a bosonic Gaussian channel $\Phi:A\to BCD$, defined by the matrices $(K,\alpha)$, the quantity $\bar{S}(\gamma) \coloneqq S(B|C)_{\Phi\qty(\oprho^\gamma_{A})}+S(B|D)_{\Phi\qty(\oprho^\gamma_{A})}$ is increasing in the covariance matrix $\gamma$ of the Gaussian state $\oprho^\gamma_{A}$.
\end{lemma}
\begin{proof}
    From \lref{lem:CompDisp}, \lref{lem:heat-evol}, the linearity of quantum channels and exploiting the concavity and invariance under local unitaries of the conditional entropy, it holds
    \begin{equation}
        \begin{split}
            \bar{S}\qty\Big(\Phi\qty\big(\oprho^\gamma_{A}))
            &= \bar{S}\qty(\Phi\qty(\int_{\mathbb{R}^n} \opD_\x\,\oprho^\beta_{A}\, \opD_\x^\dagger\, \dd{\mu_{\mathds{X}}(\x)}))\\
            &= \bar{S}\qty(\int_{\mathbb{R}^n} \Phi\qty\Big(\opD_\x\,\oprho^\beta_{A}\, \opD_\x^\dagger) \dd{\mu_{\mathds{X}}(\x)})\\
            &\geq \int_{\mathbb{R}^n} \bar{S}\qty(\Phi\qty\Big(\opD_\x\,\oprho^\beta_{A}\, \opD_\x^\dagger))\dd{\mu_{\mathds{X}}(\x)}\\
            &= \int_{\mathbb{R}^n} \bar{S}\qty\Big(\opD_{K\,\x}\,\Phi\qty(\oprho^\beta_{A})\,\opD_{K\,\x}^\dagger)\dd{\mu_{\mathds{X}}(\x)}\\
            &= \int_{\mathbb{R}^n} \bar{S}\qty\Big(\Phi\qty(\oprho^\beta_{A}))\dd{\mu_{\mathds{X}}(\x)}\\
            &= \bar{S}\qty\Big(\Phi\qty(\oprho^\beta_{A}))
            \,,
        \end{split}
    \end{equation}
    where $\oprho^\beta_{A}$ and $\mathds{X}$ are defined as in \lref{lem:heat-evol}.
\end{proof}

\begin{lemma}\label{lem:H(ABC)}
    Given a Gaussian state $\oprho^\gamma_{ABC}$ of the tripartite bosonic quantum system $ABC$ with covariance matrix $\gamma$, the function
    \begin{equation}
        \bar{S}(\gamma) \coloneqq S(A|B)_{\oprho^\gamma_{ABC}} + S(A|C)_{\oprho^\gamma_{ABC}}
    \end{equation}
    is increasing in the covariance matrix, \emph{i.e.}, $\gamma\geq\beta$ implies $\bar{S}(\gamma)\geq\bar{S}(\beta)$.
\end{lemma}
\begin{proof}
    This is a direct consequence of \pref{lem:Hbar-monotonicity}, assuming the bosonic Gaussian channel $\Phi:ABC\to ABC$ being the identity.
\end{proof}

\begin{proposition}\label{prop:GaussStMax}
    Let $A$, $B$ and $C$ be three bosonic quantum systems and let $\Phi:A\to BC$ be a bosonic Gaussian channel.
    Let $\oprho_A$ be a state of $A$ with finite average energy, and let $\opomega_A$ be the Gaussian state with the same first and second moments as $\oprho_A$.
    Let $\opomega_{BC}\coloneqq\Phi_{A\to BC}(\opomega_A)$ and $\oprho_{BC}\coloneqq\Phi_{A\to BC}(\oprho_A)$.
    Then,
    \begin{equation}
        S(B|C)_{\opomega_{BC}} \geq S(B|C)_{\oprho_{BC}}\,.
    \end{equation}
\end{proposition}
\begin{proof}
    We first verify that for each density operator $\oprho_A$ with finite average energy and first and second moments equal to those of a Gaussian state $\opomega_A$ it holds
    \begin{equation}\label{eq:relatEntr}
        S(B|C)_{\opomega_{BC}} - S(B|C)_{\oprho_{BC}} = S(\oprho_{BC}||\opomega_{BC}) - S(\oprho_C||\opomega_C)\,.
    \end{equation}
    From the definitions of these quantities, it is easy to verify that this equality \eqref{eq:relatEntr} is verified if and only if
    \begin{equation}
        \tr[\qty(\oprho_{BC}-\opomega_{BC})\qty(-\ln\opomega_{BC}-\I_B\otimes\ln\opomega_C)] = 0\,.
    \end{equation}
    This condition, being $\oprho_{BC}-\opomega_{BC}=\Phi(\oprho_A-\opomega_A)$, can be rewritten as
    \begin{equation}
        \tr[\qty(\oprho_A-\opomega_A)\Phi^\dagger_{A\to BC}\qty(-\ln\opomega_{BC}-\I_B\otimes\ln\opomega_C)] = 0\,.
    \end{equation}
    This means that the initial states $\oprho_A$ and $\opomega_A$ must have the same energy with respect to the Hamiltonian $\Phi^\dagger_{A\to BC} \qty(-\ln\opomega_{BC} - \I_B \otimes \ln\opomega_C)$. Since this Hamiltonian is a quadratic polynomial in the quadratures and since $\oprho_A$ and $\opomega_A$ have the same first and second moments then the equation above is verified.
    To conclude then
    \begin{equation}
        S(B|C)_\opomega - S(B|C)_\oprho = S(\oprho_{BC}||\opomega_{BC}) - S(\oprho_C||\opomega_C) \geq 0\,,
    \end{equation}
    where we have exploited the monotonicity of the relative entropy. Hence the thesis follows.
\end{proof}

\begin{proposition}\label{prop:K-alpha}
    Given the covariance matrix $\sigma_E\in\mathbb{R}^{2n\times2n}_{>0}$ of the environment state and the symplectic matrix $S_{SE} \coloneqq \mqty(S_{SS}&S_{SE}\\S_{ES}&S_{EE})\in\text{Sp}(4n,\mathbb{R})$ of the coupling describing a bosonic Gaussian channel, the matrices describing the channel are given by $K = S_{SS}$ and $\alpha = S_{SE}\,\sigma_E\,S_{SE}^T$.
\end{proposition}
\begin{proof}
    Knowing that the symplectic matrix describing the unitary $\opU_{SE}$ and acting on the system $SE$ is given by
    \begin{equation}
	S_{SE} \coloneqq \mqty(S_{SS}&S_{SE}\\S_{ES}&S_{EE})\,,
    \end{equation}
    and that the covariance matrix of the initial joint state is given by $\sigma_{SE}=\sigma_S\oplus\sigma_E$, where $\sigma_E$ is the covariance matrix of the environment state, the symplectic transformation acts on the overall state covariance matrix as
    \begin{equation}
	\begin{split}
            S\,\sigma_{SE}\,S_{SE}^T
            &= \mqty(S_{SS}&S_{SE}\\S_{ES}&S_{EE}) \, \mqty(\sigma_{S}&0\\0&\sigma_{E}) \, \mqty(S_{SS}^T&S_{ES}^T\\S_{SE}^T&S_{EE}^T) \\
            &= \mqty(S_{SS}\,\sigma_S\,S_{SS}^T + S_{SE}\,\sigma_E\,S_{SE}^T & * \\ * & * )
		\,.
	\end{split}
    \end{equation}
    From the equation above, we therefore have that the channel acts on the matrix $\sigma_S$ as
    \begin{equation}
	\begin{split}
            \sigma_S \longrightarrow & K\,\sigma_S\,K^T + \alpha \\
            =\,& S_{SS}\,\sigma_S\,S_{SS}^T + S_{SE}\,\sigma_E\,S_{SE}^T\,,
	\end{split}
    \end{equation}
    from which we derive that
    \begin{equation}\label{eq:K_alpha_from_S}
        K \coloneqq S_{SS}\,,
        \qquad
        \alpha \coloneqq
        S_{SE}\,\sigma_E\,S_{SE}^T\,.
    \end{equation}
\end{proof}

\begin{proposition}\label{lem:BeamSplitter}
    Let us consider a beam splitter that uniformly combines the modes of the input states, \emph{i.e.}, the attenuation parameter is equal to $\eta$ for all modes. Then, applying a symplectic to one input of that beam splitter is equivalent to applying that symplectic to the output and its inverse to the other input. Let $\Phi^{\op\phi_E}_\eta(\oprho_A)$ be the channel described by a beam splitter with a uniform attenuation parameter $\eta$ that makes the state of the system $\oprho_A$ interact with the environment state $\op\phi_E$, then it holds
    \begin{equation}
        \Phi^{\op\phi_E}_\eta(\opU_S\,\oprho_A\,\opU_S^\dagger) = \opU_S\,\Phi^{\opU_S^\dagger\,\op\phi_E\,\opU_S}_\eta(\oprho_A)\,\opU_S^\dagger
        \,.
    \end{equation}
\end{proposition}
\begin{proof}
    The proof is straightforward and proceeds as follows.
    \begin{equation}
        \begin{split}
            &\Phi^{\op\phi_E}_\eta (\opU_S\,\oprho_A\,\opU_S^\dagger)\\
            &= \tr_E\!\qty[\opU_\eta\,\qty(\opU_S\,\oprho_A\,\opU_S^\dagger\otimes\op\phi_E)\,\opU_\eta^\dagger]\\
            &= \tr_E\!\qty[\opU_\eta\qty(\opU_S\otimes\opU_S)\qty(\oprho_A\otimes\opU_S^\dagger\,\op\phi_E\,\opU_S)\qty(\opU_S\otimes\opU_S)^\dagger\opU_\eta^\dagger]\\
            &= \tr_E\!\qty[\qty(\opU_S\otimes\opU_S) \,\opU_\eta\, \qty(\oprho_A\otimes\opU_S^\dagger\,\op\phi_E\,\opU_S) \,\opU_\eta^\dagger\, \qty(\opU_S\otimes\opU_S)^\dagger]\\
            &= \opU_S\tr_E\!\qty[\opU_\eta\, \qty(\oprho_A\otimes\opU_S^\dagger\,\op\phi_E\,\opU_S) \,\opU_\eta^\dagger]\,\opU_S^\dagger\\
            &= \opU_S\,\Phi^{\opU_S^\dagger\,\op\phi_E\,\opU_S}_\eta(\oprho_A)\,\opU_S^\dagger
            \,.
        \end{split}
    \end{equation}
\end{proof}

\begin{proposition}\label{prop:SEinvariance}
    The squashed entanglement is invariant under local unitary transformations, \emph{i.e.},
    \begin{equation}\label{eq:SEinvariance}
        E_\mathrm{sq}\qty(\oprho_{AB}) = E_\mathrm{sq}\qty(\qty(\opU_A\otimes\opU_B) \,\oprho_{AB}\, \qty(\opU_A\otimes\opU_B)^\dagger)
        \,.
    \end{equation}
\end{proposition}
\begin{proof}
    This is true since squashed entanglement is a entanglement measure, see \cite{vedral1997quantifying}.
\end{proof}
\section{Numerical simulations}\label{app:NumSim}

In this section, the analytical results of \secref{sec:simulations} are discussed.

\subsection[Extremality condition]{Extremality condition}\label{sec:TestExtremality}

In this subsection we verify that the channel defined in Equation \eqref{eq:canale_eta1eta2} fulfils the condition of being extreme, following the algorithm proposed in \secref{sec:ExtremalityCond}.
First, we derive the matrices $K$ and $\alpha$ that describe the channel we wish to treat.  The symplectic matrix describing the unitary $\opU_{SE}^{(\eta_1,\eta_2)}$ and acting on the system $SE$ is given by
\begin{equation}\label{eq:SSEeta12}
    S^{(\eta_1,\eta_2)}_{SE} \!=\!
    {\setlength{\arraycolsep}{1pt}
    \mqty(
    \sqrt{\eta_1}\,I_2 & 0 & \sqrt{1-\eta_1}\,I_2 & 0  \\
    0 & \sqrt{\eta_2}\,I_2 & 0 & \sqrt{1-\eta_2}\,I_2  \\
    -\sqrt{1-\eta_1}\,I_2 & 0 & \sqrt{\eta_1}\,I_2 & 0 \\
    0 & -\sqrt{1-\eta_2}\,I_2 & 0 & \sqrt{\eta_2}\,I_2
    )
    }
    \,,
\end{equation}
while the covariance matrix $\sigma_E$ of the environment state is the covariance matrix of the two-mode squeezed vacuum state $\op\phi_E^\kappa$,
\begin{equation}
    \sigma_E = \mqty(\qty(\kappa-\dfrac12)\,I_2&\sqrt{\kappa(\kappa-1)}\,Z_2\\ \sqrt{\kappa(\kappa-1)}\,Z_2&\qty(\kappa-\dfrac12)\,I_2)\,.
\end{equation}
From \alref{alg:ExtremalityCondition} above, we therefore have that the channel is described by the matrices
\begin{equation}
    K = \mqty(\sqrt{\eta_1}\,I_2 & 0 \\
               0 & \sqrt{\eta_2}\,I_2 )
\end{equation}
and
\begin{equation}
    \alpha \!=\!
    {\setlength{\arraycolsep}{-4pt}
    \mqty(
    \qty(\kappa\!-\!\dfrac12)\qty(1\!-\!\eta_1)\,I_2 & \sqrt{\kappa(\kappa\!-\!1)(1\!-\!\eta_1)(1\!-\!\eta_2)}\,Z_2 \\
    \sqrt{\kappa(\kappa\!-\!1)(1\!-\!\eta_1)(1\!-\!\eta_2)}\,Z_2 & \qty(\kappa\!-\!\dfrac12)\qty(1\!-\!\eta_2)\,I_2
    )
    }
    \,.
\end{equation}
Finally, it is easy to verify in our case that $-\qty(-\alpha\,\qty(\Delta-K\,\Delta\,K^T)^{-1})^2$ is equal to $I_4/4$\,.

\subsection{Lower bound}\label{sec:ExampleLowerBound}

In this section we apply the algorithm of \secref{sec:SE-lower-bound} to compute the lower bound of the extreme bosonic Gaussian channel $\mathcal{N}^{(\eta_1,\eta_2)}_{S\to S}$ we are considering. Knowing that the squashed entanglement of the channel is defined by taking the $\sup$ over all possible input states on a bipartite system of the squashed entanglement of the state obtained by applying the channel to the reduced state on one of the two initial systems, a simple way to find a lower bound to this quantity is to fix the input state. Let us then take as the input state the tensor product of $2$ two-mode squeezed vacuum states $\op\phi^{\En_1}_{A_1B_1}\otimes\op\phi^{\En_2}_{A_2B_2}$ on the bipartite system $AB$, with average number of photons $\En_1$ and $\En_2$ respectively. Let us then apply the considered channel $\N^{(\eta_1,\eta_2)}_{B\to D}$ on the system $B$ and compute the squashed entanglement of the resulting state. Knowing that the lower bound of the squashed entanglement of a state (in our case the channel's output state) is a function of the blocks of the symplectic matrix that diagonalises the covariance matrix of the considered state, we must find the symplectic matrix that diagonalises the covariance matrix of the channel's output state, which is given by
\begin{equation}
    \begin{split}
    &\qty(\1_A\otimes\N_B) \qty(\op\phi^{\En_1}_{A_1B_1}\otimes\op\phi^{\En_2}_{A_2B_2})\\
    &= \tr_E\qty[\opU^{(\eta_1,\eta_2)}_{BE} \qty\Big(\qty(\op\phi^{\En_1}\otimes\op\phi^{\En_2})_{AB} \otimes \op\phi^{\en_3}_E) {\opU_{BE}^{(\eta_1,\eta_2)}}^{\dagger}]
    \,.
    \end{split}
\end{equation}

First, we compute the covariance matrix $\sigma_{CD}$ of the output state $\qty(\1_A\otimes\N_B)\qty(\op\phi^{\En_1}\otimes\op\phi^{\En_2})_{AB}$ of the channel. In particular, the symplectic matrix associated with the unitary transformation $\I_A\otimes\opU^{(\eta_1,\eta_2)}_{BE}$ is given by
\begin{equation}
    S_{ABE}^{\vb*{\eta}} = \mqty(I_4&0\\0&S_{\vb*{\eta}})\,,
\end{equation}
where $S_{\vb*{\eta}}\coloneqq S^{(\eta_1,\eta_2)}_{BE}$ is defined in Equation \eqref{eq:SSEeta12} and the covariance matrix of the overall input state $\qty(\op\phi^{\En_1}\otimes\op\phi^{\En_2})_{AB} \otimes \op\phi^{\en_3}_E$ is given by
\begin{equation}
    \sigma_{ABE} = \sigma_{AB} \oplus \sigma_{E}\,,
\end{equation}
where
\begin{equation}
    \sigma_{AB} \!=\! 
    {\setlength{\arraycolsep}{-8pt}
    \mqty(\!
    \qty(\!\En_1\!+\!\frac12\!) I_2 & 0 & \sqrt{\En_1(\!\En_1\!+\!1\!)} Z_2 & 0 \\
    0 & \qty(\!\En_2\!+\!\frac12\!) I_2 & 0 & \sqrt{\En_2(\!\En_2\!+\!1\!)} Z_2 \\
    \sqrt{\En_1(\!\En_1\!+\!1\!)} Z_2 & 0 & \qty(\!\En_1\!+\!\frac12\!) I_2 & 0 \\
    0 & \sqrt{\En_2(\!\En_2\!+\!1\!)} Z_2 & 0 & \qty(\!\En_2\!+\!\frac12\!) I_2
    \!)
    }
\end{equation}
and
\begin{equation}\label{eq:sigmaE}
    \sigma_{E} = 
    \mqty(
    \qty(\en_3+\frac12)\,I_2 & \sqrt{\en_3(\en_3+1)}\,Z_2 \\
    \sqrt{\en_3(\en_3+1)}\,Z_2 & \qty(\en_3+\frac12)\,I_2
    )\,.
\end{equation}
The matrix $\sigma_{CD}$ is then derived by considering the appropriate submatrix of the covariance matrix of the overall channel's output state
\begin{equation}
    \sigma_{CDE} = S_{ABE}\,\sigma_{ABE}\,S_{ABE}
    \,.
\end{equation}
The matrix $\sigma_{CD}$ is derived in \aref{app:cov_mat_CD} and its explicit form can be found in Equation \eqref{eq:sigma_CD_explicit}. Now, from \alref{alg:LowerBound1} we derive the lower bound by applying Equation \eqref{eq:C_LB} to the symplectic matrix which diagonalises $\sigma_{CD}$\,.

\subsection{Upper bound}\label{sec:ExampleUpperBound}

The channel we are going to consider is the one in Equation \eqref{eq:canale_eta1eta2}. As it is not an easy task to optimise over all possible squashing channels, let us consider a specific family of squashing channels: a two-mode extreme bosonic Gaussian attenuator with transmissivities $(\tau_1,\tau_2)=(1/2,1/2)$ and the environment state a two-mode squeezed vacuum state, with average number of photons per single mode $\en_s$. For what is stated in \pref{prop:GaussStMax}, the optimal state $\oprho_{A}$ in Equation \eqref{eq:SqEntChannel} is a Gaussian state. Moreover, fixed the squashing channel, \lref{lem:H(ABC)} implies that the quantity $S(B|E'')_{\op\psi} + S(B|F')_{\op\psi}$ in Equation \eqref{eq:SqEntChannel} is an increasing function in the covariance matrix. This means that the supremum on the states can be taken by considering any sequence of states with covariance matrix whose eigenvalues go to infinite. For convenience we therefore choose a sequence of thermal states with covariance matrix
\begin{equation}
    \sigma_{A} =
        \mqty(
            \qty(t+\frac12)\,I_2 & 0 \\
            0 & \qty(t+\frac12)\,I_2
        )\,.
\end{equation}
Since, in the physical representation of the channel, the state of the environment we are going to consider is a two-mode squeezed vacuum state, with average number of photons per single mode $\en_3$, and since the environment on which the squashing channel acts is the two-mode squeezed vacuum state, the covariance matrix of the initial state in the system $AEF$ is given by $\sigma_{A}\oplus\sigma_{E}\oplus\sigma_{F}$, where
\begin{equation}
    \sigma_{E} = \mqty(
    \qty(\en_3+\frac12)\,I_2 & \sqrt{\en_3(\en_3+1)}\,Z_2\\
    \sqrt{\en_3(\en_3+1)}\,Z_2 & \qty(\en_3+\frac12)\,I_2
    )
\end{equation}
and
\begin{equation}
    \sigma_{F} = \mqty(
    \qty(\en_s+\frac12)\,I_2 & \sqrt{\en_s(\en_s+1)}\,Z_2\\
    \sqrt{\en_s(\en_s+1)}\,Z_2 & \qty(\en_s+\frac12)\,I_2
    )\,.
\end{equation}
The covariance matrix of the state $\op\psi_{BE''F'}$ is therefore given by
\begin{equation}\label{eq:sigmaBEF}
    \sigma_{BE''F'} = S_\ttt\,S_\ee\,\qty\Big(\sigma_{A}\oplus\sigma_{E}\oplus\sigma_{F})\,S_\ee^T\,S_\ttt^T\,,
\end{equation}
where $S_\ttt \coloneqq S^{(\tau_1,\tau_2)}_{E'F\to E''F'}$ and $S_\ee \coloneqq S^{(\eta_1,\eta_2)}_{AE\to BE'}$ are defined as
\begin{equation}
    S_\ee \!=\!
    {\setlength{\arraycolsep}{-2pt}
    \mqty(
        \sqrt{\eta_1}\,I_2 & 0 & \sqrt{1-\eta_1}\,I_2 & 0 & 0\phantom{mm} & 0 \\
        0 & \sqrt{\eta_2}\,I_2 & 0 & \sqrt{1-\eta_2}\,I_2\phantom{m} & 0\phantom{mm} & 0 \\
        -\sqrt{1-\eta_1}\,I_2 & 0 & \sqrt{\eta_1}\,I_2 & 0 & 0\phantom{mm} & 0 \\
        0 & -\sqrt{1-\eta_1}\,I_2 & 0 & \sqrt{\eta_1}\,I_2 & 0\phantom{mm} & 0 \\
        0 & 0 & 0 & 0 & I_2\phantom{mm} & 0 \\
        0 & 0 & 0 & 0 & 0\phantom{mm} & I_2
    )
    }
\end{equation}
and
\begin{equation}
    S_\ttt \!=\!
    {\setlength{\arraycolsep}{-2pt}
    \mqty(
        I_2\phantom{mm} & 0 & 0 & 0 & 0 & 0 \\
        0\phantom{mm} & I_2 & 0 & 0 & 0 & 0 \\
        0\phantom{mm} & 0 & \sqrt{\tau_1}\,I_2 & 0 & \sqrt{1-\tau_1}\,I_2 & 0 \\
        0\phantom{mm} & 0 & 0 & \sqrt{\tau_2}\,I_2 & 0 & \sqrt{1-\tau_2}\,I_2 \\
        0\phantom{mm} & 0 & \phantom{m}-\sqrt{1-\tau_1}\,I_2 & 0 & \sqrt{\tau_1}\,I_2 & 0 \\
        0\phantom{mm} & 0 & 0 & -\sqrt{1-\tau_1}\,I_2 & 0 & \sqrt{\tau_1}\,I_2
    )
    }
    \,.
\end{equation}
From the covariance matrix $\sigma_{BE''F'}$ in Equation \eqref{eq:sigmaBEF}, the covariance matrices of the marginal states on the systems $E''$, $F'$, $BE''$ and $BF'$ can be obtained. Simple calculations show that these covariance matrices are linear in the parameter $t$, \emph{i.e.}, they can be written as $t\,U+V$, where $U$ and $V$ are two constant symmetric matrices dependent on the other parameters involved. It is also possible to show that $U$ is a strictly positive matrix. In what follows we will be interested in the symplectic eigenvalues of those matrices, \emph{i.e.}, matrices of the form $t\,U+V$ where $U$ is strictly positive and $V$ symmetric. For the computation of the upper bound we are interested in functions of the type $g\qty(\nu_{t\,U+V}^i-\frac12)$ (see \alref{alg:UpperBound}), in the limit of $t$ going to infinity. For this purpose, the \pref{prop:g_function_symplectic_relation} assures us that, in this limit, we can restrict ourselves to considering only the part of $t\,U+V$ proportional to $t$, and thus the following matrices:
\begin{subequations}
    \begin{align}
        &\sigma_{E''} = \mqty(t\,\tau_1\,\bar{\eta}_1\,I_2 & 0 \\ 0 & t\,\tau_2\,\bar{\eta}_2\,I_2)\,,\\
        &\sigma_{F'} = \mqty(t\,\bar{\tau}_1\,\bar{\eta}_1\,I_2 & 0 \\ 0 & t\,\bar{\tau}_2\,\bar{\eta}_2\,I_2)\,,\\
        &\sigma_{BE''} = {\setlength{\arraycolsep}{-6pt}\mqty(t\,\eta_1 I_2 & 0 & -t\,\sqrt{\eta_1 \tau_1 \bar{\eta}_1} I_2 & 0\\
        0 & t\,\eta_2 I_2 & 0 & -t\,\sqrt{\eta_2 \tau_2 \bar{\eta}_2} I_2\\
        -t \sqrt{\eta_1 \tau_1 \bar{\eta}_1} I_2 & 0 & t\,\tau_1 \bar{\eta}_1 I_2 & 0\\
        0 & -t \sqrt{\eta_2 \tau_2 \bar{\eta}_2} I_2 & 0 & t\,\tau_2 \bar{\eta}_2 I_2)}\,,\\
        &\sigma_{BF'} = {\setlength{\arraycolsep}{-4pt}\mqty(t\,\eta_1\,I_2 & 0 & t\,\sqrt{\eta_1\,\bar{\tau}_1\,\bar{\eta}_1}\,I_2 & 0\\
        0 & t\,\eta_2\,I_2 & 0 & t\,\sqrt{\eta_2\,\bar{\tau}_2\,\bar{\eta}_2}\,I_2\\
        t\,\sqrt{\eta_1\,\bar{\tau}_1\,\bar{\eta}_1}\,I_2 & 0 & t\,\bar{\tau}_1\,\bar{\eta}_1\,I_2 & 0\\
        0 & t\,\sqrt{\eta_2\,\bar{\tau}_2\,\bar{\eta}_2}\,I_2 & 0 & t\,\bar{\tau}_2\,\bar{\eta}_2\,I_2)}\,,
    \end{align}
\end{subequations}
where, for typographical reasons, we have defined $\bar{\eta}=1-\eta$ and $\bar{\tau}=1-\tau$. The symplectic eigenvalues of these covariance matrices are $\nu^{\sigma_{E''}} = \qty{t\,\tau_1\bar{\eta}_1, t\,\tau_2\bar{\eta}_2}$, $\nu^{\sigma_{F'}} = \qty{t\,\bar{\tau}_1\bar{\eta}_1, t\,\bar{\tau}_2\bar{\eta}_2}$, $\nu^{\sigma_{BE''}} = \qty{0, 0, t\,\qty(\tau_1+\bar{\tau}_1\,\eta_1), t\,\qty(\tau_2+\bar{\tau}_2\,\eta_2)}$ and $\nu^{\sigma_{BF'}} = \qty{0, 0, t\,\qty(\bar{\tau}_1 + \tau_1\,\eta_1), t\,\qty(\bar{\tau}_2 + \tau_2\,\eta_2)}$. From the formula \eqref{eq:entropiaGauss} for the entropy of a bosonic Gaussian state, in the limit of infinite average number of photons $t$, it holds
\begin{equation}\label{eq:HH-funct}
    \begin{split}
    S(B|E'') + S(B|F') = \sum_{i=1}^2 &\Bigl[g\qty(t\,\qty(\tau_i + \bar{\tau}_i\,\eta_i)) - g\qty(t\,\tau_i\,\bar{\eta}_i) +\\
    &+ g\qty(t\,\qty(\bar{\tau}_i + \tau_i\,\eta_i))- g\qty(t\,\bar{\tau}_i\,\bar{\eta}_i)\Bigr]
    \,.
    \end{split}
\end{equation}
Having chosen $\tau_1=\tau_2=1/2$, an upper bound to the squashed entanglement of the channel $\N$, in the limit $t\to\infty$, is therefore
\modifica{
\begin{equation}
    \begin{split}
        E_\text{sq}(\mathcal{N}) &\leq \lim_{t\to+\infty} \sum_{i=1}^2 \qty[g\qty(t\,\frac{1+\eta_i}{2}) - g\qty(t\,\frac{1-\eta_i}{2})] \\
        &= \sum_{i=1}^2 \ln(\frac{1+\eta_i}{1-\eta_i}) \,.
    \end{split}
\end{equation}
}
\begin{remark}
    We note that in the limit we have considered, \emph{i.e.}, in which the states on which the quantum channel acts are a family of thermal states with the number of photons going to infinity, we have obtained a straightforward multi-mode extension of the result for the one mode pure-loss bosonic channel \cite[Appendix A]{takeoka2014squashed}.
\end{remark}

\section*{Acknowledgements}

GDP has been supported by the HPC Italian National Centre for HPC, Big Data and Quantum Computing - Proposal code CN00000013 - CUP J33C22001170001 and by the Italian Extended Partnership PE01 - FAIR Future Artificial Intelligence Research - Proposal code PE00000013 - CUP J33C22002830006 under the MUR National Recovery and Resilience Plan funded by the European Union - NextGenerationEU.
Funded by the European Union - NextGenerationEU under the National Recovery and Resilience Plan (PNRR) - Mission 4 Education and research - Component 2 From research to business - Investment 1.1 Notice Prin 2022 - DD N. 104 del 2/2/2022, from title ``understanding the LEarning process of QUantum Neural networks (LeQun)'', proposal code 2022WHZ5XH – CUP J53D23003890006.
GDP is a member of the ``Gruppo Nazionale per la Fisica Matematica (GNFM)'' of the ``Istituto Nazionale di Alta Matematica ``Francesco Severi'' (INdAM)''.

\bibliography{biblio}
\bibliographystyle{unsrt}

\begin{IEEEbiographynophoto}{Alessandro Falco} was born in Cuneo, Italy, in 1998. He received the B.S. and M.S. degrees in physics from the University of Pisa in 2020 and 2022, respectively, and the Diploma di Licenza degree in physics from Scuola Normale Superiore in 2023. From 2022, he holds an associate researcher position with the Technology Innovation Institute, Abu Dhabi.
\end{IEEEbiographynophoto}

\begin{IEEEbiographynophoto}{Giacomo De Palma} was born in Lanciano, Italy, in 1990. He received the B.S. and M.S. degrees in physics from the University of Pisa in 2011 and 2013, respectively, and the Diploma di Licenza and Ph.D. degrees in physics from Scuola Normale Superiore in 2014 and 2016, respectively. From 2016 to 2018, he held a post-doctoral research position with the University of Copenhagen. From 2018 to 2019, he was a Marie-Sklodowska Curie Fellow with the University of Copenhagen. From March 2019 to March 2021, he was a Post-Doctoral Associate with MIT. From March 2021 to December 2021, he was a Tenure-Track Assistant Professor in mathematical physics with Scuola Normale Superiore. In December 2021, he became Associate Professor in mathematical physics with the University of Bologna, where since October 2024 he is Full Professor. He is the author of 52 scientific articles. His research interests include all aspects of quantum information and computation. He is a member of the International Association of Mathematical Physics (IAMP), of the Italian Mathematical Union (UMI) and of the Istituto Nazionale di Alta Matematica “Francesco Severi” (INdAM). He was a recipient of the Best Italian Researcher in Denmark (BIRD) Award in 2018.
\end{IEEEbiographynophoto}

\vfill

\end{document}